\documentclass[useAMS,usenatbib,referee]{biom}

\usepackage{amsmath,bbm,amssymb}

\usepackage{filecontents}

\usepackage{xcolor}
\usepackage{graphicx, graphics}
\usepackage{algorithmic}
\usepackage{algorithm}
\usepackage{url}
\usepackage{enumitem}
\usepackage{psfrag,pst-node,rotating, amsmath, bbm, amssymb, setspace, picture, epsfig, amsfonts, upgreek}
\usepackage{mathtools}
\usepackage{helvet}
\usepackage[eulergreek]{sansmath}
\usepackage{booktabs}
\usepackage{enumitem}
\usepackage{ifthen}
\usepackage{tikz} 
\usetikzlibrary{shapes,backgrounds,shapes.misc, positioning,shapes.geometric,arrows,matrix,fit,calc,arrows.meta, trees, hobby}

\newcommand{\blind}{1}

\usepackage{caption}
\usepackage{subcaption}

\usepackage{multirow}

\usepackage{wrapfig}

\usepackage{arydshln}

\newtheorem{proposition}{Proposition}

\newcommand{\argmin}{\operatornamewithlimits{argmin}}

\def\coef{\boldsymbol \beta}

\def\bbeta{\boldsymbol \beta}

\def\bstheta{\boldsymbol \theta}
\def\balpha{\boldsymbol \alpha}

\def\bnu{\boldsymbol \nu}
\def\bxi{\boldsymbol \xi}

\def\bfX{\mathbf X}
\def\bsbeta{\boldsymbol \beta}

\def\bsz{\boldsymbol z}

\def\bsC{\boldsymbol C}

\def\bfZ{{\boldsymbol Z}}
\def\bfz{{\boldsymbol z}}
\def\bfa{{\boldsymbol a}}

\def\bfx{\mathbf x}

\def\wtbbeta{\widetilde{\boldsymbol \beta}}

\def\P{\mathrm{P}}

\def\vec{\mathrm{vec}}
\def\vecl{\mathrm{vecl}}

\def\transp{\top}
\def\transp{\mathsf{T}}

\def\gobblestop#1#2{#1}
\def\killstop{%
	\aftergroup\gobblestop
}

  \tikzset{
  mybackground/.style={execute at end picture={
        \begin{scope}[on background layer]
          \draw[draw=blue!10,fill=blue!10,rounded corners=1.5ex] (current bounding box.south west)
                    rectangle (current bounding box.north east);
          \node[draw,anchor=west,inner sep=1pt,minimum width=4ex] at (current bounding box.north
                   west){#1};
        \end{scope}
    }},
}

\tikzset{
    buffer/.style={
        isosceles triangle,
        isosceles triangle apex angle=66,
        shape border rotate=90,
        draw,
        thick,
        fill=blue!20,
        node distance=5cm,
        rounded corners=60pt,
        opacity=0.6,
        minimum height=6.25cm
    }
}

\pgfdeclarelayer{foreground} 
\pgfdeclarelayer{background}
   \pgfsetlayers{background,%
                 main,%
                 foreground%
                 }

\title{\vspace{-0.5in}\bf Sufficient Dimension Reduction for Populations with Structured Heterogeneity}
\date{}

\if1\blind
{
	\author{Jared D. Huling$^{*}$\email{huling@umn.edu}\\
		Division of Biostatistics, University of Minnesota, Minneapolis, Minnesota
		\and
		Menggang Yu$^{*}$\email{meyu@biostat.wisc.edu}\\
		Department of Biostatistics and Medical Informatics, 
		University of Wisconsin-Madison
} \fi

\begin{document}

\label{firstpage}

\begin{abstract}
	A key challenge in building effective regression models for large and diverse populations is accounting for patient heterogeneity. An example of such heterogeneity is in health system risk modeling efforts where different combinations of comorbidities fundamentally alter the relationship between covariates and health outcomes. Accounting for heterogeneity arising combinations of factors can yield more accurate and interpretable regression models. Yet, in the presence of high dimensional covariates, accounting for this type of heterogeneity can exacerbate estimation difficulties even with large sample sizes. To handle these issues, we propose a flexible and interpretable risk modeling approach based on semiparametric sufficient dimension reduction. The approach accounts for patient heterogeneity, borrows strength in estimation across related subpopulations to improve both estimation efficiency and interpretability, and can serve as a useful exploratory tool or as a powerful predictive model. In simulated examples, we show that our approach often improves estimation performance in the presence of heterogeneity and is quite robust to deviations from its key underlying assumptions. We demonstrate our approach in an analysis of hospital admission risk for a large health system and demonstrate its predictive power when tested on further follow-up data. 
\end{abstract}


\maketitle

\newpage

\section{Introduction}\label{sec:intro}

Sufficient dimension reduction (SDR) is a powerful tool for regression modeling with many covariates \citep{Li1991, cook2007fisher}. It assumes that the underlying relationship between covariates and outcome can be effectively modeled based on a few linear combinations of all covariates. The immediate goal is to estimate the dimension reduction subspace formed by these linear combinations. In this paper we focus on estimation of the conditional mean -- in this context, the dimension reduction subspace of modeling interest is called the central mean subspace \citep{Cook2002}. The recent book by \citet{BingLi2018} gives a comprehensive review for many attractive features of SDR, which include a high degree of flexibility in modeling in the form of single-index \citep{stoker1986consistent} or multiple-index models \citep{samarov1993exploring} and the ability to construct concise visualizations of high-dimensional relationships between covariates and outcome \citep{cook1998regression}. 

This article focuses on the construction of regression models from heterogeneous populations in the presence of high dimensional data. Our specific motivating example is in modeling the risk of hospital admission for Medicare patients across a health system using both electronic health records (EHRs) and Medicare claims. In this setting, population heterogeneity due to differences in patient comorbidity profiles is particularly pronounced and not properly accounting for it can lead to ineffective or misleading models. We develop an SDR estimation framework designed to address some of the particular challenges that arise in this setting. While our motivation is on heterogeneity defined by chronic conditions, the notion of heterogeneity resulting from binary factors has analogues in a wide variety of areas and the framework we propose is highly general with wide applicability.

 A  challenging issue researchers face in health system-wide risk modeling is the heterogeneity of the study population.  Medicare patients often have multiple chronic conditions -- different combinations of which can result in fundamentally different risks and care needs. In our motivating study, we are faced with the subpopulations defined by the presence of \textit{specific combinations} of these  conditions. The subpopulations are depicted by a Venn diagram shown in Figure \ref{fig:strata_data}.  In particular, three prevalent chronic conditions, congestive {heart} failure (C{\bf H}F), {\bf d}iabetes, and chronic obstructive {pulmonary} disease (CO{\bf P}D), denoted as $H$, $D$, and $P$, respectively, are highly related to health outcomes and complicate health care.  It is thus important to build reliable risk prediction models for the subpopulations with these conditions. The inherent differences among  those with different chronic conditions can complicate regression relationships. For example, patients with both CHF and diabetes need to be considered separately from patients with just CHF due to the particular complications arising from having both conditions together, e.g. the diabetes medications rosiglitazone and pioglitazone are known to cause or worsen CHF \citep{wooltorton2002rosiglitazone} and thus their usage by those with both CHF and diabetes may exacerbate risk relative to those with only diabetes.

{
\begin{figure}[!ht]
	\centering
	\begin{subfigure}[b]{0.475\textwidth}
		\centering
		\tikzstyle{background rectangle}=
		[draw=blue!8,fill=gray!40,rounded corners=1.5ex]
		\resizebox{.95\textwidth}{!}{
			\begin{tikzpicture}[font=\sffamily\sansmath,show background rectangle]
			\tikzset{venn circle/.style={draw,circle,minimum width=6.2cm,fill=#1,opacity=0.5}}
			
			\node [venn circle = red][label=below left:{\LARGE C{\underline H}F}] (A) at (0,0) {\Large {}};
			\node [venn circle = yellow][label=below:{\LARGE {\underline D}iabetes}] (C) at (0:4cm) {\Large {}};
			\node [venn circle = blue][label=above left:{\LARGE CO{\underline P}D}] (B) at (60:4cm) {\Large {}};
			\node[left,text=white] at (barycentric cs:A=0.9/3,B=1/2 ) {\large $n = 448$}; 
			\node[below] at (barycentric cs:A=1/2,C=1/2 ) {\large $n = 1715$};   
			\node[right,text=white] at (barycentric cs:B=1/2,C=0.9/3 ) {\large $n = 327$};   
			\node[below,text=white] at (barycentric cs:A=0.9/3,B=1/3,C=0.9/3 ){\large $n = 276$};
			\node[below left] at (A.center) {{\large $n=3628$}}; 
			\node[below right] at (C.center)  {{ \large $n = 7048$}}; 
			\node[above,text=white] at (B.center) {{\large $n = 1227$}};  
			
			\node[label=right:\Large$None$ ] (Non) at (5.5,5)  {}; 
			\node[right] at (5.5,4) {{\large $n = 36203$}};  
			
			
			\path(-3.5,0) --(7.5,0);
			\path(0,-3.5) --(0,6.9);
			\end{tikzpicture}   
		}
		\caption{Displayed are the sample sizes for each subpopulation included in our study of risk of hospital admissions across a large health system.}
		\label{fig:strata_data}
	\end{subfigure}%
	\hfill
	\begin{subfigure}[b]{0.475\textwidth}
		\centering
		\resizebox{.785\textwidth}{!}{
			\begin{tikzpicture}[font=\sffamily\sansmath,
			vc/.style = {
				circle, fill=#1, 
				minimum width=26mm, opacity=0.45},
			vcO/.style = {
				circle, fill=#1, 
				minimum width=26mm, opacity=0.75},
			vcnocol/.style = {
				circle, draw,  opacity=0.5, 
				minimum width=26mm},
			bigvcnocol/.style = {
				circle, draw, opacity=0.5, 
				minimum width=42mm},
			bigvc/.style = {
				circle, fill=#1, 
				minimum width=42mm, opacity=0.7},
			vg/.style args = {#1/#2}{
				minimum height=32mm,
				minimum width=#1+\pgfkeysvalueof{/pgf/minimum height},
				thick, 
				draw, rounded corners=\pgfkeysvalueof{/pgf/minimum height}/2, 
				fill=#2, opacity=0.65,
				sloped}, 
			]
			\begin{pgfonlayer}{foreground}
			
			\node (HPD)[bigvc=teal, text=white]  at (0,-5.25)       {};
			\node (H) [vcO=red,text=white]  at (-0.5,-5.0)  {};
			\node (P) [vc=blue,text=white]  at (0,-5.85)   {};
			\node (D) [vc=yellow,text=white]  at (0.5,-5.0)   {};

			\node (HD)[bigvcnocol, text=white]  at (0,-5.25)       {};
			\node (DD) [vcnocol,text=white]  at (0.5,-5.0)   {};
			\node (HH) [vcnocol,text=white]  at (-0.5,-5.0)   {};
			\node (PP) [vcnocol,text=white]  at (0,-5.85)   {};
			
			\node (D) [text=white]  at (0.9,-4.8)   {\normalsize ${\mathcal{S}_{HP}}$};
			\node (H) [text=white]  at (-0.9,-4.8)  {\normalsize ${\mathcal{S}_{DH}}$};
			\node (P) [text=white]  at (0,-6.25)  {\normalsize ${\mathcal{S}_{DP}}$};
			\node (HDlab) [text=white]  at (0,-4.55)  {\normalsize ${\mathcal{S}_{H}}$};
			\node (HPlab) [text=white]  at (-0.6,-5.6)  {\normalsize ${\mathcal{S}_{D}}$};
			\node (PDlab) [text=white]  at (0.625,-5.6)  {\normalsize ${\mathcal{S}_{P}}$};
			\node (HDlabel)[text=white]  at (0,-3.45)       {\normalsize ${\mathcal{S}_{DHP}}$};
			\node (HDlabel)[text=white]  at (0,-5.25)       {\normalsize ${\mathcal{S}_{none}}$};
			\end{pgfonlayer}
			\end{tikzpicture}
		}
		\caption{Displayed are the relationships among the central mean subspaces under our hierarchical assumption.  The red circle in its entirety is $\mathcal{S}_{DH}$, the yellow circle in its entirety is $\mathcal{S}_{HP}$, $\mathcal{S}_{H}$ is the intersection of $\mathcal{S}_{HP}$ and $\mathcal{S}_{DH}$, and $\mathcal{S}_{none}$ is the intersection of $\mathcal{S}_{D}$, $\mathcal{S}_{H}$, and $\mathcal{S}_{P}$. This diagram seemingly appears as the inverse of the Venn diagram shown in Figure \ref{fig:strata_data}, however this structure captures how the regressions for each subpopulation should be related: $\mathcal{S}_{none}$ should comprise a portion of every other subspace, and $\mathcal{S}_H$ should comprise a portion of any subspace relating to patients with $H$ and any other condition.}
		\label{fig:hierarchical_subspaces_none}
\end{subfigure}	
\caption{Subpopulations and corresponding dimension reduction subspaces.}
\label{fig:subpops_and_subspaces}
\end{figure}
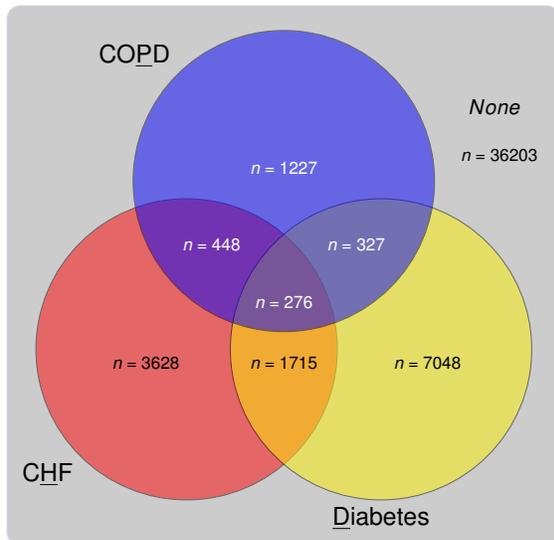
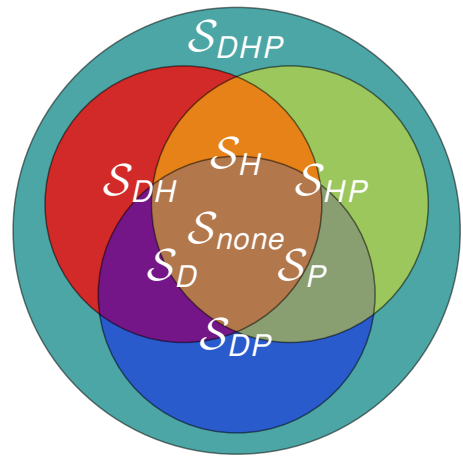
}

In the context of regression modeling, this implies that the relationship between covariates and outcome may be different for different subpopulations, necessitating stratification of regression models by binary factors.  The vast majority of existing SDR methods assume one central mean subspace for the entire population and thus ignore this heterogeneity. The partial SDR methods of \citet{Chiaromonte2002} and \citet{li2003dimension} move away from this assumption and allow for SDR for multiple subpopulations defined by a qualitative predictor. 
These approaches  can be directly applied, resulting in separate central mean subspaces for different subpopulations. Yet, these central mean subspaces may overlap arbitrarily, when in practice the overlapping of subspaces may be {expected} \textit{a priori} to follow certain patterns.

When multiple binary variables (e.g. chronic conditions) stratify the population, the resulting subpopulations are aligned by the common underlying structure of \textit{shared}  stratifying factors.  For example, the subpopulation with only $D$ and the subpopulation with both $D$ and $H$ have important characteristics in common.  Restricting the central mean subspaces to overlap in a manner that aligns with the structural relationships between subpopulations 
 can improve estimation performance and interpretability. This is especially important for subpopulations defined by multiple chronic conditions as they are sicker and  have higher hospitalization rates. Yet, these subpopulations with more conditions have smaller sample sizes, necessitating the borrowing of information from larger, related subpopulations.

Consider a population stratified on two factors $D$ and $H$, resulting in the following subpopulations: patients with both conditions,  denoted as $DH$, patients with just $H$, patients with just $D$, and patients with neither, denoted as $none$. Their corresponding dimension reduction subspaces, to be defined more specifically later, are ${\cal S}_{M}$ for $M \in \{none, D, H, DH\}$.  To borrow strength across subpopulations, in this paper we introduce the assumption that the dimension reduction subspaces are \textit{hierarchical} in the sense that
\begin{equation}\label{eqn:hierarchical_assumption_none_example}
{\cal S}_{none} \subseteq {\cal S}_{D} \subseteq {\cal S}_{DH} \supseteq {\cal S}_{H} \supseteq {\cal S}_{none}.
\end{equation} Visually, this assumption under three chronic conditions is represented in Figure \ref{fig:hierarchical_subspaces_none}. Since $\mathcal{S}_M$ carries the regression information for subpopulation $M$, \eqref{eqn:hierarchical_assumption_none_example} encourages the sharing of information across related subpopulations. We propose to estimate $\mathcal{S}_M$ such that they possess the property \eqref{eqn:hierarchical_assumption_none_example}, which we will demonstrate in Section \ref{sec:hierarchical_subspace_assumption} results in interpretable parameters.

\citet{Ma2012, Ma2014} recently developed a general semiparametric estimation approach for estimation of the central mean subspace. Their unifying framework includes many existing estimation techniques as special cases. The semiparametric approach is attractive, as it obviates the need for the commonly-utilized linearity assumption on the covariates \citep{Ma2012} in methods such as the principal Hessian directions (PHD) \citep{li1992principal} and the approach presented in \citet{Cook2002}. 
We build on the framework of \citet{Ma2012} and show it can be extended to the scenario with models stratified on binary factors and can be further modified to estimate hierarchical subspaces. 
We focus on this semiparametric framework due to its fairly extensive unification of methods for sufficient dimension reduction; working within this framework allows one to develop methods general enough to naturally extend to other sufficient dimension reduction techniques.  However, we acknowledge that this framework is not the only effective approach to dimension reduction. Numerous other works have focused on flexible estimation in SDR.  \citet{wang2020aggregate} introduce an estimation method based on aggregating localized estimators and is able to flexibly circumvent the linearity condition. The MAVE approach of \citet{Xia2002} is another flexible estimator and has been shown to be highly effective in estimating the central mean subspace. More recent approaches such as \citet{Luo2014} and \citet{luo2016new} are also quite promising and have connections to the framework of \citet{Ma2012}. In particular, \citet{luo2016new} proposes an efficient estimator based on \citet{Luo2014} that refines and improves the performance of semiparametric estimators of the central mean subspace. The principles we develop in this manuscript can be straightforwardly extended to incorporate the particular estimation methods of these works, such as \citet{Xia2002}, \citet{Luo2014}, and \citet{luo2016new}. We view this extensibility as an advantage of our proposal.

Our investigations reveal that our hierarchical assumption can be cast as a constraint on parameters.  Imposing constraints to incorporate prior information in classical SDR estimation was proposed in \citet{Naik2005}. The connection of our hierarchical assumption with parameter constraints is not obvious and it demonstrates how to incorporate  prior information from a different perspective, i.e. by making assumptions on $\mathcal{S}_M$ instead of explicit construction of parameter constraints. 
The approach of \citet{Naik2005} is based on classical SDR estimation approaches such as sliced inverse regression or PHD and thus requires strict distributional assumptions on the covariates. 
Our estimation framework is a natural semiparametric approach for the incorporation of prior information in SDR.

The remainder of this paper is organized as follows. In Section \ref{sec:model_assumptions} we lay out our modeling assumptions and develop a preliminary semiparametric estimation framework for SDR in the vein of \citet{Ma2012} to our setting, but without the hierarchical assumption. In Section \ref{sec:method} we formalize our hierarchical assumption and derive a corresponding estimator; we also investigate consequences of violations of the hierarchical assumption. In Section \ref{sec:sim} we demonstrate the utility of our approach with numerical studies. Finally, in Section \ref{sec:analysis} we utilize our approach to analyze a study of hospital admissions in a large health system and performance is evaluated on validation data from the same population.


\section{Notation and Preliminaries}\label{sec:model_assumptions}

In this section we formalize the notion of SDR for subpopulations defined by a set of binary factors and provide  preliminary results for semiparametric estimation of the subpopulation-specific central mean subspaces. The results we develop in this section will serve as a foundation for our proposed approach for borrowing strength across subpopulations.

In general, we consider scenarios where subpopulations are determined by $C$ different binary stratifying factors, resulting in $2^C$ subpopulations. For example, $C=3$ in our motivating example and the population is thus stratified into 8 subpopulations as depicted in Figure \ref{fig:strata_data}. To facilitate our presentation, we introduce a random vector $\bfZ$ that indicates the different subpopulations. That is,  $\bfZ$ is of length  $C$ with its $j$th element taking value of 1 or 0, according to the presence or absence of the $j$th factor for $j=1, \ldots, C$.  In the motivating example, $\bfZ$ takes the value $000$ to indicate the $none$ subpopulation,  $100$ to indicate the subpopulation with only chronic condition $D$, $010$ with $H$, $001$ with $P$, $110$ with $DH$,  $011$ with only $HP$, $101$ with only $DP$, and $111$ with $DHP$.  We denote $\mathcal{Z}^C$ as the set of all possible values of $\bfZ$; in this example  $\mathcal{Z}^3 = \{000, 001, 010, 100, 011, 101, 110, 111\}$. 

Consider regression models for an outcome $Y$ based on a $p$-dimensional covariate vector $\bfX\in \mathcal{X}$ with population heterogeneity based on $\bfZ$.  Our aim is to estimate $p\times d_\bfz$ matrices $\coef_{\bfz}^*$ for  $\bfz \in \mathcal{Z}^C$ such that 
$
E(Y | \bfX, \bfZ = \bfz) =  E(Y | \bfX^\transp\coef_{\bfz}^*, \bfZ = \bfz).
$
The matrices $\coef_{\bfz}^*$ are clearly not identifiable, however the central mean subspaces $\mathcal{S}_{\bfz}$, defined as the linear span of the column vectors of $\coef_{\bfz}^*$, are identifiable.  
The central mean subspace has an invariance property \citep{Cook2002, Ma2014}, which enables us to assume {without loss of generality} in this paper that $E(\bfX) = \boldsymbol 0$ and $\text{cov}(\bfX) = \boldsymbol I$.
In alignment with the mean-preservation goal above and assuming an additive error structure, we posit the following model:
\begin{equation}\label{eqn:mean_models}
Y =  \sum_{\bfz \in \mathcal{Z}^C}I(\bfZ = \bfz)\ell_\bfz(\bfX^\transp\coef_\bfz^* ) + \epsilon, 
\end{equation}
where $\epsilon$ is an error term with $E(\epsilon|\bfX, \bfZ) = 0$, $I(\cdot)$ is the indicator function, and $\ell_{\bfz}$ is some function corresponding to subpopulation $\bfz$.

The observable data based on a sample of size $n$ are $(Y_i, \bfX_i, \bfZ_i)_{i=1}^n$, where the sample size for subpopulation $\bfz$ is denoted as $n_\bfz = \sum_{i=1}^nI(\bfZ_i = \bfz)$. In Web Appendix B, we derive the nuisance tangent space and its orthogonal complement corresponding to the model \eqref{eqn:mean_models}, allowing us to characterize all possible estimating functions.  
We define quantities relating to the resulting estimating functions and introduce a class of estimators based on these. We combine $\bbeta_\bfz$ for all $\bfz \in \mathcal{Z}^C$ into a big matrix via column combination. That is, we denote $\bbeta = (\bbeta_\bfz : \bfz \in \mathcal{Z}^C)$, where the order of $\bbeta_\bfz$ in $\bbeta$ is of no real consequence but is assumed to be consistent throughout this paper. From our derivation of the nuisance tangent space,
we show that the corresponding population-level estimating equation is
\begin{align}
	\Psi_0(\bbeta) = {} &  E\left(\vec\left[\left\{  Y - {E}(Y | \bfX^\transp{\coef}_{\bfZ}, \bfZ)  \right\} \left\{  \balpha(\bfX,\bfZ) - {E}(\balpha(\bfX,\bfZ) | \bfX^\transp{\coef}_{\bfZ}, \bfZ)  \right\} \right]\right) \label{eqn:pop_est_eqn},
\end{align}
where $\vec(\boldsymbol A)$ indicates the vectorization of a matrix $\boldsymbol A$, and $\balpha(\bfx, \bfz)$ is any square-integrable function defined on $\mathcal{X}\times\mathcal{Z}$. Typically, $\balpha(\bfx, \bfz)$ will be matrix-valued.
Similar to \citet{Ma2012}, \eqref{eqn:pop_est_eqn} enjoys a double-robustness property in the sense that consistency for zero at the true parameters still holds when either ${E}(\balpha(\bfX,\bfZ) | \bfX^\transp{\coef}_{\bfZ}, \bfZ)$ or ${E}(Y | \bfX^\transp{\coef}_{\bfZ}, \bfZ)$  is replaced by an arbitrary term.
The corresponding empirical estimating equation is 
\begin{align}
	\hat{\Psi}_n(\bbeta) = {} &  \sum_{i = 1}^n\vec\left[\big\{  Y_i - \widehat{E}(Y_i | \bfX_i^\transp{\coef}_{\bfZ_i}, \bfZ_i)  \big\} \big\{  \balpha(\bfX_i,\bfZ_i) - \widehat{E}(\balpha(\bfX_i,\bfZ_i) | \bfX_i^\transp{\coef}_{\bfZ_i}, \bfZ_i)  \big\}\right], \label{eqn:sample_est_eqn} 
\end{align}
where $\widehat{E}(\cdot| \bfX^\transp{\coef}_{\bfZ}, \bfZ)$ are nonparametric kernel estimates of ${E}(\cdot| \bfX^\transp{\coef}_{\bfZ}, \bfZ)$. 
 As in \citet{Ma2013} and \citet{Ma2015}, to facilitate our asymptotic studies and to ensure identifiability of the parameters, the upper $d_\bfz\times d_\bfz$ block of each ${\coef}_{\bfz}^*$ (and any estimates thereof) is assumed to be an identity matrix. 
 For extended discussion on this parameterization, see \citet{Ma2013}. 
We propose to estimate $\bsbeta^*$ via the following
\begin{equation}\label{eqn:est_unconstr}
\wtbbeta = \argmin_{\bbeta = (\bsbeta_\bfz:\bfz \in \mathcal{Z}^C)}\frac{1}{2n}\hat{\Psi}_n(\bbeta)^\transp\boldsymbol W_n \hat{\Psi}_n(\bbeta)
\end{equation}
where $\boldsymbol W_n$ is a user-specified weight function that converges in probability to a symmetric positive semi-definite matrix $\boldsymbol W$ and the upper block of each $\bsbeta_\bfz$ is enforced to be identity. For simplicity of presentation and computation, in this paper we simply use the identity matrix for ${\boldsymbol W}_n$, however we discuss other choices in Web Appendix A. 

There is a wide variety of possible choices for the function $\balpha(\bfx,\bfz)$ and each choice results in a different estimator, similar to \citet{Ma2012} and \citet{Ma2014}. 
As described in later sections, in this paper we focus on a particular choice that corresponds to a semiparametric PHD approach, similar to that described in Section 4.2 of \citet{Ma2012}. 
The central partial mean subspace of \citet{li2003dimension} can be seen as a special case of \eqref{eqn:est_unconstr} and we explore this connection in the Web Appendix A.

We now proceed to studying the asymptotic properties of estimators from the resulting estimating functions. To streamline our presentation, let $\vecl(\boldsymbol A)$ be the operator which vectorizes the lower $(a - b)\times b$ block of an $a\times b$ matrix $\boldsymbol A$ when $a\ge b$. Denote $\vecl{(\bsbeta)}^\transp \equiv (\vecl{(\bsbeta_\bfz)}^\transp:\bfz \in \mathcal{Z}^C)$ as the lower block vectorization of $\bsbeta \equiv (\bsbeta_\bfz:\bfz \in \mathcal{Z}^C)$. In Web Appendix A, we show that $\sqrt{n}\vecl(\widetilde{\coef} - \coef^*)$ is  asymptotically normal with mean zero and variance $\boldsymbol \Sigma_1 \boldsymbol G_1^\transp\boldsymbol W \boldsymbol V_1 \boldsymbol W\boldsymbol G_1 \boldsymbol \Sigma_1$, where the terms in the variance are described in Web Appendix A.
An important point is that the asymptotic convergence is for fixed dimensions $d_\bfz$ similar to \citet{Ma2012}. In practice these dimensions need to be estimated. We present an approach for such determination in Web Appendix A, which is a special case of the dimension determination approach we propose in Section \ref{sec:method}. We discuss selection of the bandwidth parameters in Web Appendix A.

\section{Constrained estimation for populations with structured heterogeneity}\label{sec:method}

In this section we build on the results of Section \ref{sec:model_assumptions} by introducing a natural assumption that aligns with the inherent structure of a population stratified by binary factors. We then derive a corresponding estimator based on the estimating equations derived in Section \ref{sec:model_assumptions} and provide a means for consistently selecting the structural dimensions. We also investigate consequences due to violations of our assumption. We show a connection to structural constraints on parameters, suggesting broad applicability of our proposed estimation approach. 

The patterns of shared characteristics across subpopulations represent important information that can be leveraged in estimation. Consider $D$ and $H$ as binary stratifying factors. The subpopulation with both $D$ and $H$, denoted by $11$, is inherently related to both the subpopulation with only $D$ present, denoted by $10$, and the subpopulation with only $H$ present, denoted by $01$. This relationship is due to the presence of \textit{shared} binary factors (e.g. chronic conditions). 
Yet, subpopulation $01$ is not as obviously related to subpopulation $10$ due to the lack of shared factors. Yet, it would be unreasonable to assume $10$ and $01$ have no relation whatsoever. 
In estimating the central mean subspaces, we aim to borrow strength in estimation in a manner which respects these relationships. 
To avoid any ambiguity, whenever we refer to chronic conditions, we treat them as stratifying factors which comprise  $\mathcal{Z}^C$.

\subsection{The hierarchical dimension reduction subspace assumption}\label{sec:hierarchical_subspace_assumption}

We now formalize our assumption introduced in the Section \ref{sec:intro} and in Figure \ref{fig:hierarchical_subspaces_none} using the terminology of partially ordered set (poset) theory \citep{davey2002introduction, tatsuoka2003sequential}. 
A poset is defined for a set and an ordering property $(Q; \preceq)$. Two distinct elements, $a$ and $b$ in $Q$, are comparable if $a\preceq b$ or $b\preceq a$; otherwise, they are incomparable. 
A simple poset is an order set of $\{0, 1\}$, such as $0\leq 1$. 
In our setting, the poset $( \mathcal{Z}^C, \preceq)$  is a very natural representation of the subpopulations and how they relate to each another. In this poset, partial order is defined by subpopulation relationships. That is, 
for two subpopulations $\bsz$ and $\bsz'$, we say that $\bfz  \preceq \bfz'$ if $\bfz[j] \leq \bfz'[j]$ for all $j=1, \dots, C$. For example, when $C=2$, $01  \preceq 11$ because $11$ possesses all factors that $01$ has; similarly, $00  \preceq 01$, $00  \preceq 10$, and $10  \preceq 11$.  However $01$ and $10$ are not comparable since they share no conditions.  If we view $\bfz[j]=1$ or 0 as the presence or absence of the $j$th stratifying factor, then  $\bfz  \preceq \bfz'$ if and only if all stratifying factors present in $\bfz$ are also present in $\bfz'$.

For $\bfz, \bfz' \in \mathcal{Z}^C$ we denote the greatest lower infimum of $\bfz$ and $\bfz'$ as $\bfz\wedge \bfz' \in  \mathcal{Z}^C$, which, loosely speaking, describes the most similar population linking $\bfz$ and $\bfz'$. The greatest lower infimum of two subpopulations $\bfz$ and $\bfz'$ represents the subpopulation that has as much similarity as possible with both $\bfz$ and $\bfz'$ in terms of the binary stratifying variables, but does not have any stratifying variables that are not present in either $\bfz$ or $\bfz'$.  Formally, the greatest lower infimum is defined through the following two conditions: (1) $\bfz\wedge \bfz'  \preceq \bfz$ and $\bfz\wedge \bfz'  \preceq \bfz'$, (2) for any  $\bfa \in \mathcal{Z}^C$, if $\bfa  \preceq \bfz$ and $\bfa   \preceq \bfz'$,  then $\bfa  \preceq \bfz\wedge \bfz'$.  For example, $00 \wedge 01 = 00$, $01 \wedge 11 = 01$, and $10 \wedge 01 = 00$.

Using this notation, we proposed to extend the structural assumption \eqref{eqn:hierarchical_assumption_none_example} on the central mean space to the following two conditions. The first condition is that
\begin{equation}\label{eqn:hierarchical_assumption_none}
{\cal S}_{\bfz} \subseteq {\cal S}_{\bfz'} \text{ for all } \bfz  \preceq \bfz', \text{ where } \bfz, \bfz' \in \mathcal{Z}^C.
\end{equation}
The second condition is that
\begin{equation}\label{eqn:none_assumption2}
{\cal S}_{\bfz} \cap {\cal S}_{\bfz'} =  {\cal S}_{\bfz \wedge \bfz' } \;\;\;\;\;\text{ for all } \bfz, \bfz' \in \mathcal{Z}^C.
\end{equation} 
Assumptions \eqref{eqn:hierarchical_assumption_none} and \eqref{eqn:none_assumption2} when combined generalize the structural assumption \eqref{eqn:hierarchical_assumption_none_example}. 

The hierarchical assumption \eqref{eqn:hierarchical_assumption_none} encourages the sharing of regression information across subpopulations of patients with similar chronic conditions and therefore enables borrowing strength across related subpopulations.
Thus, for subpopulations $\bfz$ and $\bfz'$ with $\bfz \preceq \bfz'$, the data from $\bfz$ can be utilized to estimate a portion of the central mean subspace for subpopulation $\bfz'$, while still allowing ${\cal S}_{\bfz}$ to be distinguished from ${\cal S}_{\bfz'}$.
Assumption \eqref{eqn:none_assumption2} implies a clear separation of the parameters that comprise the dimension reduction spaces, ensuring that a set of parameters are explicitly devoted to each subpopulation and each chronic condition in particular. This separation allows for concise interpretation of the directions. It further demonstrates how seemingly unrelated subpopulations should be related. In particular, with our example of subpopulations $10$ and $01$, \eqref{eqn:none_assumption2} implies that ${\cal S}_{10} \cap {\cal S}_{01} = {\cal S}_{00}$. Thus, two subpopulations with no shared factors are related through the subpopulation that is as similar as possible to them without having any additional factors, in this case, the subpopulation with no chronic conditions. 
In the following, we consider estimation based on assumptions \eqref{eqn:hierarchical_assumption_none} and \eqref{eqn:none_assumption2}. We will also investigate the consequences of possible violations of this assumption from both theoretical and numerical perspectives. 

The following proposition is crucial to understanding how to estimate the central mean subspaces that meet \eqref{eqn:hierarchical_assumption_none} and \eqref{eqn:none_assumption2} and shows it  can be achieved by constraints on parameters. 
\vspace{-1.25cm}
\begin{proposition}\label{thm:achieving_hier_central}
	The hierarchical central mean subspaces $\mathcal{S}_{\bfz}$, for $\bfz \in \mathcal{Z}^C$, that meet \eqref{eqn:hierarchical_assumption_none} and \eqref{eqn:none_assumption2} can be expressed as $\mathcal{S}_{\bfz} = \text{span}(\bsbeta^0_{\bfz})$ for some parameters $\bsbeta^0_{\bfz}$  that have the following decomposition 
\begin{align}
\bsbeta^0_{\bfz} = (\bnu^0_{\bfz'} : \bfz' \preceq  \bfz, \bfz' \in \mathcal{Z}^C) \label{eqn:equality_condition}
\end{align}
where $\bnu^0_{\bfz'}$ is a matrix of dimension $p\times k^0_ {\bfz'}$ with $k^0_ {\bfz'} \geq 0$,  $\bsbeta^0_{\bfz}$ is a {\em full rank} matrix with dimension $p\times d^0_ {\bfz}$ and $d^0_ {\bfz}=\sum_{ \bfz' \preceq  \bfz} k^0_ {\bfz'}$, and $\text{span}(\cdot)$ denotes the column space spanned by the columns of a matrix.  
\end{proposition}
The key purpose of Proposition \ref{thm:achieving_hier_central} is to show how conditions \eqref{eqn:hierarchical_assumption_none} and \eqref{eqn:none_assumption2} can be enforced. The implication of Proposition \ref{thm:achieving_hier_central} is that the hierarchical central mean subspaces under assumptions \eqref{eqn:hierarchical_assumption_none} and \eqref{eqn:none_assumption2} can be characterized by enforcing \eqref{eqn:equality_condition} in estimation. 
We require $\bsbeta^0_{\bfz}$ to be of rank $d^0_{\bfz}$, which indicates $\bnu^0_{\bfz'}$ should be full rank and none of its columns can be  linearly dependent of other  $\bnu^0_{\bfz''}$ for $\bfz''\in \mathcal{Z}^C$.
Clearly, the representation \eqref{eqn:equality_condition} is not unique without further constraints on $\bsbeta^0_{\bfz}$, but uniqueness can be achieved by constraining the upper blocks of each $\bnu^0_{\bfz}$ to be identity, as in the previous section. However, we note that this constraint implies that one should re-order the covariates so that covariate known to impact the response for all subpopulations are first in the design matrix. In applications where nothing is known about the covariate-response relationship, if unimportant variables happen to be ordered first, then other coefficients corresponding to important variables will be likely be very large -- in this case the variables can simply be re-sorted so that such variables with large coefficients are ordered first.

\begin{remark}
	If $\mathcal{Z}^C$ contains all possible subpopulations induced by the $C$ factors, then all parameters parameters $\bnu^0_{\bfz}$ for $\bfz \in \mathcal{Z}^C$ are identifiable. However, if some subpopulations are not contained in $\mathcal{Z}^C$, then it is possible some parameters may not be identifiable.  If a particular subpopulation $\bfz$ is not contained in $\mathcal{Z}^C$, then the dimension of $\bnu_\bfz$ must be set to 0 in order for all parameters to be identifiable. 
\end{remark}

An additional implication of Proposition \ref{thm:achieving_hier_central} is what \eqref{eqn:hierarchical_assumption_none} and \eqref{eqn:none_assumption2}  jointly imply for interpretation of parameters.
 The terms $\bnu^0_{\bfz'}$ can be viewed as sets of parameters that are explicitly devoted to each chronic condition or each particular combination of chronic conditions $\bfz'$.   In some sense, the $\bnu_\bfz$ parameters may be of more clinical interest than the $\bsbeta_\bfz$ parameters.

Take our motivating example, for a given value of covariates $\bfx$, the term $\bfx^\transp\bnu^0_{000}$ corresponds to directions which are shared among all patients and $\bfx^\transp\bnu^0_{100}$ can be thought of as all effects relating to the outcome that are necessary to explain the outcome for patients with only chronic condition $D$ beyond the information that is available in $\bfx^\transp\bnu^0_{000}$. Similarly, $\bfx^\transp\bnu^0_{110}$ corresponds to a set of directions that pertain to all patients with both $D$ and $H$. Thus, if there is a significant interaction between $D$ and $H$ in terms of the covariate effects on the response, $\bfx^\transp\bnu^0_{110}$ may have positive dimension  (i.e. $k^0_ {110} > 0$). Otherwise,  $k^0_ {110} = 0$ implies that no extra directions are needed for patients with both  $D$ and $H$ once the directions of $D$-only patients $\bnu^0_{100}$ and those of $H$-only patients $\bnu^0_{010}$ are included.

\vspace{-1.em}
\subsection{Semiparametric estimation of the hierarchical central mean subspaces}\label{sec:borrowing_strength_estimation}

Similar to Section \ref{sec:model_assumptions}, we now constrain the upper $k^0_ {\bfz'} \times k^0_{\bfz'}$ block of the true parameters $\bnu^0_{\bfz'}$ to be identity matrix $I_{k^0_{\bfz'}}$ and correspondingly enforce the same in estimation such that all parameters are identifiable. The decomposition \eqref{eqn:equality_condition} can be alternatively represented by a set of linear equality constraints on $\bsbeta^0 = (\bsbeta_\bfz^0 : \bfz \in \mathcal{Z}^C)$, which can be expressed as
\begin{equation}\label{eqn:hiersdr_subpop_constraints}
\bsC^\transp\vecl(\bsbeta^0) = \boldsymbol 0
\end{equation}
for some matrix $\bsC$. We show in Web Appendix A how the matrix $\bsC$ can be constructed. Here the $\vecl(\cdot)$ operator is applied to each of the  components $\bnu^0_{\bfz'}$ of $\bsbeta^0_{\bfz}$ in line with the identifiability constraints.  In light of these constraints, we propose the following estimator
\begin{align}
\widehat{\bbeta} = \argmin_{\bbeta = ({{\bbeta}_\bfz }: \bfz \in \mathcal{Z}^C)}  \biggl\{  \frac{1}{2n}\hat{\Psi}_{n}(\bbeta)^\transp\boldsymbol W_n\hat{\Psi}_{n}(\bbeta) \ \mbox{ such that } \boldsymbol \bsC^\transp \vecl(\bsbeta) = \boldsymbol 0 \biggr\},  \label{eqn:constr_min}
\end{align}
where $\boldsymbol W_n$ is a weight matrix possibly chosen for efficiency improvement and the individual components ${\bnu}_{\bfz'}$ that comprise each ${\bbeta}_\bfz$ are constrained to have identity upper blocks. Due to the constraint that $\bsC^\transp\vecl(\bsbeta^0) = \boldsymbol 0$, we enforce the hierarchical assumptions \eqref{eqn:hierarchical_assumption_none} and \eqref{eqn:none_assumption2} directly in estimation. In particular, since $\hat{\Psi}_{n}(\bbeta)$ assess fit of the parameters $\bbeta$ to the data, \eqref{eqn:constr_min} essentially finds the ``best'' fitting parameters that meet our hierarchical conditions. We present in Web Appendix A a validated information criterion (VIC) approach similar to that in \citet{Ma2015} for dimension determination and derive theoretical results that demonstrate it results in consistent dimension determination.

Similar to the procedure described in \citet{Ma2012}, we can estimate both the gradient and the Jacobian of $(2n)^{-1}\Psi_{n}(\bbeta)^\transp\boldsymbol W_n\Psi_{n}(\bbeta)$
using any numerical differentiation approach such as a finite difference approximation to optimize \eqref{eqn:constr_min} using Newton's method with equality constraints \citep[Chapter~10]{boyd2004convex}.

Conditions (C1)-(C5) below are required for our asymptotic results and are stated using a generic parameter vector $\bstheta_\bfz$.  Denote $\nu(\bstheta_\bfz)$ as the number of columns in $\bstheta_\bfz$.
\vspace{-10pt}
\begin{enumerate}[label={(C\arabic*)}]
	\item $n_\bfz/n \to p_\bfz \in (0,1)$ as $n \to \infty$ for all $\bfz \in \mathcal{Z}^C$.
	\item The univariate kernel function $K(\cdot)$ is Lipschitz and has compact support. Furthermore for all $\bfz \in \mathcal{Z}^C$ it satisfies 
$\int K(u) \mathrm{d} u = 1, \int u^iK(h_\bfz)\mathrm{d}u = 0, 1 \leq i \leq m_{\bstheta_\bfz} - 1$  and  $0 \neq \int u^{m_{\bstheta_\bfz}}K(h_\bfz)\mathrm{d}u < \infty,$
	where $m_{\bstheta_\bfz}$ is an integer related to the Lipschitz continuity of functions defined in condition (C4).
	The $\nu(\bstheta_\bfz)$-dimensional kernel function is a product of $\nu(\bstheta_\bfz)$ univariate kernel functions, i.e. $K_{h_\bfz}(\boldsymbol u) = K(\boldsymbol u / h_\bfz) / h ^ {\nu(\bstheta_\bfz)} = \prod_{j = 1}^{\nu(\bstheta_\bfz)}K(u_j/h_\bfz)/h^{\nu(\bstheta_\bfz)}$ for $\boldsymbol u = (u_1, \dots, u_{\nu(\bstheta_\bfz)})^\transp$. In the above notation, $K$ is used regardless of dimension.
	
	\item The joint density/mass functions of $(\bfX, \bfZ)$ and $(\bfX^\transp\bstheta_\bfZ, \bfZ)$, respectively denoted as $f_{\bfX,\bfZ}(\bfx, \bfz) = f_{\bfX|\bfZ}(\bfx)\pi_{\bfZ}(\bfz)$ and $f(\bfx^\transp\bstheta_\bfz, \bfz) = f_{\bfX^\transp\bstheta_\bfZ|\bfZ}(\bfx^\transp\bstheta_\bfz)\pi_{\bfZ}(\bfz)$ are bounded from below and above, where $f_{\bfX|\bfZ}(\bfx)$ is the density function of $\bfX$ given $\bfZ$, $\pi_{\bfZ}(\bfz)$ is the mass function of $\bfZ$, and $f_{\bfX^\transp\bstheta_\bfZ|\bfZ}(\bfx^\transp\bstheta_\bfz)$ is the density function of $\bfX^\transp\bstheta_\bfZ$ given $\bfZ$. Furthermore, $E[Y^2 | \bfX^\transp\bstheta_\bfz = \bxi_\bfz, \bfZ = \bfz]$ and each entry of $E[\boldsymbol \alpha(\bfX, \bfz)\boldsymbol \alpha(\bfX, \bfz)^\transp | \bfX^\transp\bstheta_\bfz = \bxi_\bfz, \bfZ = \bfz]$  are locally Lipschitz-continuous and bounded from above as a function of $\bxi_\bfz$ for every $\bfz \in \mathcal{Z}^C$.
	
	\item Let $\boldsymbol r_{1}(\bxi_\bfz, \bfz) = E[\boldsymbol\alpha(\bfX, \bfz) | \bfX^\transp\bstheta_\bfz = \bxi_\bfz,\bfZ = \bfz]f(\bxi_\bfz, \bfz)$ and $\boldsymbol r_2(\bxi_\bfz, \bfz) = E[Y | \bfX^\transp\bstheta_\bfz = \bxi_\bfz,\bfZ = \bfz]f(\bxi_\bfz, \bfz).$ The $m_{\bstheta_\bfz}$th derivatives of $\boldsymbol r_{1}(\bxi_\bfz, \bfz)$, $\boldsymbol r_{2}(\bxi_\bfz, \bfz)$, and $f(\bxi_\bfz, \bfz)$ are locally Lipschitz-continuous in $\bxi_\bfz$ for every $\bfz \in \mathcal{Z}^C$.
	
	\item For all $\bfz \in \mathcal{Z}^C$ the bandwidths $h_\bfz = O(n^{-\kappa_{\bstheta_\bfz}})$ for $1 / (4m_{\bstheta_\bfz}) < \kappa_{\bstheta_\bfz} < 1/(2\nu(\bstheta_\bfz))$ where $\kappa_{\bstheta_\bfz} > 0$ is a constant.
\end{enumerate}
Condition (C1)  indicates that the relative sample size for any subpopulation should not vanish. Condition (C2) is a standard requirement in nonparametric kernel regression estimation. Conditions (C3) and (C4) are standard smoothness and boundedness assumptions. Condition (C5) is a standard bandwidth rate assumption.

\vspace{-1.5em}
 \begin{theorem} \label{thm1}
	Assume conditions (C1)-(C5) listed above hold for $\bsbeta_{\bfz}^0 = (\bnu_{\bfz'}^0 : \bfz' \preceq \bfz, \bfz' \in \mathcal{Z}^C)$ for all $\bfz \in \mathcal{Z}^C$ and that assumptions \eqref{eqn:hierarchical_assumption_none} and \eqref{eqn:none_assumption2} hold. Let
	\begin{align*}
	\boldsymbol G_2 = {} & E\big(  \partial\vec\big[\left\{  Y - {E}(Y| \bfX^\transp{\coef}_{\bfZ}^0, \bfZ)  \right\} \left\{  \balpha(\bfX,\bfZ) - {E}(\balpha(\bfX,\bfZ) | \bfX^\transp{\coef}_{\bfZ}^0, \bfZ)  \right\}\big] / \partial  \vecl(\coef)^\transp  \big) \\
	\boldsymbol V_2  = {} &  \mbox{cov}\big(  \vec\big[\left\{  Y - {E}(Y| \bfX^\transp{\coef}_{\bfZ}^0, \bfZ)  \right\} \left\{  \balpha(\bfX,\bfZ) - {E}(\balpha(\bfX,\bfZ) | \bfX^\transp{\coef}_{\bfZ}^0, \bfZ)  \right\}\big]   \big).
	\end{align*}
	Further assume $\boldsymbol W_n$ converges in probability to a symmetric positive semi-definite matrix $\boldsymbol W$ and that $\boldsymbol \Sigma_2 = \{\boldsymbol G_2^\transp \boldsymbol W \boldsymbol G_2\}^{-1}$ is nonsingular. Then the solution $\widehat{\bbeta}$ of the constrained minimization \eqref{eqn:constr_min} satisfies
	\begin{equation} \label{asymptotic distribution of beta hat}
	n^{1/2}\vecl(\widehat{\bbeta} - \bbeta^0) \xrightarrow{\mathcal{L}} \mathcal{N}(\boldsymbol 0, (\boldsymbol I - \boldsymbol P)\boldsymbol \Sigma_2 \boldsymbol \Xi_2 \boldsymbol \Sigma_2(\boldsymbol I - \boldsymbol P) ),
	\end{equation}
	where $\boldsymbol P ={\boldsymbol \Sigma_2}\bsC(\bsC^\transp {\boldsymbol \Sigma_2}\bsC)^{-1}\bsC^\transp$, $\boldsymbol \Xi_2 = \boldsymbol G_2^\transp\boldsymbol W\boldsymbol V_2 \boldsymbol W\boldsymbol G_2$, and $\bbeta^0 = ({{\bbeta}_\bfz^0 }: \bfz \in \mathcal{Z}^C)$.
\end{theorem}
 In Web Appendix A we present an asymptotic investigation into the consequences of misspecifying the hierarchical assumptions and provide a discussion.
 
\vspace{-.5em}
 A sensible choice for ${\boldsymbol W}_n$ would be $\widehat{\boldsymbol V}_{2n}^{-1}$, where $\widehat{\boldsymbol V}_{2n}$ is any consistent estimator of ${\boldsymbol V}_{2}$, however, for simplicity and computational ease we simply use the identity matrix for ${\boldsymbol W}_n$.

\vspace{-1.0em}

\section{Simulation studies}\label{sec:sim}

In this section we investigate the finite sample performance of our estimator, in both settings when the structural dimensions are known and when they are unknown and estimated with our proposed VIC approach described in Web Appendix C. As the primary focus of this paper is with respect to central mean subspaces, we  focus on variants of the PHD method due to its wide use \cite[Chapter~8]{BingLi2018} and generally good performance across a wide variety of data-generating scenarios. The first variant of PHD we consider is the standard PHD method, which solves the empirical version of $\Lambda_{PHD} - \boldsymbol P \bfX\bfX^\transp\Lambda_{PHD} = \boldsymbol 0$ for each subpopulation, where $\Lambda_{PHD} = E[\{  Y - E(Y)\}\bfX\bfX^\transp]$ and $\boldsymbol P = \coef (\coef^\transp\coef)^{-1}\coef^\transp$. For this approach, population heterogeneity is handled by estimating the central mean subspace for each subpopulation separately. The second variant of PHD we consider is the semiparametric PHD (sPHD) defined as the minimizer of \eqref{eqn:est_unconstr} with $\hat{\Psi}_n(\coef) = \sum_{i = 1}^n\vec\big\{ [Y_i - \widehat{E}(Y_i|\bfX_i^\transp\coef_{\bfZ_i}, \bfZ_i)]{\boldsymbol \alpha(\bfX_i, \bfZ_i)} \big\}$ and ${\boldsymbol W}_n = {\boldsymbol I}$, where $\boldsymbol\alpha(\bfX_i, \bfZ_i)$ is a block diagonal matrix whose block corresponding to $\bfZ_i$ is $\bfX_i\bfX_i^\transp$.  We consider the PHD version of our proposed method with the hierarchical assumptions \eqref{eqn:hierarchical_assumption_none} and \eqref{eqn:none_assumption2} (hier sPHD), which is defined as the solution to \eqref{eqn:constr_min} with $\hat{\Psi}_n(\coef)$ defined in the same way as in sPHD. In addition,  we also evaluate the performance  of hier sPHD with the structural dimensions estimated by the VIC approach outlined in the Suppement, labeled as as ``hier sPHD VIC''. The last approach used is the PHD version of the \citet{Naik2005} (labeled ``N\&T'') with the implied constraints \eqref{eqn:hiersdr_subpop_constraints} imposed.

\subsection{Data generating scenarios}

The data-generating scenarios we consider are when subpopulations are based on two conditions. In the following simulations responses are generated from the following three models of different structural dimensions, which are motivated by the simulation studies of \citet{Ma2012, Ma2013, Ma2015}. 

We consider two binary factors and generate the outcomes  according to \eqref{eqn:mean_models} using three models. That is, we set $Y_\bfz =  \ell_\bfz(\bfX^\transp\coef_\bfz^0 ) + \epsilon$ where $\epsilon$ is an error term generated from independent $N(0, 1)$ random variables. The mean functions are described in Web Appendix C and are generated so as to mimic differing scenarios with heterogeneity across subpopulations.

The conditions \eqref{eqn:hierarchical_assumption_none} and \eqref{eqn:none_assumption2} are satisfied for all the three data generating models. Throughout the simulations, all of the true parameters  $\bbeta^0_{00}, \bnu^0_{10}$, and $\bnu^0_{11}$ in the mean models  are $p\times 1$ vectors. Half of the elements of each of $\bbeta^0_{00}, \bnu^0_{10}$, and $\bnu^0_{11}$ are from a uniform distribution on $[-0.25, 0.25]$ and the rest from a uniform distribution on $[-0.5, 0.5]$. The true dimensions are $\boldsymbol d^0 = (d^0_{00}, d^0_{10}, d^0_{01}, d^0_{11}) = (1, 1, 1, 1)$  for Model 1, $\boldsymbol d^0 = (1, 1, 1, 2)$  for Model 2, and  $\boldsymbol d^0 = (1, 2, 1, 2)$  for Model 3. For $n_\bfz$ and $p$, we consider scenarios with all combinations of $n_\bfz\in \{200, 600, 1200\}$ and $p\in\{10,20\}$. 
We generate covariates that do not meet the linearity condition or the constant variance condition required by many SDR estimation methods and describe the generation process in Web Appendix C.
Simulations are replicated independently 250 times for all settings. We consider a scenario with only continuous covariates and a scenario with more complicated heterogeneity, which are presented in Web Appendix C.

\vspace{-.5em}
\subsection{Estimation results}

Here the true structural dimensions are used for PHD, sPHD,  N\&T, and hier sPHD, except for ``hier sPHD VIC" which estimated structural dimensions. To evaluate the performance of the VIC criterion for structural dimension determination for the ``hier sPHD VIC" approach, we considered all combinations of dimensions such that the largest total dimension is less than 5. As in \citet{Ma2012} and  \citet{Ma2015} we compare the performance of all approaches using the average norm
$\frac{1}{2^C}\sum_{\bfz \in \mathcal{Z}^C}|| \widehat{\bbeta}_\bfz (\widehat{\bbeta}_\bfz^\transp\widehat{\bbeta}_\bfz)^{-1}\widehat{\bbeta}_\bfz^\transp - \bbeta^0_\bfz({\bbeta^0}_\bfz^\transp\bbeta^0_\bfz)^{-1}{\bbeta^0}_\bfz^\transp  ||_2.$
We also report the angle in degrees between estimated and true subspaces, as used in past literature \citep{cook2007fisher}, to evaluate  estimating subspaces. Smaller angles indicate better subspace estimation, with 0 indicating perfect recovery of the subspaces and 90 a complete miss. Average angles over the subpopulations are provided as a summary of subspace estimation performance across the entire population. Results within each subpopulation  are displayed in Web Appendix C. Also in Web Appendix C are two sets of simulations under varying degrees of hierarchy misspecification. From these studies, the proposed hier sPHD is  robust to moderate misspecification of the hierarchical assumption.

 The results are displayed in Figure \ref{fig:sim_model_123_norm_none}.  Under all scenarios, using the common underlying structure via hier sPHD results in better estimation, as indicated by smaller difference norms on average. The proposed method of this paper outperforms the standard PHD, the sPHD method, and the PHD version of the \citet{Naik2005} approach in terms of recovering the central mean subspaces for all subpopulations. We note that the performance of the ``hier sPHD VIC" approach is better in terms of the angle because the difference norm metric penalizes an estimator of a subspace if the dimension is over-estimated, whereas the angle can be zero if the subspace is fully recovered but with an over-estimated dimension. Simulation results illustrating the probability of correct structural dimension selection are displayed in Web Appendix C and show that the probability of the VIC approach selecting the correct dimensions tends to 1 as the sample size increases.

{
\begin{figure}[!ht]
 	\centering
 	\includegraphics[width=0.95\textwidth]{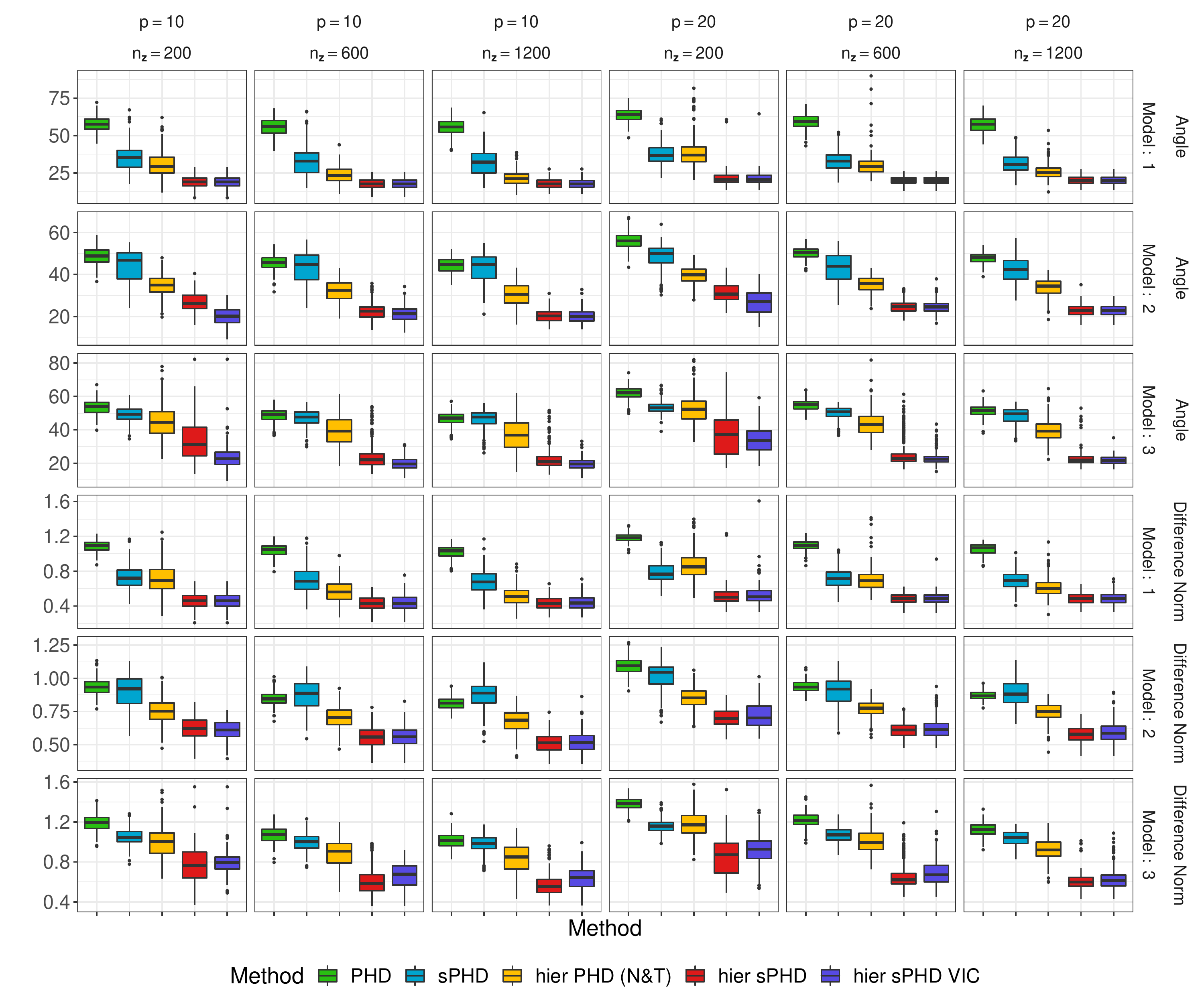}
 	\caption{Displayed are the difference norms and angles between estimated and true subspaces for each method over a variety of simulation settings over 250 datasets under Models 1, 2, and 3 when the covariates are generated under the mixed discrete and continuous setting. All metrics are averages over all subpopulations.}
 	\label{fig:sim_model_123_norm_none}
\end{figure}
}

\vspace{-2em}

\section{Analysis of the hospital admission risk study}\label{sec:analysis}

In 2016 the United States spent 18\% of its GDP on healthcare. The average amount spent on healthcare per person doubled that in comparable countries according to a recent report from the Kaiser Family Foundation \citep{WinNT}. Hospital spending comprised 33\% of total health spending in the United States in 2017, and as health care costs continue to rise and the population ages, policymakers are increasingly concerned about the growing burden of hospital-based medical care expenses on insurers, patients, and employers. As a result, there is an urgent need to build predictive models for hospital (re)admissions so that hospitals and health care systems can intervene to improve care and reduce costly admissions. 

In this study we seek to construct models for the risk of hospital admission for a large population of Medicare ACO beneficiaries. We consider heterogeneity defined by the presence of specific combinations of three prevalent chronic conditions: diabetes ($D$), CHF ($H$), and COPD ($P$). As can be seen in Table \ref{table:auc_hospitalization_with_none}, the subpopulations of patients with different combinations of these chronic conditions have markedly different hospitalization risk. 

To build risk models, we utilize data  with patient covariate information collected from a baseline period of 12 months and outcomes measured as whether or not each patient was admitted to the hospital within a 3 month period following the baseline period. Since the data are collected regularly each month, we can construct a natural validation dataset by looking back 12 months prior to the initial baseline period such that the baseline period for the training data does not overlap with the baseline period for the validation data. We construct our models using the training period and evaluate the predictive performance of the models using the validation data from the year prior. 
The subpopulation sizes for both the training and validation data are displayed in Table \ref{table:auc_hospitalization_with_none}.
Our analysis includes the 31 covariates relating to lab measurements, demographic covariates, pharmaceutical information, and medical information about diagnoses and related health issues.

We used the training data to fit models predicting hospital admission given the 31 covariates. All methods considered in Section \ref{sec:sim} were used to construct directions for each subpopulation. The structural dimensions were selected for the hier sPHD approach using the VIC criterion presented in Web Appendix A. The structural dimensions selected by VIC for ``hier sPHD VIC" were $d_{none} = d_{D} = d_{H} = d_{P} = d_{DH} = d_{DP} = 1$ and $d_{HP}  = d_{DHP} = 2$, indicating that an extra dimension is required to explain the relationship between covariates and outcome for subpopulations that have both CHF and COPD. For the sPHD approach, the VIC approach outlined in Web Appendix A was used to select the structural dimensions. For the PHD approach, the structural dimensions for each subpopulation are selected using the \texttt{dr} package. For comparison and assessment of the need to consider heterogeneity from the three chronic conditions, a semiparametric PHD approach was used on the entire population. This approach is denoted as ``Single'' and the VIC approach of \citet{Ma2015} is used to select its structural dimension. 
The estimated directions from all methods were then utilized as covariates in logistic generalized additive models with a functional ANOVA decomposition model form for each subpopulation fit using the \texttt{mgcv} package. We then evaluated the performance of prediction of the risk of hospitalization on the validation  set in terms of area under the receiver operating characteristic curve (AUC).

\begin{figure}[!ht]
	\centering
	\includegraphics[width=0.7\textwidth]{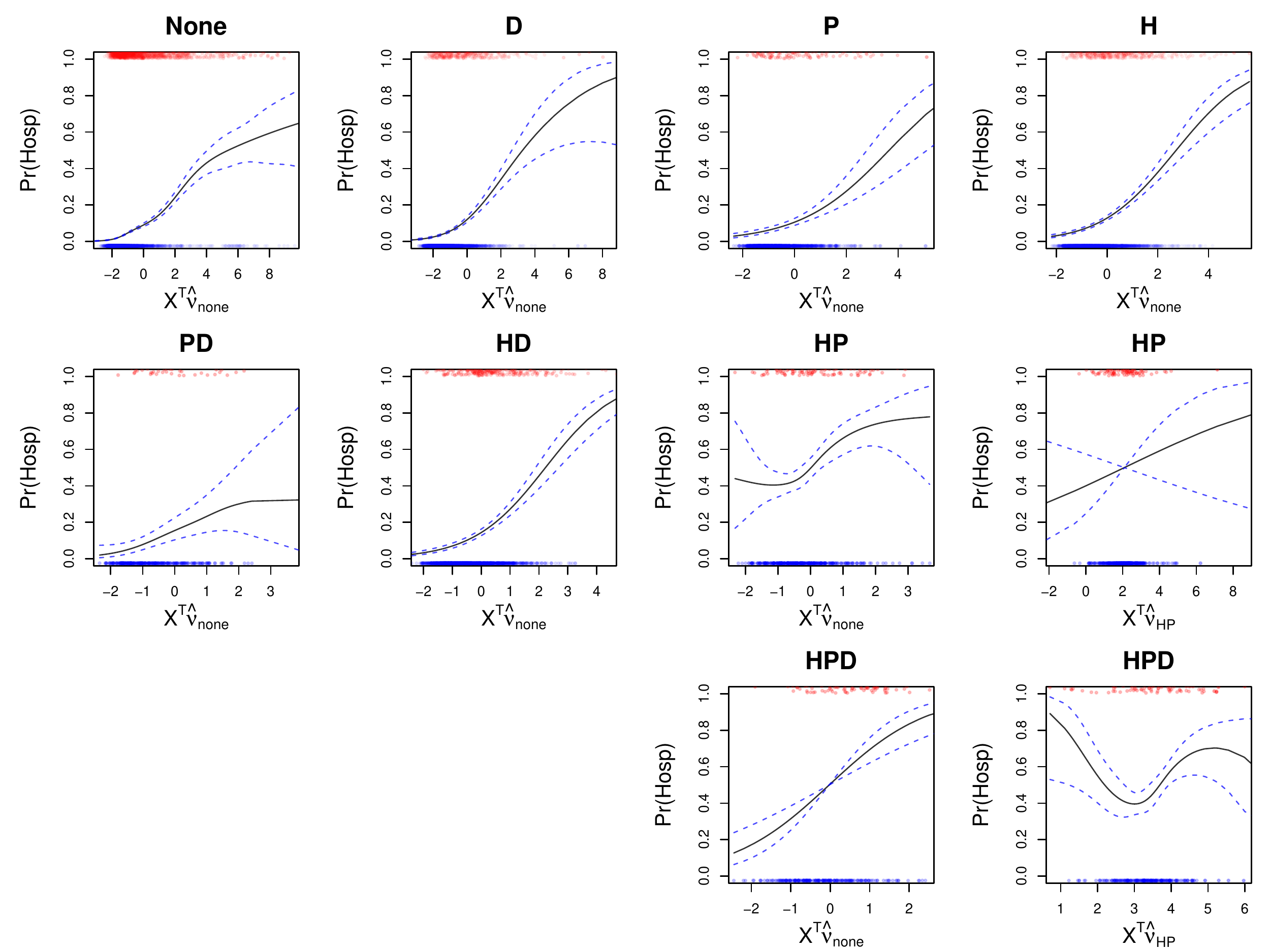}
	\caption{Plots of the predicted probabilities of hospitalization given each estimated direction. Confidence intervals are at the $95\%$ level. For the $HP$ and $HPD$ subpopulations, the plots are marginal on each of the first order terms with the functional ANOVA interaction term left out. The red dots represent patients with a hospitalization or death within 90 days and blue dots represent patients with no such hospitalization or death. This figure appears in color in the electronic version of this article, and any mention of color refers to that version.
	}
	\label{fig:fitted_models_smoothed_terms}
\end{figure}

The AUCs of each method for each subpopulation on the validation dataset are displayed in Table \ref{table:auc_hospitalization_with_none}. The proposed approach yields increased AUCs over the other SDR approaches for most of the subpopulations. Further, the performance of the proposed approach evaluated on the entire population is also improved relative to other approaches. A subset of the estimated coefficients from the hier sPHD method along with their standard errors estimated via a nonparametric bootstrap are displayed in Table \ref{table:some_coefs}. 
The predicted probabilities of hospitalization given the directions estimated by hier sPHD are displayed in Figure \ref{fig:fitted_models_smoothed_terms}. This figure appears in color in the electronic version of this article, and any mention of color refers to that version. 
Noting that the predicted probabilities of admission increase marginally in  both $\bfx^\transp\hat{\bnu}_{none}$ and $\bfx^\transp\hat{\bnu}_{HP}$, the estimated coefficients have a clear interpretation. In particular, the number of diagnoses related to chronic conditions has a positive coefficient for both $\hat{\bnu}_{none}$ and $\hat{\bnu}_{HP}$, indicating that more such diagnoses result in a larger probability of hospitalization for all patients and that for patients with both CHF and COPD (HP), the impact of the number of such diagnoses is significantly larger than for patients without both such conditions. A similar pattern holds also for those who had palliative care in the baseline, end stage renal disease (ESRD), and the number of times GFR was ordered in the baseline. The number of unique drugs in the baseline period  increases the hospitalization risk for all, however this increase in risk does not change for those with H and P. On the other hand, the number of secondary cancer sites is associated with a large increase in hospitalization risk for those with H and P, but not significantly so for those without both H and P. Other variables that significantly increase risk for those with H and P but not for other patients include the number of Emergency Room related payments in the baseline, whether the patient had home health care in the baseline, and the number of days since the last hospitalization. 

\begin{table}
\caption{AUCs of the predicted probability of hospitalization for the 8 subpopulations in addition to the entire population. The structural dimension for Single is 3, the structural dimensions for sPHD for all populations are 3, 1, 1, 1, 1, 1, 2, 1 in the respective order of the populations above and the structural dimensions for PHD are all 1.}
\label{table:auc_hospitalization_with_none}
\centering
\resizebox{0.9\textwidth}{!}{
	\begin{tabular}{crcrcccccc}
		\\[-1.8ex]
		\toprule \\[-1.8ex]
		& \multicolumn{2}{c}{Training Data} & \multicolumn{2}{c}{Validation Data}  & \multicolumn{5}{c}{Validation AUC} \\
		\cmidrule{2-3} \cmidrule{4-5} \cmidrule{6-10}
		Subpopulation & $n$ & $\frac{1}{n}\sum_{i=1}^nY_i$ & $n$ & $\frac{1}{n}\sum_{i=1}^nY_i$ & hier sPHD & Single & sPHD & PHD & N\&T \\
		\midrule
		None & 36203 & 0.040 & 29409 & 0.031 & \textbf{0.763} & 0.716 & 0.739 & 0.649 & 0.636 \\
		D & 7048 & 0.060 & 5679 & 0.051 & \textbf{0.734} & 0.702 & 0.666 & 0.646 & 0.595 \\
		H & 3628 & 0.121 & 2641 & 0.111 & \textbf{0.695} & 0.658 & 0.610 & 0.619 & 0.629 \\
		P & 1227 & 0.089 & 1023 & 0.071 & \textbf{0.778} & 0.706 & 0.701 & 0.654 & 0.646 \\
		DH &  1715 & 0.140 & 1294 & 0.150 & \textbf{0.698} & 0.661 & 0.656 & 0.630 & 0.569 \\
		DP & 327 & 0.107 & 277 & 0.094 & 0.696 & 0.667 & 0.589 & \textbf{0.739} & 0.630 \\
		HP & 448 & 0.194 & 347 & 0.196 & 0.638 & 0.625 & \textbf{0.672} & 0.558 & 0.630 \\
		DHP & 276 & 0.243 & 217 & 0.171 & \textbf{0.697} & 0.655 & 0.561 & 0.604 & 0.679 \\
		\midrule
		All & 50872 & 0.056 & 40887 & 0.046 & \textbf{0.786} & 0.745 & 0.759 & 0.714 & 0.710 \\
		\bottomrule \\[-1.8ex]
	\end{tabular}
}	
\end{table}

\begin{table}
	\caption{Displayed are a subset of the estimated coefficients from the hier sPHD method applied to the hospital admission data. Standard errors are estimated via a nonparametric bootstrap with 100 iterations. }
	\label{table:some_coefs}
	\centering 
	\resizebox{1\textwidth}{!}{
		\begin{tabular}{cccccl}
			\\[-1.8ex]
			\toprule \\[-2.8ex]
			$\large{\hat{\bnu}_{none}}$ & $\large{\hat{\bnu}_{HP}}$ && $\text{SE}(\large{\hat{\bnu}_{none}})$ & $\text{SE}(\large{\hat{\bnu}_{HP}})$ &  Variable \\
			\midrule
			\hphantom{-}0.170 & \hphantom{-}0.356 && 0.108 & 0.178 & Num. diagnoses related to chronic conditions  \\
			\hphantom{-}0.055 & -0.008                    && 0.017 & 0.016 &  Num. unique drugs taken in baseline \\
			\hphantom{-}0.626 & \hphantom{-}0.480 && 0.334 & 0.184 &  Any palliative care in baseline \\
			\hphantom{-}0.043 & \hphantom{-}0.494 && 0.075 & 0.311 &  Num. secondary cancer sites \\
			\hphantom{-}0.033 & \hphantom{-}0.057 && 0.018 & 0.027 &  Num. times GFR lab ordered (tests kidney function) \\
			\hphantom{-}0.526 & \hphantom{-}1.445 && 0.228 & 0.407 &  ESRD in baseline \\
			\hphantom{-}0.060 & \hphantom{-}0.185 && 0.058 & 0.092 &  Num. of Emergency Room related payments \\
			\hphantom{-}0.107 & \hphantom{-}0.024 && 0.043 & 0.055 &  Num. of Acute Care related payments \\
			\hphantom{-}0.000 & \hphantom{-}0.006 && 0.001 & 0.002 &  Any home health care in baseline \\
			\hphantom{-}0.001 & \hphantom{-}0.004 && 0.001 & 0.001 &  Days since last hospitalization \\
			-0.051 & \hphantom{-}0.104 && 0.029 & 0.025 &  Num. months covered by Medicare in baseline \\
			\bottomrule \\[-1.8ex]
		\end{tabular}
	}
\end{table}

\vspace{-2.5em}

\section{Discussion}\label{sec:discussion}

In this paper we introduced a semiparametric estimation framework to estimate central mean subspaces for heterogeneous populations defined by binary factors. We developed a class of estimators that borrow strength across related subpopulations, and proposed an approach for the determination of the structural dimensions. Although we did not explicitly deal with estimation of central subspaces, the methodology and results can be generalized to target central subspaces such as in \citet{Ma2012}. Web Appendix B contains a derivation of the orthogonal complement of the nuisance tangent space for this setting. 

An interesting extension of the proposed framework would be the development of locally efficient estimators of the central mean subspaces such as in \citet{Ma2014}. Semiparametric analysis when the central mean subspaces are hierarchically nested as in \eqref{eqn:hierarchical_assumption_none} may bring interesting challenges and may yield estimators with improved performance.

\vspace{-1.5em}

\section*{Data Availability Statement}

 The data that support the findings in this paper are not publicly available due to privacy or ethical restrictions.

\vspace{-1.5em}

{
	\bibliographystyle{biom}
	\bibliography{sdr_hierarchical_main_biom}
}

\vspace{-2.5em}

\section*{Supporting Information}

The Web Appendices containing (a) a method for dimension determination (b) additional theoretical results and derivations,  (c) proofs of theoretical results, and (d) additional simulation results referenced in Sections 2,3,4,5, and 6 are available with this paper at the Biometrics website on Wiley Online Library. The R code implementing the method is available on the same website and as the R package hierSDR at \url{https://github.com/jaredhuling/hierSDR}. \vspace*{-8pt}

\makeatletter\@input{xx.tex}\makeatother
\end{document}


\maketitle

\section{Additional results}

\subsection{Selection of the bandwidth parameter}

Another issue in practice is the selection of the bandwidth in the nonparametric regressions for the terms in \eqref{eqn:sample_est_eqn}. As is noted in \citet{Ma2014}, the estimator is not critically sensitive to this choice and can be chosen via cross validation or other standard approaches. Throughout this paper use a simple but effective subpopulation-specific choice of the bandwidth, $c_\bfz(3n_\bfz/4)^{-1/(d_\bfz + 4)}$  \citep{Ma2012}, where $c_\bfz$ is the average standard deviation of $(\bfX_i^\transp\bsbeta_\bfz: i=1, \dots, n, \bfZ_i = \bfz)$. 

\subsection{Asymptotic results for unconstrained estimator $\widetilde{\bbeta}$ in \eqref{eqn:est_unconstr}}

Recall that we let $\vecl(\boldsymbol A)$ be the operator which vectorizes the lower $(a - b)\times b$ block of an $a\times b$ matrix $\boldsymbol A$ when $a\ge b$. Denote $\vecl{(\bsbeta)}^\transp \equiv (\vecl{(\bsbeta_\bfz)}^\transp:\bfz \in \mathcal{Z}^C)$ as the lower block vectorization of $\bsbeta \equiv (\bsbeta_\bfz:\bfz \in \mathcal{Z}^C)$. 
\begin{theoremsupp}\label{thm:thm1a}
	Assume regularity conditions (C1)-(C5) listed in Section \ref{sec:regularity_conditions} hold for $\bsbeta_\bfz^*:\bfz \in \mathcal{Z}^C$. Further assume that ${\boldsymbol W}_n\xrightarrow{p}\boldsymbol W$ as $n\to\infty$, where $\boldsymbol W$ is a symmetric positive semi-definite matrix. Let
	\vspace{-0.1in}
	\begin{align*}
		\boldsymbol G_1 = {} & E\big(  \partial\vec\big[\left\{  Y - {E}(Y| \bfX^\transp{\coef}_{\bfZ}^*, \bfZ)  \right\} \left\{  \balpha(\bfX,\bfZ) - {E}(\balpha(\bfX,\bfZ) | \bfX^\transp{\coef}_{\bfZ}^*, \bfZ)  \right\}\big ] / \partial  \vecl(\coef)^\transp  \big).
	\end{align*}
	Further assume that $\boldsymbol \Sigma_1 = \{\boldsymbol G_1^\transp \boldsymbol W \boldsymbol G_1\}^{-1}$ is nonsingular. Then the minimizer $\wtbbeta$ of \eqref{eqn:est_unconstr} satisfies 
	\begin{equation} \label{asymptotic distribution of beta tilde}
		\sqrt{n}\vecl(\widetilde{\coef} - \coef^*) \xrightarrow{\mathcal{L}}  \mathcal{N}(\boldsymbol 0, \boldsymbol \Sigma_1 \boldsymbol \Xi_1 \boldsymbol \Sigma_1),
	\end{equation} where $\xrightarrow{\mathcal{L}}$ is convergence in distribution, $\coef^* = (\coef^*_\bfz : \bfz \in \mathcal{Z}^C)$, 
	$\boldsymbol \Xi_1 = \boldsymbol G_1^\transp\boldsymbol W \boldsymbol V_1 \boldsymbol W\boldsymbol G_1$, and 
	\vspace{-0.1in}
	\begin{align*}
		\boldsymbol V_1  = {} &  \mbox{cov} \big (  \vec \big [\left\{  Y - {E}(Y| \bfX^\transp{\coef}_{\bfZ}^*, \bfZ)  \right\} \left\{  \balpha(\bfX,\bfZ) - {E}(\balpha(\bfX,\bfZ) | \bfX^\transp{\coef}_{\bfZ}^*, \bfZ)  \right\} \big ] \big  ).
	\end{align*}
\end{theoremsupp}

An important point is that Theorem \ref{thm:thm1a} is for fixed dimensions $d_\bfz$ similar to \citet{Ma2012}. In practice these dimensions need to be estimated. We present an approach for such determination in the following section, which is a special case of the dimension determination approach we propose in Section \ref{sec:method}. 
Based on \eqref{asymptotic distribution of beta tilde}, a sensible choice for ${\boldsymbol W}_n$ would be $\widehat{\boldsymbol V}_{1n}^{-1}$, where $\widehat{\boldsymbol V}_{1n}$ is any consistent estimator of ${\boldsymbol V}_{1}$. 

Note that $\boldsymbol G_2 \neq \boldsymbol G_1$ and $\boldsymbol V_2 \neq \boldsymbol V_1$ due to the difference in how the upper blocks of parameters are imposed in $\widehat{\bbeta}$ versus $\widetilde{\bbeta}$, indicating a differing number of \textit{free} parameters. However, comparing the variance terms between \eqref{asymptotic distribution of beta tilde} and \eqref{asymptotic distribution of beta hat}, we would generally expect improvement in efficiency due to the product term $(\boldsymbol I - \boldsymbol P)$ in   \eqref{asymptotic distribution of beta hat}.

\subsection{Relationship of $\wtbbeta$ with \citet{li2003dimension}}

In our estimator, the emphasis is on estimating the central mean subspaces \textit{conditional} on subpopulations, as opposed to the \textit{central partial mean subspace}, which is the direct sum of the conditional mean subspaces \citep{li2003dimension}.  

We note that the central partial mean subspace of \citet{li2003dimension} can be seen as a special case of \eqref{eqn:est_unconstr} of the main text. To see this, let $\balpha(\bfx,\bfz)$ be the vector that takes all values equal to zero except in the $p$ locations corresponding to the terms $\bsbeta_\bfz$ in $\bsbeta$ where it takes values equal to $\bfx$. Further, take ${E}(Y| \bfX^\transp{\coef}_{\bfZ}^*, \bfZ) = 0$ and ${E}(\bfX| \bfX^\transp{\coef}_{\bfZ}^*, \bfZ) = 0$, which requires additional assumptions about $\bfX$ for consistent estimation. The approach of \citet{li2003dimension}  is based on ordinary least squares (OLS) regressions for each subpopulation, limiting the dimension of the central mean subspace for each subpopulation to be 1. Thus, to compare \eqref{eqn:est_unconstr} with this approach, assume that each $d_\bfz = 1$, and further take ${\boldsymbol W}_n = {\boldsymbol I}$. Then \eqref{eqn:est_unconstr} reduces to fitting OLS regressions for each subpopulation. Thus, if we use the span of the resulting subpopulation-specific estimates $\wtbbeta_\bfz$ to recover the central partial mean subspace, \eqref{eqn:est_unconstr} relates directly to the approach of \citet{li2003dimension}. Due to the general form of \eqref{eqn:pop_est_eqn}, pooled OLS-type estimators are also possible, for one possibility among many, one could take a common $\balpha(\bfx,\bfz) = \text{rep}(\bfx, 2^C)$ for all subpopulations, where $\text{rep}(\cdot)$ forms a vector by repeating $\bfx$ for $2^C$ times.

\subsection{Example of construction of the constraint matrix $\bsC$}

Here, we show how the constraint matrix $\bsC$ is constructed to enforce the hierarchical conditions. 
For illustrative purposes we consider a simplified scenario with models stratified by two factors and a specific choice of structural dimensions, but the principle applies more generally. 
Let $\mathcal{Z}^C = \{00,01,10,11\}$, $d_{00}=1$, $d_{10}=2$, $d_{01}=2$, and $d_{11}=4$, so that the combined unconstrained parameter matrix $\bsbeta = (\bsbeta_{\bfz}:\bfz\in \mathcal{Z}^C)$ can be expressed as
\begin{align*}
	\bsbeta = ( \underbrace{\bnu_{00}}_{\bsbeta_{00}},  \underbrace{\bnu^1_{00}, \bnu_{01}}_{\bsbeta_{01}},  \underbrace{\bnu^2_{00}, \bnu_{10}}_{\bsbeta_{10}}, \underbrace{\bnu^3_{00}, \bnu^1_{01}, \bnu^1_{10},\bnu_{11}}_{\bsbeta_{11}}),
\end{align*}
where it is not necessarily the case that $\bnu_{00}$ equals $\bnu^1_{00}$, $\bnu^2_{00}$, or $\bnu^3_{00}$ and not necessarily the case that $\bnu_{01}=\bnu^1_{01}$ or $\bnu_{10}=\bnu^1_{10}$, but we use this naming convention as our goal in constructing $\bsC$ is to make $\bnu_{01}=\bnu^1_{01}$, $\bnu_{10}=\bnu^1_{10}$, and so on. Here, the dimension of all $\bnu$ terms is 1, so  the constraint that the upper blocks of the $\bnu$ terms is identity implies that the free parameters in each are the lower $p-1$ elements. We denote the free parameters of $\bnu_{00}$ as  $\bnu_{00,L}$, the free parameters of $\bnu^1_{00}$ as $\bnu^1_{00,L}$, and so on. Using this notation, 
here, $\vecl(\bsbeta) = (\bnu_{00,L}, \bnu^1_{00,L}, \bnu_{01,L}, \bnu^2_{00,L}, \bnu_{10,L}, \bnu^3_{00,L}, \bnu^1_{01,L}, \bnu^1_{10,L},\bnu_{11,L})^T$, since the dimension of each $\bnu$ term here is 1. To enforce the hierarchical constraints, we need 
\begin{align*}
	\bsC^T\vecl(\bsbeta) = 
	\begin{pmatrix}
		\bnu_{00,L}-\bnu^1_{00,L} \\
		\bnu_{00,L}-\bnu^2_{00,L} \\
		\bnu_{00,L}-\bnu^3_{00,L} \\
		\bnu_{01,L}-\bnu^1_{01,L} \\
		\bnu_{10,L}-\bnu^1_{10,L} \\
	\end{pmatrix} = \bzero,
\end{align*}
which is a $5p\times 1$ matrix. The constraint matrix that yields this is
\[
\bsC^T = 
\begin{pmatrix}
	\bsI_{p-1} & -\bsI_{p-1} & \bzero & \bzero & \bzero & \bzero & \bzero & \bzero & \bzero \\
	\bsI_{p-1} & \bzero & \bzero & -\bsI_{p-1} & \bzero & \bzero & \bzero & \bzero & \bzero \\
	\bsI_{p-1} & \bzero & \bzero & \bzero & \bzero & -\bsI_{p-1} & \bzero & \bzero & \bzero \\
	\bzero & \bzero & \bsI_{p-1} & \bzero & \bzero & \bzero & -\bsI_{p-1} & \bzero & \bzero \\
	\bzero & \bzero & \bzero & \bzero & \bsI_{p-1} & \bzero & \bzero & -\bsI_{p-1} & \bzero \\
\end{pmatrix},
\]
where $\bsI_{p-1}$ are $(p-1)\times(p-1)$ identity matrices.

The key principle here is to construct a $\bsC$ that applies the appropriate contrasts that, when set to zero, enforce the desired constraints. Of course, if the terms in $\bsbeta$ are switched around, $\bsC$ will be different, but these differences are unimportant and is not consequential in any way. When the dimension $k$ of any $\bnu$ term is larger than 1, an additional set of rows of constraints is necessary, but the dimension of each set of constraints would be $p-k$ instead of $p-1$.

\subsection{Determination of structural dimensions}\label{sec:dimsel}

Determining the structural dimension in SDR in the semiparametric framework is a challenging problem. The challenges inherent in dimension determination are amplified for the determination of multiple dimensions, especially when the different dimensions are relevant across multiple subpopulations. 
In the setting of this paper, we are interested in estimating the structural dimensions of the true parameters $\bnu^0_\bfz$ for $\bfz \in \mathcal{Z}^C$. Due to \eqref{eqn:equality_condition}, the set of dimensions of $\bnu_\bfz$ for $\bfz \in \mathcal{Z}^C$ uniquely determines the set of dimensions of $\bsbeta_\bfz = (\bnu_{\bfz'} : \bfz' \preceq \bfz, \bfz' \in \mathcal{Z}^C)$ for $\bfz \in \mathcal{Z}^C$; thus we focus on the former. 

\citet{Ma2015} proposed a simple, effective, and computationally tractable information criterion-based approach, called the validated information criterion (VIC).
We extend their approach to our setting with the imposed hierarchical assumption \eqref{eqn:hierarchical_assumption_none} and \eqref{eqn:none_assumption2}.

Denote a given set of candidate dimensions as ${\boldsymbol k\ } \equiv (k_\bfz :\bfz\in \mathcal{Z}^C)$ for the terms $\bnu_{\bfz}$  and denote the true dimensions as ${{\boldsymbol  k\ }\!\!}^0 \! = (k_\bfz^0 : \bfz \in \mathcal{Z}^C)$ of the parameters ${\bnu}_\bfz^0$. Our proposal arises by noting that the population estimating equations should be zero when $\sum_{ \bfz' \preceq \bfz} k_ {\bfz'} \equiv {d}_\bfz \geq {d}_\bfz^0 \equiv \sum_{ \bfz' \preceq \bfz} k_ {\bfz'}^0$ for all $\bfz$ and should be nonzero when ${d}_\bfz < {d}_\bfz^0$ for any $\bfz$. 
Denote ${\boldsymbol d}= ({d}_\bfz : \bfz \in \mathcal{Z}^C)$, ${\boldsymbol d}^0= ({d}_\bfz^0 : \bfz \in \mathcal{Z}^C)$, and $\bbeta_{(\boldsymbol d)} = (\bbeta_{\bfz} : \bfz \in \mathcal{Z}^C)$ where each $\bbeta_{\bfz}$ is of dimension $p\times d_\bfz$ and defined through \eqref{eqn:equality_condition}. 
We estimate $\bbeta_{(\boldsymbol d)} $ similar to \eqref{eqn:constr_min} by solving
\begin{align*}
	\whbbeta_{(\boldsymbol d)}= \argmin_{\bbeta_{(\boldsymbol d)} = (\bbeta_{\bfz} : \bfz \in \mathcal{Z}^C)} {} \biggl\{&\frac{1}{2n}\hat{\Psi}_{n}(\bbeta_{(\boldsymbol d)})^\transp\boldsymbol W_n\hat{\Psi}_{n}(\bbeta_{(\boldsymbol d)}) \mbox{ s.t. } \boldsymbol \bsC_{(\boldsymbol d)}^\transp \vecl(\bsbeta_{(\boldsymbol d)}) = \boldsymbol 0 \biggr\},  
\end{align*}
where  $\bsC_{(\boldsymbol d)}$ is constructed similarly to $\bsC$ but accounting for different dimensions of the terms in $\bsbeta_{(\boldsymbol d)}$ from those under true dimensions ${\boldsymbol d}^0$. Similarly denote the population version as
\begin{align*}
	\bbeta^0_{(\boldsymbol d)} = \argmin_{\bbeta_{(\boldsymbol d)} = (\bbeta_{\bfz} : \bfz \in \mathcal{Z}^C)} {} \biggl\{ &\frac{1}{2}\Psi_{0}(\bbeta_{(\boldsymbol d)})^\transp\boldsymbol W_n\Psi_{0}(\bbeta_{(\boldsymbol d)}) 
	\mbox{ s.t. } \boldsymbol \bsC_{(\boldsymbol d)}^\transp \vecl(\bsbeta_{(\boldsymbol d)}) = \boldsymbol 0 \biggr\}. 
\end{align*}
Note that under the true structural dimensions $\boldsymbol d^0$, $\bbeta^0_{(\boldsymbol d)}$ should yield estimated subspaces that satisfy \eqref{eqn:hierarchical_assumption_none} and \eqref{eqn:none_assumption2} if they are indeed correct, in which case $\bbeta^0_{(\boldsymbol d)} = \bbeta^0$. 

Since the estimates $\whbbeta$ are constructed by minimizing the norm of the estimating equations, evaluation of the estimating equations at $\whbbeta$ does not necessarily provide a valid means of assessing the fit of the model. As a workaround, similar to \citet{Ma2015}, the basic idea of the approach is to construct expansions  of the parameters $\bnu_{\bfz}$ for each population from dimension $k_\bfz$ to $k_\bfz+1$. The expanded parameters are used as a means to assess model fit.
In particular, write ${\bnu_{\bfz}^0}^\transp = (I_{k_\bfz}, {\bnu^{0,U}_{\bfz}}^\transp, {\bnu^{0,L}_{\bfz}}^\transp )$ and ${\widehat{\bnu}_{\bfz}}^\transp = (I_{k_\bfz}, {{}\widehat{\bnu}^{U}_{\bfz} }^\transp, {{}\widehat{\bnu}^{L}_{\bfz}}^\transp)$,
where $\bnu^{0,U}_{\bfz}$ and $\widehat{\bnu}^{U}_{\bfz}$ are $1 \times k_\bfz$ vectors, 
$\bnu^{0,L}_{\bfz}$ and $\widehat{\bnu}^{L}_{\bfz}$ are $(p - 1 - k_\bfz)\times k_\bfz$ matrices where for simplicity of presentation we drop explicit reference to the dimension in the notation. The  matrices $\bnu^0_{\bfz}(v)$ and $\widehat{\bnu}_{\bfz}(v)$ are constructed so that for any $v$, 
$\bnu^0_{\bfz}(v) (I_{k_\bfz},
\bnu^{0,U}_{\bfz})^\transp = \bnu^{0}_{\bfz}$ and
$\widehat{\bnu}_{\bfz}(v) (I_{k_\bfz},
\widehat{\bnu}^{U}_{\bfz})^\transp=\widehat{\bnu}_{\bfz}$. 
Therefore the spaces spanned by the columns of $\bnu^0_{\bfz}(v)$ and $\widehat{\bnu}_{\bfz}(v)$ contain the spaces spanned by the columns of $\bnu^{0}_{\bfz}$ and $\widehat{\bnu}_{\bfz}$ respectively. 
In particular, write ${\bnu_{\bfz}^0}^\transp = (I_{k_\bfz}, {\bnu^{0,U}_{\bfz}}^\transp, {\bnu^{0,L}_{\bfz}}^\transp )$ and ${\widehat{\bnu}_{\bfz}}^\transp = (I_{k_\bfz}, {{}\widehat{\bnu}^{U}_{\bfz} }^\transp, {{}\widehat{\bnu}^{L}_{\bfz}}^\transp)$,
where $\bnu^{0,U}_{\bfz}$ and $\widehat{\bnu}^{U}_{\bfz}$ are $1 \times k_\bfz$ vectors, 
$\bnu^{0,L}_{\bfz}$ and $\widehat{\bnu}^{L}_{\bfz}$ are $(p - 1 - k_\bfz)\times k_\bfz$ matrices where for simplicity of presentation we drop explicit reference to the dimension in the notation. Then define
\begin{equation*}
	\bnu^0_{\bfz}(v) = 
	\begin{pmatrix}   I_{k_\bfz} & \boldsymbol 0_{k_\bfz\times 1}\\ 
		\boldsymbol 0_{1\times k_\bfz}  & 1 \\
		\bnu^{0,L}_{\bfz} - \boldsymbol v \bnu^{0,U}_{\bfz} & \boldsymbol v  \end{pmatrix},
	\widehat{\bnu}_{\bfz}(v) = 
	\begin{pmatrix}   I_{k_\bfz} & \boldsymbol 0_{k_\bfz\times 1}\\ 
		\boldsymbol 0_{1\times k_\bfz}  & 1 \\
		\widehat{\bnu}^{L}_{\bfz} - \boldsymbol v \widehat{\bnu}^{U}_{\bfz} & \boldsymbol v  \end{pmatrix},
\end{equation*}
where $\boldsymbol v$ is a conformable vector with all elements equal to $v$. It is easy to verify that for any $v$, 
$\bnu^0_{\bfz}(v) (I_{k_\bfz},
\bnu^{0,U}_{\bfz})^\transp = \bnu^{0}_{\bfz}$ and
$\widehat{\bnu}_{\bfz}(v) (I_{k_\bfz},
\widehat{\bnu}^{U}_{\bfz})^\transp=\widehat{\bnu}_{\bfz}$. The corresponding $\bbeta^0_{\bfz}(v)$ is defined through \eqref{eqn:equality_condition} with $\bnu^0_{\bfz}(v)$ in place of $\bnu_{\bfz}^0$ and $\bbeta^0_{(\boldsymbol d)}(v) = (\bbeta^0_{\bfz}(v) :  \bfz \in \mathcal{Z}^C)$.

The corresponding $\bbeta^0_{\bfz}(v)$ is defined through \eqref{eqn:equality_condition} with $\bnu^0_{\bfz}(v)$ in place of $\bnu_{\bfz}^0$ and $\bbeta^0_{(\boldsymbol d)}(v) = (\bbeta^0_{\bfz}(v) :  \bfz \in \mathcal{Z}^C)$. Similarly, $\widehat{\bbeta}_{\bfz}(v)$ is also defined through \eqref{eqn:equality_condition} with $\widehat{\bnu}^0_{\bfz}(v)$ in place of $\widehat{\bnu}_{\bfz}^0$ and $\widehat{\bbeta}_{(\boldsymbol d)}(v) = (\widehat{\bbeta}_{\bfz}(v) :  \bfz \in \mathcal{Z}^C)$.

Intuitively, if ${d}_{\bfz} < {d}_{\bfz}^0$ for any $\bfz \in \mathcal{Z}^C$, then $n^{-1}\Psi_n(\widehat{\bbeta}_{(\boldsymbol d)}(v))$ is in general inconsistent for zero. On the other hand, if ${d}_{\bfz} \ge {d}_{\bfz}^0$ for all $\bfz \in \mathcal{Z}^C$, then $n^{-1}\hat{\Psi}_n(\widehat{\bbeta}_{(\boldsymbol d)}(v))$ will still be consistent for zero even if ${d}_{\bfz'} > {d}_{\bfz'}^0$ for some $\bfz' \in \mathcal{Z}^C$. Thus, our proposed criterion for  dimension determination is
\begin{align}
	\mbox{VIC}({{\boldsymbol k}}) = {} & \frac{1}{rn}\sum_{j = 1}^{r}\lVert\hat{\Psi}_n(\widehat{\bbeta}_{(\boldsymbol d)}(v_j)) \rVert^2 + \log(n)p \sum_{\bfz \in \mathcal{Z}^C}   {d}_{\bfz}. \label{eqn:vic_criterion}
\end{align} 
We have proposed to use $r$ different values $v_1, \ldots, v_r$ for $v$. Although a particular $v$ should work but in theory, if $\bbeta^0_{(\boldsymbol d)}(v) = \bbeta^0_{(\boldsymbol d + 1)}$, then the VIC criterion for $r=1$ with $v$ will fail to work for dimension determination. Using $r$ different values for  $v_j$, it is possible that $\bbeta^0_{(\boldsymbol d)}(v_j) = \bbeta^0_{(\boldsymbol d + 1)}$ for any $r-1$ values $j \in \{1,\dots, r\}$. In practice, choosing a moderate number such as $r=5$ tends to perform well. 

To estimate the structural dimensions, one would evaluate \eqref{eqn:vic_criterion} for all valid sets of candidate dimensions. The estimates of the structural dimensions for $\bnu^{0}_{\bfz}$ for $\bfz \in \mathcal{Z}^C$, denoted as $\hat{{\boldsymbol  k\ }} = (\hat{k}_\bfz : \bfz \in \mathcal{Z}^C)$, is the minimizer of \eqref{eqn:vic_criterion}. Further, let the estimated structural dimensions for $\bbeta^{0}_{\bfz}$ for $\bfz \in \mathcal{Z}^C$ be denoted as $\hat{\boldsymbol d\ } = ( \sum_{\bfz' \preceq \bfz}\hat{k}_{\bfz'} : \bfz \in \mathcal{Z}^C)$. It is important to note that  $p \sum_{\bfz \in \mathcal{Z}^C}   {d}_{\bfz} $, instead of $p \sum_{\bfz \in \mathcal{Z}^C}   k_{\bfz} $,  should be used as a penalty in the VIC criterion.

\begin{theoremsupp}\label{thm:vic_consistency}
	Define $\boldsymbol d \leq {\boldsymbol d}^0$  to mean $d_\bfz \leq d_\bfz^0$ for any $\bfz \in \mathcal{Z}^C$.  	Assume conditions (C1)-(C5)  in the Supplement hold for $\bstheta_\bfz \in (\bbeta^0_{(\boldsymbol d)}, \bbeta^0_{(\boldsymbol d)}(v_j): \boldsymbol d \leq {\boldsymbol d}^0, j = 1, \dots, r)$ and assume that $E(\Psi_{0}(\bbeta_{(\boldsymbol d)}))=0$ has at most $r$ solutions for $\boldsymbol d \leq {\boldsymbol d}^0$. 
	Then $\mbox{Pr}(\hat{\boldsymbol d\ } = \boldsymbol d^0) \to 1$ as $n \to \infty$.
\end{theoremsupp}

Thus, minimization of \eqref{eqn:vic_criterion} leads to consistent determination of the structural dimensions. Our result shows that despite the constraints on the parameters, we are still able to develop a VIC-based approach to consistently estimate the structural dimensions. The presence of these constraints and the choice of the level of penalty make when using these constraints prevents direct application of the results of \citet{Ma2015} to our setting for dimension determination.

\subsection{Structural dimension determination for $\widetilde{\bbeta}$}

Similar to $\whbbeta_{(\boldsymbol d)}$ from Section \ref{sec:dimsel} of the main text, we define

\begin{align*}
	\wtbbeta_{(\boldsymbol d)} = \argmin_{\bbeta_{(\boldsymbol d)} = (\bbeta_{\bfz(d_\bfz)} : \bfz \in \mathcal{Z}^C)} {} & \frac{1}{2n}\hat{\Psi}_{n}(\bbeta_{(\boldsymbol d)})^\transp\boldsymbol W_n\hat{\Psi}_{n}(\bbeta_{(\boldsymbol d)}), 
\end{align*}
where $\wtbbeta_{(\boldsymbol d)} $ can be expressed as  $(\wtbbeta_{\bfz(d_\bfz)} : \bfz \in \mathcal{Z}^C)$.
Using this, we define the non-stochastic expansions $\wtbbeta_{\bfz(d_\bfz)}(v)$ of $\wtbbeta_{\bfz(d_\bfz)}$ similarly as for $\widehat{\bnu}_{(d_\bfz)}$. Similarly as for $\wtbbeta_{(\boldsymbol d)}(v)$, we define $\wtbbeta_{(\boldsymbol d)}(v) = (\wtbbeta_{\bfz(d_\bfz)}(v): \bfz\in \mathcal{Z}^C)$.
The proposed criterion for structural dimension determination is then
\begin{align}
	\mbox{VIC}(\boldsymbol d) = {} & \left\{  \frac{1}{rn}\sum_{j = 1}^{r}\lVert\hat{\Psi}_n(\wtbbeta_{(\boldsymbol d)}(v_j)) \rVert^2 + \log(n)p\sum_{\bfz \in \mathcal{Z}^C}d_\bfz  \right\}. \label{eqn:vic_criterion_unconstrained}
\end{align} 
To estimate the structural dimensions, one would evaluate \eqref{eqn:vic_criterion_unconstrained} for all valid sets of candidate dimensions and choose the dimensions that minimize \eqref{eqn:vic_criterion_unconstrained} as the estimated dimensions. A selection consistency result like Theorem \ref{thm:vic_consistency} can be shown for this minimizer in a similar fashion as the proof of Theorem \ref{thm:vic_consistency}.

 \vspace{-0.5cm}
\subsection{Implications of the hierarchy assumption misspecification}

We now consider implications of misspecifying the hierarchical assumption. 
\vspace{-0.5cm}
\begin{corollarysupp}\label{thm:corollary1}
	Under the assumptions of Theorem \ref{thm1}, when the hierarchical assumptions \eqref{eqn:hierarchical_assumption_none} and \eqref{eqn:none_assumption2} are not satisfied, i.e. $\bsC^\transp\vecl(\coef^0) \neq \boldsymbol 0$, the solution of \eqref{eqn:constr_min} can be represented as
	$$\vecl(\whbbeta) = (\boldsymbol I - \boldsymbol P)\vecl(\coef^0) + o_p(1) \text{ and} $$ 
	\begin{equation}
		n^{1/2}\big\{ \vecl(\widehat{\bbeta}) - (\boldsymbol I - \boldsymbol P)\vecl(\bbeta^0)\big\} \xrightarrow{\mathcal{L}} \mathcal{N}(\boldsymbol 0, (\boldsymbol I - \boldsymbol P)\boldsymbol \Sigma_2 \boldsymbol \Xi_2 \boldsymbol \Sigma_2(\boldsymbol I - \boldsymbol P) ),
	\end{equation}
	where $\boldsymbol P$, ${\boldsymbol \Sigma_2}$, and $\boldsymbol \Xi_2$ are as defined in Theorem \ref{thm1}.
\end{corollarysupp}

Thus, there can be possible bias due to the hierarchical assumption violation. The bias term $\boldsymbol P\vecl(\bbeta^0)={\boldsymbol \Sigma_2}\bsC(\bsC^\transp {\boldsymbol \Sigma_2}\bsC)^{-1}\bsC^\transp \vecl(\bbeta^0)$ is closely related to the hierarchical assumption equation, $\bsC^\transp \vecl(\bbeta^0)$. When the hierarchical assumption is not correct, but holds approximately, i.e. $\bsC^\transp \vecl(\bbeta^0) \approx {\boldsymbol 0}$, then the potential reduction in variance from assuming subspace hierarchy may still result in smaller mean squared error than the unconstrained estimator. 

Another implication of the corollary is that even when the assumptions \eqref{eqn:hierarchical_assumption_none} and \eqref{eqn:none_assumption2} do not hold for the true central mean subspaces themselves (i.e. the smallest dimension reduction subspace), forcing the assumption usually leads to a set of dimension reduction subspaces that are slightly larger than the true central mean subspaces. However with those larger subspaces, this assumption can hold. 
While such expansion leads to incorrect structural dimensions, its impact on consequent model building may not be critical as the extra directions will likely be unimportant. We will explore this further through numerical studies in Section \ref{sec:sim}.

\section{Proofs and derivations}\label{sec:proofs}

\subsection{Regularity conditions}\label{sec:regularity_conditions}

Conditions (C1)-(C5) below are stated using a generic parameter vector $\bstheta_\bfz$. For Theorem \ref{thm:thm1a}, it refers to $\bsbeta_\bfz^*$. For Theorem \ref{thm1} and Corollary \ref{thm:corollary1} it refers to $\bsbeta_{\bfz}^0$. For Theorem \ref{thm:vic_consistency}, it refers to elements in $\bstheta_\bfz \in \{\bbeta^0_{\boldsymbol d},  \bbeta^0_{\boldsymbol d}(v_j): \boldsymbol d \preceq {\boldsymbol d}^0, j=1, \ldots, r\}$. Denote $\nu(\bstheta_\bfz)$ as the number of columns in $\bstheta_\bfz$.

\begin{enumerate}[label={(C\arabic*)}]
	\item $n_\bfz/n \to p_\bfz \in (0,1)$ as $n \to \infty$ for all $\bfz \in \mathcal{Z}^C$.
	\item The univariate kernel function $K(\cdot)$ is Lipschitz and has compact support. Furthermore for all $\bfz \in \mathcal{Z}^C$ it satisfies 
	\begin{align*}
		& \int K(u) \mathrm{d} u = 1, \int u^iK(h_\bfz)\mathrm{d}u = 0, 1 \leq i \leq m_{\bstheta_\bfz} - 1, \\ &\mbox{ and } 0 \neq \int u^{m_{\bstheta_\bfz}}K(h_\bfz)\mathrm{d}u < \infty,
	\end{align*}
	where $m_{\bstheta_\bfz}$ is an integer related to the Lipschitz continuity of functions defined in condition (C4).
	The $\nu(\bstheta_\bfz)$-dimensional kernel function is a product of $\nu(\bstheta_\bfz)$ univariate kernel functions, i.e. $K_{h_\bfz}(\boldsymbol u) = K(\boldsymbol u / h_\bfz) / h ^ {\nu(\bstheta_\bfz)} = \prod_{j = 1}^{\nu(\bstheta_\bfz)}K(u_j/h_\bfz)/h^{\nu(\bstheta_\bfz)}$ for $\boldsymbol u = (u_1, \dots, u_{\nu(\bstheta_\bfz)})^\transp$. In the above notation, $K$ is used regardless of dimension.
	
	\item The joint density/mass functions of $(\bfX, \bfZ)$ and $(\bfX^\transp\bstheta_\bfZ, \bfZ)$, respectively denoted as $f_{\bfX,\bfZ}(\bfx, \bfz) = f_{\bfX|\bfZ}(\bfx)\pi_{\bfZ}(\bfz)$ and $f(\bfx^\transp\bstheta_\bfz, \bfz) = f_{\bfX^\transp\bstheta_\bfZ|\bfZ}(\bfx^\transp\bstheta_\bfz)\pi_{\bfZ}(\bfz)$ are bounded from below and above, where $f_{\bfX|\bfZ}(\bfx)$ is the density function of $\bfX$ given $\bfZ$, $\pi_{\bfZ}(\bfz)$ is the mass function of $\bfZ$, and $f_{\bfX^\transp\bstheta_\bfZ|\bfZ}(\bfx^\transp\bstheta_\bfz)$ is the density function of $\bfX^\transp\bstheta_\bfZ$ given $\bfZ$. Furthermore, $E[Y^2 | \bfX^\transp\bstheta_\bfz = \bxi_\bfz, \bfZ = \bfz]$ and each entry of $E[\boldsymbol \alpha(\bfX, \bfz)\boldsymbol \alpha(\bfX, \bfz)^\transp | \bfX^\transp\bstheta_\bfz = \bxi_\bfz, \bfZ = \bfz]$  are locally Lipschitz-continuous and bounded from above as a function of $\bxi_\bfz$ for every $\bfz \in \mathcal{Z}^C$, where a function $\boldsymbol g(\bfx)$ from $\mathcal{S}\subset \mathbbm{R}^k$ to $\mathbbm{R}^r$ is locally Lipschitz-continuous if for every point $\bfx\in\mathcal{S}$, there is a constant $L>0$ such that $\boldsymbol g$ is Lipschitz-continuous on the open ball $\mathcal{B}_L(\bfx)$ with center $\bfx$ and radius $L$. A function $\boldsymbol g$ is Lipschitz continuous on $\mathcal{B}$ if there is a constant $M>0$ such that $\| \boldsymbol g(\bfx) - \boldsymbol g(\bfy) \| \leq M\|\bfx - \bfy\|$ for all $\bfx, \bfy \in \mathcal{B}$.
	
	\item Let $$\boldsymbol r_{1}(\bxi_\bfz, \bfz) = E[\boldsymbol\alpha(\bfX, \bfz) | \bfX^\transp\bstheta_\bfz = \bxi_\bfz,\bfZ = \bfz]f(\bxi_\bfz, \bfz)$$ and $$\boldsymbol r_2(\bxi_\bfz, \bfz) = E[Y | \bfX^\transp\bstheta_\bfz = \bxi_\bfz,\bfZ = \bfz]f(\bxi_\bfz, \bfz).$$ The $m_{\bstheta_\bfz}$th derivatives of $\boldsymbol r_{1}(\bxi_\bfz, \bfz)$, $\boldsymbol r_{2}(\bxi_\bfz, \bfz)$, and $f(\bxi_\bfz, \bfz)$ are locally Lipschitz-continuous in $\bxi_\bfz$ for every $\bfz \in \mathcal{Z}^C$.
	
	\item For all $\bfz \in \mathcal{Z}^C$ the bandwidths $h_\bfz = O(n^{-\kappa_{\bstheta_\bfz}})$ for $1 / (4m_{\bstheta_\bfz}) < \kappa_{\bstheta_\bfz} < 1/(2\nu(\bstheta_\bfz))$ where $\kappa_{\bstheta_\bfz} > 0$ is a constant.
\end{enumerate}

Condition (C1)  indicates that the sample size for any subpopulation should not vanish with respect to the overall sample size. Condition (C2) is a standard requirement in nonparametric kernel regression estimation for multivariate kernels and merely ensures the kernel used is valid. Conditions (C3) and (C4) are standard smoothness and boundedness assumptions that are necessary to ensure order of operations can be exchanged and appropriate terms can be differentiated. Condition (C5) is a bandwidth assumption necessary to ensure root-$n$  convergence of the parameters.

\subsection{Derivation of the orthogonal complement of the nuisance tangent space for the conditional mean}\label{derivation}

Recall the model from the main text:
\begin{equation}\label{eqn:mean_models}
Y =  \sum_{\bfz \in \mathcal{Z}^C}I(\bfZ = \bfz)\ell_\bfz(\bfX^\transp \coef_\bfz^* ) + \epsilon, 
\end{equation}
where 
$\epsilon$ is an error term with $E(\epsilon|\bfX, \bfZ) = 0$. Note that 
$\ell_{\bfz}(\bfX^\transp\coef_{\bfz}^*)$ is the mean regression function for subpopulation $\bfz$. Denote $\eta_{\bfX|\bfZ}$ as the density of $\bfX$ given $\bfZ$, $\pi_\bfZ$ as the probability mass function (pmf) of $\bfZ$, and  $\eta_\epsilon$ as the density of $\epsilon$ conditional on $\bfX$ and $\bfZ$ with respect to some dominating measure. We use the notation $\pi$ for pmfs and $\eta$ for density functions.
We begin by writing down the likelihood of a single observation:
\begin{equation}\label{eqn:lik_obs}
\eta_{\epsilon}\left(Y - \sum_{\bfz \in \mathcal{Z}^C}I(\bfZ = \bfz) \ell_{\bfz}(\bfX^\transp\coef_{\bfz}^*) \right)\eta_{\bfX|\bfZ}(\bfX, \bfZ)\pi_{\bfZ}(\bfZ)
\end{equation}

In a similar vein as \cite{Ma2012}, in order to characterize all estimation functions associated with \eqref{eqn:mean_models}, we seek to identify the orthogonal complement of the nuisance tangent space in order to derive the space  of all estimating functions for estimation of $\coef^* = (\coef_\bfz^*: \bfz\in \mathcal{Z}^C)$ in (\ref{eqn:mean_models}). Denote as the mean zero Hilbert space $\mathcal{H} = \{\bff(\epsilon, \bfX, \bfZ) : E(\bff) = 0, E(\bff^\transp\bff) < \infty, \bff \text{ is measurable} \}$ with inner product between any two functions $\bff$ and $\bfg$ defined as $E(\bff^\transp \bfg)$. 
The tangent spaces corresponding to the infinite dimensional nuisance parameters $\eta_{\bfX|\bfZ}$, $\pi_\bfZ$, and $\eta_\epsilon$ are 
\begin{align*}
\Lambda_\bfZ = & {} \{\bff(\bfX) \in \mathcal{H} : E(\bff) = 0 \} \\
\Lambda_{\bfX|\bfZ} = & {} \{   \bff(\bfX, \bfZ) \in \mathcal{H} : E(\bff | \bfZ) = 0  \} \\
\Lambda_\epsilon = & {} \{   \bff(\epsilon, \bfX, \bfZ) \in \mathcal{H} : E(\bff|\bfX, \bfZ) = 0 \text{ and } E(\bff\epsilon|\bfX, \bfZ) = 0  \}
\end{align*}
The tangent space of the unknown mean functions $\boldsymbol{\ell} = (\ell_\bfz : \bfz \in \mathcal{Z}^C)$ is 
\begin{align*}
\Lambda_{\boldsymbol{\ell}}= & \left\{  \frac{\eta_{\epsilon, 1}'(\epsilon, \bfX, \bfZ)}{\eta_{\epsilon}(\epsilon, \bfX, \bfZ)} \sum_{\bfz \in \mathcal{Z}^C}I(\bfZ = \bfz) h_{\bfz}(\bfX^\transp\coef_{\bfz}^*), 
\forall  h_{\bfz}(\bfX^\transp\coef_{\bfz}^*) : \bfz \in \mathcal{Z}^C,  h_{\bfz}(\bfX^\transp\coef_{\bfz}^*) : \mathcal{X} \mapsto \mathbbm{R}   \right\}
\end{align*}

$\Lambda_\epsilon$ represented as the intersection of the following two linear subspaces
\begin{align}
\Lambda_{\epsilon a} = & \{\bff(\epsilon, \bfX, \bfZ) \in \mathcal{H}: E(\bff|\bfX, \bfZ) = 0\} \mbox{ and}\\
\Lambda_{\epsilon b} = & \{\bff(\epsilon, \bfX, \bfZ) \in \mathcal{H}: E(\bff\epsilon|\bfX, \bfZ) = 0\}.
\end{align}

$\Lambda_\bfZ \perp \Lambda_{\bfX|\bfZ}$ since for $\bff \in \Lambda_\bfZ$ and $\bfg \in \Lambda_{\bfX|\bfZ}$, $E(\bff\bfg) = E(\bff E[\bfg|\bfZ]) = 0$ and similarly $\Lambda_\bfZ \perp \Lambda_{\epsilon a}$ since for $\bff \in \Lambda_\bfZ$ and $\bfg \in \Lambda_{\epsilon a}$, $E(\bff\bfg) = E(\bff E[\bfg|\bfX, \bfZ]) = 0$. 

We want to express $\Lambda_{\epsilon b}^\perp$. Take $\bff \in \Lambda_{\epsilon b}$, then for any $\alpha(\bfX, \bfZ) \in \mathcal{H}$,  $E(\bff\boldsymbol\alpha(\bfX, \bfZ)\epsilon) = E(\boldsymbol\alpha(\bfX, \bfZ) E[\bff\epsilon|\bfX, \bfZ])) = 0$. Suppose $\bff_{\epsilon b} = \bff(\epsilon, \bfX, \bfZ) - \epsilon \boldsymbol\alpha(\bfX, \bfZ) \in \Lambda_{\epsilon b}$. Thus
\begin{align*}
0 = & E[\{ \bff(\epsilon, \bfX, \bfZ) - \epsilon \boldsymbol\alpha(\bfX, \bfZ) \} \epsilon | \bfX, \bfZ] \\
= & E(\epsilon \bff(\epsilon, \bfX, \bfZ)|\bfX, \bfZ) - \boldsymbol\alpha(\bfX, \bfZ) E(\epsilon^2|\bfX, \bfZ).
\end{align*}
Hence $\boldsymbol\alpha(\bfX, \bfZ)$  is such that $\boldsymbol\alpha(\bfX, \bfZ) = E[\epsilon^2|\bfX, \bfZ]^{-1}E[\epsilon \bff(\epsilon, \bfX, \bfZ)|\bfX, \bfZ]$.

Now define $\mathcal{S}_1 = \{ \epsilon \boldsymbol\alpha(\bfX, \bfZ) \mbox{ for all } \boldsymbol\alpha(\bfX, \bfZ) \text{ that map from } \mathcal{X} \times \mathcal{Z}^C   \}$ and \\ $\mathcal{S}_2 = \{ \boldsymbol\gamma(\bfX, \bfZ) \mbox{ for all } \boldsymbol\gamma(\bfX, \bfZ) \in \mathcal{H} \}$. 
Then $\mathcal{S}_1 \perp \mathcal{S}_2$ because $E(\epsilon|\bfX, \bfZ) = 0$. Hence we have showed $\mathcal{S}_1 = (\Lambda_\bfX + \Lambda_\bfZ + \Lambda_\epsilon)^\perp$. Finally, we need to ensure that $\epsilon\boldsymbol\alpha(\bfX, \bfZ) \in \Lambda_{\boldsymbol{\ell}}^{\perp}$ in order to characterize $\Lambda^\perp$, since if $\bff \in \mathcal{S}_1$ and $\bff \in \Lambda_{\boldsymbol{\ell}}^{\perp}$, then $\bff \in \Lambda^\perp$. We use $\bfX^\transp\coef^*_\bfZ$ to denote $\sum_{\bfz \in \mathcal{Z}^C}I(\bfZ = \bfz)\bfX^\transp\coef_{\bfz}^*$.

To construct such an $\boldsymbol\alpha$, we note that
\begin{align*}
0 = & \sum_{\bfz \in \mathcal{Z}^C} \int \epsilon \boldsymbol \alpha(\bfx, \bfz) \frac{\eta_{\epsilon, 1}'(\epsilon, \bfx, \bfz)}{\eta_{\epsilon}(\epsilon, \bfx, \bfz)} h_{\bfz}(\bfx^\transp\coef_{\bfz}^*)\eta_{\bfX|\bfZ}(\bfx, \bfz)\pi_{\bfZ}(\bfz) \eta_\epsilon(\epsilon, \bfx, \bfz)\mathrm{d}\mu_\bfz(\bfx)\mathrm{d}\mu(\epsilon) \\
= & \sum_{\bfz \in \mathcal{Z}^C} \int \epsilon \boldsymbol \alpha(\bfx, \bfz) h_{\bfz}(\bfx^\transp \coef_{\bfz}^*)\eta_{\bfX|\bfZ}(\bfx, \bfz) \left\{  \int \epsilon \eta_{\epsilon, 1}'(\epsilon, \bfx, \bfz) \mathrm{d} \mu(\epsilon)  \right\} \pi_\bfZ(\bfz)\mathrm{d}\mu_\bfz(\bfx) \\
& = - E\left[ \boldsymbol\alpha(\bfX, \bfZ) \sum_{\bfz \in \mathcal{Z}^C} I(\bfZ = \bfz)h_\bfz(\bfX^\transp\coef_{\bfz}^*, \bfz)\right],
\end{align*}
where $\mu_\bfz$ is the measure of $\bfX$ with respect to population $\bfz$.
Hence $E[\boldsymbol\alpha(\bfX, \bfZ) | \bfX^\transp\coef_{\bfZ}^*, \bfZ] = 0$. Then together, this implies
\begin{equation*}
\Lambda^\perp = \big \{  [Y - E(Y | \bfX^\transp\coef_{\bfZ}^*, \bfZ )] [\, \boldsymbol \alpha(\bfX, \bfZ) - E[\boldsymbol\alpha(\bfX, \bfZ) | \bfX^\transp\coef_{\bfZ}^*, \bfZ]\, ]  \mbox{ for all } \boldsymbol \alpha \in \mathcal{H}  \big \}.
\end{equation*}

\subsection{Derivation of the orthogonal complement of the nuisance tangent space for the conditional distribution}\label{derivation_distr}

If the goal of sufficient dimension reduction is with respect to the full conditional distribution of the response while considering population heterogeneity based on $\bfZ$, the aim is to estimate matrices $\coef_{\bfz}^*$ such that 
\begin{equation}\label{eqn:sdr_mean_equivalences}
F(y | \bfX, \bfZ = \bfz) =  F(y | \bfX^\transp\coef_{\bfz}^*, \bfZ = \bfz) \text{ for } y \in \mathbb{R},
\end{equation}
where $F(y | \bfX, \bfZ = \bfz) = Pr(Y \leq y |  \bfX, \bfZ = \bfz)$ is the conditional distribution function of $Y$ given $\bfX$ and $\bfZ$. The likelihood of one observation is then
\begin{equation}\label{eqn:lik_obs_distr}
\eta_{Y}\left(Y, \sum_{\bfz \in \mathcal{Z}^C}I(\bfZ = \bfz) \bfX^\transp\coef_{\bfz}^*, \bfZ\right)\pi_{\bfZ}(\bfZ)\eta_{\bfX|\bfZ}(\bfX, \bfZ),
\end{equation}
where $\eta_{\bfX|\bfZ}$ is the probability mass function of $\bfX$ given $\bfZ$, probability density function of $\bfX$ given $\bfZ$, or a mixture distribution, $\pi_{\bfZ}$ is the probability mass function of $\bfZ$, and $\eta_{Y}$ is the conditional probability mass/density function of $Y$ given $\bfX$ and $\bfZ$. 

The nuisance tangent spaces of $\eta_\bfZ$, $\pi_{\bfX|\bfZ}$, and $\eta_Y$ are denoted as, respectively, $\Lambda_\bfZ$, $\Lambda_{\bfX|\bfZ}$, and $\Lambda_Y$.
They are 
\begin{align*}
\Lambda_\bfZ = & {} \{\bff(\bfX) \in \mathcal{H} : E(\bff) = 0 \} \\
\Lambda_{\bfX|\bfZ} = & {} \{   \bff(\bfX, \bfZ) \in \mathcal{H} : E(\bff | \bfZ) = 0  \} \\
\Lambda_Y = & {} \left\{   \bff\left(Y, \bfX^\transp\coef^*_\bfZ, \bfZ\right) : \text{ for all } \bff \text{ such that }  \right. \\ & \left. E(\bff |\bfX, \bfZ) = E\left(\bff \vert \bfX^\transp\coef^*_\bfZ, \bfZ\right) = 0 \right\}.
\end{align*}

The following derivation largely follows \citet{Ma2012}. It is trivial to see that $\Lambda_\bfZ \perp \Lambda_{\bfX|\bfZ}$, $\Lambda_\bfZ \perp \Lambda_Y$, and $\Lambda_{\bfX|\bfZ}\perp \Lambda_Y$ and thus $\Lambda = \Lambda_\bfZ \oplus \Lambda_{\bfX|\bfZ} \oplus \Lambda_Y$. Further, note that  $(\Lambda_\bfZ + \Lambda_{\bfX|\bfZ})^\perp = \{\bff(Y, \bfX, \bfZ): E(\bff|\bfX, \bfZ) = 0\} \supseteq \Lambda_Y$. We will now demonstrate that $\Lambda_Y^\perp = \{\bff(Y, \bfX, \bfZ): E(\bff|\bfX^\transp\bsbeta^*_\bfZ, \bfZ, Y) \text{ is a function of only } (\bfX^\transp\bsbeta^*_\bfZ, \bfZ)\}$.

We first show that all elements $\bff \in \Lambda_Y^\perp$ must be such that $E(\bff|\bfX^\transp\bsbeta^*_\bfZ, \bfZ, Y)$ is a function of only $(\bfX^\transp\bsbeta^*_\bfZ, \bfZ)$. Consider any $\bff \in \Lambda_Y^\perp$ and denote $\bfg = E(\bff|\bfX^\transp\bsbeta^*_\bfZ, \bfZ, Y) - E(\bff|\bfX^\transp\bsbeta^*_\bfZ, \bfZ)$. $E(\bfg|\bfX, \bfZ) = E(\bfg|\bfX^\transp\coef^*_\bfZ, \bfZ) = 0$, so $\bfg \in \Lambda_Y$, which indicates by definition that $E(\bfg^\transp\bff) = 0$. Expanding on this, we have
\begin{align*}
0 = E(\bfg^\transp\bff) = {} & E[\bfg^\transp E(\bff|\bfX^\transp\coef^*_\bfZ, \bfZ, Y)] \\ 
= {} & E(\bfg^\transp\bfg) + E[\bfg^\transp E(\bff|\bfX^\transp\coef^*_\bfZ, \bfZ)] \\
= {} &  E(\bfg^\transp\bfg) + E[E(\bfg|\bfX^\transp\coef^*_\bfZ, \bfZ)^\transp E(\bff|\bfX^\transp\coef^*_\bfZ, \bfZ)] \\
= {} & E(\bfg^\transp\bfg),
\end{align*}
which indicates that $\bfg = 0$. The last equality on the first line holds since $\bfg$ is a function of only $\bfX^\transp\coef^*_\bfZ, \bfZ, Y$, the second line's equality holds due to the definition of $\bfg$, the third line's equality holds by conditioning on $\bfX^\transp\coef^*_\bfZ, \bfZ$, and the last equality holds by the tower property of conditional expectation. Thus, it holds for $\bff$ that $E(\bff|\bfX^\transp\bsbeta^*_\bfZ, \bfZ, Y)$ is a function of only $(\bfX^\transp\bsbeta^*_\bfZ, \bfZ)$.

We now aim to show that $\Lambda^\perp = (\Lambda_\bfZ + \Lambda_{\bfX|\bfZ})^\perp\cap \Lambda_Y^\perp = \{\bff(Y, \bfX, \bfZ) - E(\bff|\bfX^\transp\coef^*_\bfZ, \bfZ, Y): E(\bff|\bfX, \bfZ) = E(\bff|\bfX^\transp\coef^*_\bfZ, \bfZ) \text{ for all } \bff \}$. We denote the set $\mathcal{A} = \{\bff(Y, \bfX, \bfZ) - E(\bff|\bfX^\transp\coef^*_\bfZ, \bfZ, Y): E(\bff|\bfX, \bfZ) = E(\bff|\bfX^\transp\coef^*_\bfZ, \bfZ) \text{ for all } \bff \}$, which is contained in $\Lambda_Y^\perp$ by the construction of elements in $\mathcal{A}$. Then to show that $\mathcal{A} \subseteq (\Lambda_\bfZ + \Lambda_{\bfX|\bfZ})^\perp$, we have
\begin{align*}
& E[\bff(Y, \bfX, \bfZ) - E(\bff|\bfX^\transp\coef^*_\bfZ, \bfZ, Y) | \bfX, \bfZ] \\
= {} & E(\bff|\bfX^\transp\coef^*_\bfZ, \bfZ) - E[E(\bff|\bfX^\transp\coef^*_\bfZ, \bfZ, Y) | \bfX^\transp\coef^*_\bfZ, \bfZ] \\
= {} & E(\bff|\bfX^\transp\coef^*_\bfZ, \bfZ) - E(\bff|\bfX^\transp\coef^*_\bfZ, \bfZ) = 0,
\end{align*}
and thus $\mathcal{A} \subseteq (\Lambda_\bfZ + \Lambda_{\bfX|\bfZ})^\perp$ and $\mathcal{A} \subseteq \Lambda^\perp$. Now it remains to show that $\Lambda^\perp \subseteq  \mathcal{A}$. Now take any $\bff\in\Lambda^\perp\in \Lambda_Y^\perp$. We thus have $E(\bff| \bfX^\transp\coef^*_\bfZ, \bfZ, Y) = \bfa(\bfX^\transp\coef^*_\bfZ, \bfZ)$ for some function $\bfa$. The integral representation of this equality is 
\begin{align*}
\bfa(\bfx^\transp\coef^*_\bfz, \bfz) = {} & \frac{\sum_{\bfz' \in \mathcal{Z}^C, \bfz' = \bfz}\int_{\bfX:\bfX^\transp\coef^*_{\bfz'} = \bfx^\transp\coef^*_{\bfz'} }  \bff(Y, \bfX, \bfz')  \pi_\bfZ(\bfz') \eta_{\bfX|\bfZ}(\bfX, \bfz') \eta_Y(Y, \bfX^\transp\coef^*_{\bfz'}, \bfz') \mathrm{d}\mu_\bfz(\bfX) }{\sum_{\bfz' \in \mathcal{Z}^C, \bfz' = \bfz}\int_{\bfX:\bfX^\transp\coef^*_{\bfz'} = \bfx^\transp\coef^*_{\bfz'} }   \pi_\bfZ(\bfz') \eta_{\bfX|\bfZ}(\bfX, \bfz') \eta_Y(Y, \bfX^\transp\coef^*_{\bfz'}, \bfz') \mathrm{d}\mu_\bfz(\bfX)} \\
= {} & \frac{\sum_{\bfz' \in \mathcal{Z}^C, \bfz' = \bfz}\int_{\bfX:\bfX^\transp\coef^*_{\bfz'} = \bfx^\transp\coef^*_{\bfz'} }  \bff(Y, \bfX, \bfz') \pi_\bfZ(\bfz') \eta_{\bfX|\bfZ}(\bfX, \bfz')  \mathrm{d}\mu_\bfz(\bfX) }{\sum_{\bfz' \in \mathcal{Z}^C, \bfz' = \bfz}\int_{\bfX:\bfX^\transp\coef^*_{\bfz'} = \bfx^\transp\coef^*_{\bfz'} }   \pi_\bfZ(\bfz') \eta_{\bfX|\bfZ}(\bfX, \bfz')  \mathrm{d}\mu_\bfz(\bfX)}.
\end{align*}
Using this we have 
\begin{align*}
&\bfa(\bfx^\transp\coef^*_\bfz, \bfz) \\
&=  \int \bfa(\bfx^\transp\coef^*_\bfz, \bfz)  \eta_Y(Y, \bfX^\transp\coef^*_{\bfz'}, \bfz') \mathrm{d}\mu(Y) \\
&=  \int \frac{\sum_{\bfz' \in \mathcal{Z}^C, \bfz' = \bfz}\int_{\bfX:\bfX^\transp\coef^*_{\bfz'} = \bfx^\transp\coef^*_{\bfz'} }  \bff(Y, \bfX, \bfz')  \pi_\bfZ(\bfz') \eta_{\bfX|\bfZ}(\bfX, \bfz')  \mathrm{d}\mu_\bfz(\bfX) }{\sum_{\bfz' \in \mathcal{Z}^C, \bfz' = \bfz}\int_{\bfX:\bfX^\transp\coef^*_{\bfz'} = \bfx^\transp\coef^*_{\bfz'} }  \pi_\bfZ(\bfz') \eta_{\bfX|\bfZ}(\bfX, \bfz') \mathrm{d}\mu_\bfz(\bfX)}  \eta_Y(Y, \bfX^\transp\coef^*_{\bfz'}, \bfz') \mathrm{d}\mu(Y)  \\
&=   \frac{\sum_{\bfz' \in \mathcal{Z}^C, \bfz' = \bfz}\int \int_{\bfX:\bfX^\transp\coef^*_{\bfz'} = \bfx^\transp\coef^*_{\bfz'} }  \bff(Y, \bfX, \bfz')  \pi_\bfZ(\bfz') \eta_{\bfX|\bfZ}(\bfX, \bfz') \eta_Y(Y, \bfX^\transp\coef^*_{\bfz'}, \bfz') \mathrm{d}\mu_\bfz(\bfX)  \mathrm{d}\mu(Y)  }{\sum_{\bfz' \in \mathcal{Z}^C, \bfz' = \bfz}\int_{\bfX:\bfX^\transp\coef^*_{\bfz'} = \bfx^\transp\coef^*_{\bfz'} }   \pi_\bfZ(\bfz') \eta_{\bfX|\bfZ}(\bfX, \bfz') \mathrm{d}\mu_\bfz(\bfX)} \\
&=   \frac{\sum_{\bfz' \in \mathcal{Z}^C, \bfz' = \bfz}\int_{\bfX:\bfX^\transp\coef^*_{\bfz'} = \bfx^\transp\coef^*_{\bfz'} } \int  \bff(Y, \bfX, \bfz')  \pi_\bfZ(\bfz') \eta_{\bfX|\bfZ}(\bfX, \bfz') \eta_Y(Y, \bfX^\transp\coef^*_{\bfz'}, \bfz')  \mathrm{d}\mu(Y)  \mathrm{d}\mu_\bfz(\bfX) }{\sum_{\bfz' \in \mathcal{Z}^C, \bfz' = \bfz}\int_{\bfX:\bfX^\transp\coef^*_{\bfz'} = \bfx^\transp\coef^*_{\bfz'} }   \pi_\bfZ(\bfz') \eta_{\bfX|\bfZ}(\bfX, \bfz')  \mathrm{d}\mu_\bfz(\bfX)} \\
= {} & 0
\end{align*}
since $\bff \in (\Lambda_\bfZ + \Lambda_{\bfX|\bfZ})^\perp$ and thus  $E(\bff|\bfX, \bfZ) = 0$ which means elements in $\Lambda^\perp$ have the form $\bff(Y, \bfX, \bfZ) - E(\bff|\bfX^\transp\coef^*_\bfZ, \bfZ, Y)$. Then since $\bff \in (\Lambda_\bfZ + \Lambda_{\bfX|\bfZ})^\perp$, $0 = E(\bff|\bfX, \bfZ) = E(\bff|\bfX^\transp\coef^*_\bfZ, \bfZ)$ by the law of total expectation and thus we have the second condition of $\mathcal{A}$. Hence $\Lambda^\perp = \mathcal{A}$.

Similar to \citet{Ma2012}, given any functions $\bfg(Y, \bfX^\transp\coef^*_\bfZ, \bfZ)$ and $\boldsymbol \alpha(\bfX, \bfZ)$ we can take $\bff(Y, \bfX, \bfZ)$ to be 
\[
[\bfg(Y, \bfX^\transp\coef^*_\bfZ, \bfZ) - E\{\bfg(Y, \bfX^\transp\coef^*_\bfZ, \bfZ) | \bfX^\transp\coef^*_\bfZ, \bfZ\}][ \bff(Y, \bfX, \bfZ) - E\{\bff(Y, \bfX, \bfZ) | \bfX^\transp\coef^*_\bfZ, \bfZ\} ],
\]
which has the double-robustness property described in \citet{Ma2012}.

\subsection{Proofs}\label{sec:proofs_thms}

\begin{lemma}\label{thm:lemma1}
	Assume Conditions (C1)-(C5) hold. Denote the space $\Omega_{\coef} = \{ (Y, \bfx, \bfz, \widehat{\coef}) : \bfx \in \mathbb{R}^p, Y \in \mathbb{R}, \bfz \in \mathcal{Z}^C, \mbox{ and } ||\widehat{\coef} - \coef^* || \leq Kn^{-1/2} )  \}$, where $||\cdot||$ is the Euclidean norm and $K$ is a constant. Then there exist bases $\coef_{\bfz}^*$ of $\mathcal{S}_\bfz$ such that 
	\begin{align} \label{eqn:lem1a}
	& \sup_{\Omega_{\coef}}  \lvert \widehat{E}\{ \boldsymbol\alpha(\bfX, \bfZ) | \bfX^\transp\widehat{\coef}_\bfZ, \bfZ \} - \widehat{E}\{ \boldsymbol\alpha(\bfX, \bfZ) | \bfX^\transp\coef_{\bfZ}^*, \bfZ \} - {E}\{ \boldsymbol\alpha(\bfX, \bfZ) | \bfX^\transp\widehat{\coef}_\bfZ, \bfZ \} + \nonumber \\ 
	& {E}\{ \boldsymbol\alpha(\bfX, \bfZ) | \bfX^\transp\coef_{\bfZ}^*, \bfZ \}\rvert = O_p\left(\sum_{\bfz \in \mathcal{Z}^C}\left[n^{-1/2}h_\bfz^m + n^{-1}h_\bfz^{-(d_\bfz+ 1)}\log n\right]\right)
	\end{align}
	and similarly 
	\begin{align}\label{eqn:lem1b}
	& \sup_{\Omega_{\coef}}  \lvert \widehat{E}\{ Y | \bfX^\transp\widehat{\coef}_\bfZ, \bfZ \} - \widehat{E}\{ Y | \bfX^\transp\coef_{\bfZ}^*, \bfZ \} - {E}\{ Y | \bfX^\transp\widehat{\coef}_\bfZ, \bfZ \} + \nonumber \\ 
	& {E}\{ Y | \bfX^\transp\coef_{\bfZ}^*, \bfZ \}\rvert = O_p\left(\sum_{\bfz \in \mathcal{Z}^C}\left[h_\bfz^mn^{-1/2} + n^{-1}h_\bfz^{-(d_\bfz + 1)}\log n\right]\right)
	\end{align}
\end{lemma}

The proof of Lemma \ref{thm:lemma1} is similar to that of \cite{Ma2012} with  adjustments for the fact that there are differing dimensions and bandwidths for each subpopulation. We prove Lemma \ref{thm:lemma2} in full to demonstrate how this type of modification carries through.

\begin{lemma} \label{thm:lemma2}
	Assume Conditions (C1) - (C4) hold. Then 
	\begin{align}
	& \frac{1}{n}\sum_{i = 1}^n\left[ Y_i - E(Y_i | \bfX_i^\transp\coef^*_{\bfZ_i}, \bfZ_i) \right]\left[ \widehat{E}(\boldsymbol\alpha(\bfX_i, \bfZ_i)|\bfX_i^\transp\coef_{\bfZ_i}^*, \bfZ_i) - {E}(\boldsymbol\alpha(\bfX_i, \bfZ_i)|\bfX_i^\transp\coef_{\bfZ_i}^*, \bfZ_i) \right] \nonumber \\
	& = O_p\left( \sum_{\bfz \in \mathcal{Z}^C}\left[   1 / (nh_\bfz^{d_\bfz/2}) + h_\bfz^m/n^{1/2} + h_\bfz^{2m} + \log^2n/(nh_\bfz^{d_\bfz})   \right]\right) \\
	\mbox{and } & \frac{1}{n}\sum_{i = 1}^n\left[ \widehat{E}(Y_i | \bfX_i^\transp\coef_{\bfZ_i}^*, \bfZ_i) - E(Y_i | \bfX_i^\transp\coef_{\bfZ_i}^*, \bfZ_i) \right]\left[ \boldsymbol\alpha(\bfX_i, \bfZ_i) - {E}(\boldsymbol\alpha(\bfX_i, \bfZ_i)|\bfX_i^\transp\coef_{\bfZ_i}^*, \bfZ_i) \right] \nonumber \\
	& = O_p\left( \sum_{\bfz \in \mathcal{Z}^C}\left[   1 / (nh_\bfz^{d_\bfz/2}) + h_\bfz^m/n^{1/2} + h_\bfz^{2m} + \log^2n/(nh_\bfz^{d_\bfz})   \right]\right)
	\end{align}
\end{lemma}

\begin{proof}[Proof of Lemma \ref{thm:lemma2}]
	
	The result can be shown by similar arguments to those in the proof of Lemma 4 of \cite{Ma2012}.
	
	Recall from the Appendix of the main text that \newline $\bfr_1(\bfX_i^\transp\coef_{\bfZ_i}, \bfZ_i) = E(\balpha(\bfX_i, \bfZ_i) | \bfX_i^\transp\coef_{\bfZ_i}, \bfZ_i) f(\bfX_i^\transp\coef_{\bfZ_i}, \bfZ_i)$. For any $i=1,\dots,n$, let $\varepsilon_i = Y_i - E(Y_i | \bfX_i^\transp\coef_{\bfZ_i}, \bfZ_i)$, $\hat{f}(\bfX_i^\transp\coef_{\bfZ_i}, \bfZ_i) = (n - 1)^{-1}\sum_{j\neq i}I(\bfZ_j = \bfZ_i)K_{h_{\bfz_i}}(\bfX_i^\transp\coef_{\bfZ_i} - \bfX_j^\transp\coef_{\bfZ_j})$, and $\hat{\bfr}_{1}(\bfX_i^\transp\coef_{\bfZ_i}, \bfZ_i) = (n - 1)^{-1}\sum_{j\neq i}I(\bfZ_j = \bfZ_i)K_{h_{\bfZ_i}}(\bfX_i^\transp\coef_{\bfZ_i} - \bfX_j^\transp\coef_{\bfZ_j})\balpha(\bfX_j, \bfZ_j)$. Then similar to the proof of Lemma 4 in \cite{Ma2012}
	\begin{align}
	& \frac{1}{n}\varepsilon_i\left[ \widehat{E}(\boldsymbol\alpha(\bfX_i, \bfZ_i)|\bfX_i^\transp\coef_{\bfZ_i}, \bfZ_i) - {E}(\boldsymbol\alpha(\bfX_i, \bfZ_i)|\bfX_i^\transp\coef_{\bfZ_i}, \bfZ_i) \right]   \nonumber \\
	= {} &  \frac{1}{n}\varepsilon_i\left[ \frac{\hat{\bfr}_{1}(\bfX_i^\transp\coef_{\bfZ_i}, \bfZ_i)}{\hat{f}(\bfX_i^\transp\coef_{\bfZ_i}, \bfZ_i)}  - \frac{{\bfr}_{1}(\bfX_i^\transp\coef_{\bfZ_i}, \bfZ_i)}{{f}(\bfX_i^\transp\coef_{\bfZ_i}, \bfZ_i)} \right]  \nonumber\\
	= {} & \frac{1}{n}\varepsilon_i\left[  \frac{\hat{\bfr}_{1}(\bfX_i^\transp\coef_{\bfZ_i}, \bfZ_i) - {\bfr}_{1}(\bfX_i^\transp\coef_{\bfZ_i}, \bfZ_i)}{{f}(\bfX_i^\transp\coef_{\bfZ_i}, \bfZ_i)}  \right] - \frac{1}{n}\varepsilon_i\left[  \frac{ {\bfr}_{1}(\bfX_i^\transp\coef_{\bfZ_i}, \bfZ_i) \{ \hat{f}(\bfX_i^\transp\coef_{\bfZ_i}, \bfZ_i) - {f}(\bfX_i^\transp\coef_{\bfZ_i}, \bfZ_i) \}    }{{f}^2(\bfX_i^\transp\coef_{\bfZ_i}, \bfZ_i)}\right] \nonumber \\
	{} & - \frac{1}{n}\varepsilon_i\left[  \frac{ \{ \hat{\bfr}_{1}(\bfX_i^\transp\coef_{\bfZ_i}, \bfZ_i) - {\bfr}_{1}(\bfX_i^\transp\coef_{\bfZ_i}, \bfZ_i) \} \{ \hat{f}(\bfX_i^\transp\coef_{\bfZ_i}, \bfZ_i) - {f}(\bfX_i^\transp\coef_{\bfZ_i}, \bfZ_i) \}    }{{f}(\bfX_i^\transp\coef_{\bfZ_i}, \bfZ_i)\hat{f}(\bfX_i^\transp\coef_{\bfZ_i}, \bfZ_i)}\right] \nonumber\\
	{} &+ \frac{1}{n}\varepsilon_i\left[  \frac{  {\bfr}_{1}(\bfX_i^\transp\coef_{\bfZ_i}, \bfZ_i) \{ \hat{f}(\bfX_i^\transp\coef_{\bfZ_i}, \bfZ_i) - {f}(\bfX_i^\transp\coef_{\bfZ_i}, \bfZ_i) \} ^ 2    }{{f}^2(\bfX_i^\transp\coef_{\bfZ_i}, \bfZ_i)\hat{f}(\bfX_i^\transp\coef_{\bfZ_i}, \bfZ_i)}\right]  \label{eqn:lem2_expansion}
	\end{align}
	
	Since $n_\bfz/n \to c_\bfz$ for all $\bfz \in \mathcal{Z}^C$, then $\log(n_\bfz)/\log(n) \to 1$ and by the uniform convergence of nonparametric regression, the third and fourth terms in \eqref{eqn:lem2_expansion} are $O_p(\sum_{\bfz \in \mathcal{Z}^C} [   h_{\bfz}^{2m} +   \log^2(n)/(nh_\bfz^{d_\bfz}) ])$. Due to (C5), it is enough to focus on the convergence rates of the quantity $n^{-1}\sum_{i=1}^n\varepsilon_i\hat{\bfr}_{1}(\bfX_i^\transp\coef_{\bfZ_i}, \bfZ_i)$, which we now express as a second-order $U$-statistic.
	\begin{equation*}
	\frac{1}{n}\sum_{i=1}^n\hat{\bfr}_{1}(\bfX_i^\transp\coef_{\bfZ_i}, \bfZ_i)\varepsilon_i = \frac{1}{n(n-1)} \sum_{i \neq j}^nI(\bfZ_i = \bfZ_j)K_{h_{\bfZ_i}}(\bfX_i^\transp\coef_{\bfZ_i} - \bfX_j^\transp\coef_{\bfZ_j})\{  \varepsilon_i\balpha(\bfX_j, \bfZ_j) + \varepsilon_j\balpha(\bfX_i, \bfZ_i)  \}.
	\end{equation*}
	An application of Lemma 5.2.1.A of \citet[page 183]{serfling2009approximation} and noticing that the following difference is a degenerate $U$-statistic yields
	\begin{align}
	& \frac{1}{n}\sum_{i=1}^n\hat{\bfr}_{1}(\bfX_i^\transp\coef_{\bfZ_i}, \bfZ_i)\varepsilon_i - \frac{1}{n} \sum_{i=1}^n\varepsilon_iE\left\{K_{h_{\bfZ_i}}(\bfX_i^\transp\coef_{\bfZ_i} - \bfX_j^\transp\coef_{\bfZ_j}){\bfr}_{1}(\bfX_i^\transp\coef_{\bfZ_i}, \bfZ_i) | \bfX_i^\transp\coef_{\bfZ_i}, \bfZ_i\right\} \nonumber\\
	& =  O_p\left( \sum_{\bfz \in \mathcal{Z}^C}1 / (nh_\bfz^{d_\bfz/2}) \right). \label{eqn:op_residual_r}
	\end{align}
	By similar arguments as in \citet{Ma2012}, we also have 
	\begin{align}
	& \frac{1}{n} \sum_{i \neq j}^n\varepsilon_iI(\bfZ_i = \bfZ_j)\left[ E\left\{K_{h_{\bfZ_i}}(\bfX_i^\transp\coef_{\bfZ_i} - \bfX_j^\transp\coef_{\bfZ_j}){\bfr}_{1}(\bfX_i^\transp\coef_{\bfZ_i}, \bfZ_i) | \bfX_i^\transp\coef_{\bfZ_i}, \bfZ_i\right\} \right. \\ {} &  \left. \vphantom{K_{h_{\bfZ_i}}} - {\bfr}_{1}(\bfX_i^\transp\coef_{\bfZ_i}, \bfZ_i)f(\bfX_i^\transp\coef_{\bfZ_i}, \bfZ_i)\right] \nonumber \\
	& = O_p\left( \sum_{\bfz \in \mathcal{Z}^C}h_\bfz^{m} / n^{1/2} \right). \label{eqn:op_r_r}
	\end{align}
	Combining \eqref{eqn:op_residual_r} and \eqref{eqn:op_r_r}, we have
	\begin{equation*}
	\frac{1}{n}\sum_{i=1}^n\varepsilon_i\left\{\hat{\bfr}_{1}(\bfX_i^\transp\coef_{\bfZ_i}, \bfZ_i) - {\bfr}_{1}(\bfX_i^\transp\coef_{\bfZ_i}, \bfZ_i)f(\bfX_i^\transp\coef_{\bfZ_i}, \bfZ_i)\right\} = O_p\left( \sum_{\bfz \in \mathcal{Z}^C} (1 / (nh_\bfz^{d_\bfz/2}) +  h_\bfz^{m} / n^{1/2} ) \right),
	\end{equation*}
	which,  combined with \eqref{eqn:lem2_expansion}, finishes the proof.

\end{proof}

\begin{proof}[Proof of Theorem \ref{thm:thm1a}]
	
	For notational convenience, throughout this proof we refer to $\coef$, $\wtbbeta$, and $\coef^*$ by their vectorized versions and drop the use of $\vec$ and $\vecl$. We begin by decomposing $\hat{\Psi}_{n}$ as 
	\begingroup
	\allowdisplaybreaks
	\begin{align*}
	& n^{-1/2}\hat{\Psi}_n(\wtbbeta) \\ = {} &  n^{-1/2}\sum_{i = 1}^n\left[  Y_i - \widehat{E}(Y_i | \bfX_i^\transp{\wtbbeta}_{\bfZ_i}, \bfZ_i)  \right]\left[  \balpha(\bfX_i,\bfZ_i) - \widehat{E}(\balpha(\bfX_i,\bfZ_i) | \bfX_i^\transp{\wtbbeta}_{\bfZ_i}, \bfZ_i)  \right] \\
	= {} & n^{-1/2}\sum_{i = 1}^n\left[  Y_i - {E}(Y_i | \bfX_i^\transp{\wtbbeta}_{\bfZ_i}, \bfZ_i)  \right]\left[  \balpha(\bfX_i,\bfZ_i) - {E}(\balpha(\bfX_i,\bfZ_i) | \bfX_i^\transp{\wtbbeta}_{\bfZ_i}, \bfZ_i)  \right] \\
	& + n^{-1/2}\sum_{i = 1}^n\left[  Y_i - \widehat{E}(Y_i | \bfX_i^\transp{\wtbbeta}_{\bfZ_i}, \bfZ_i)  \right]\left[  {E}(\balpha(\bfX_i,\bfZ_i) | \bfX_i^\transp{\wtbbeta}_{\bfZ_i}, \bfZ_i)  - \widehat{E}(\balpha(\bfX_i,\bfZ_i) | \bfX_i^\transp{\wtbbeta}_{\bfZ_i}, \bfZ_i)  \right] \\
	& + n^{-1/2}\sum_{i = 1}^n\left[  {E}(Y_i | \bfX_i^\transp{\wtbbeta}_{\bfZ_i}, \bfZ_i)- \widehat{E}(Y_i | \bfX_i^\transp{\wtbbeta}_{\bfZ_i}, \bfZ_i)  \right]\left[  \balpha(\bfX_i,\bfZ_i) - {E}(\balpha(\bfX_i,\bfZ_i) | \bfX_i^\transp{\wtbbeta}_{\bfZ_i}, \bfZ_i)  \right] \\
	& + n^{-1/2}\sum_{i = 1}^n\left[  {E}(Y_i | \bfX_i^\transp{\wtbbeta}_{\bfZ_i}, \bfZ_i)- \widehat{E}(Y_i | \bfX_i^\transp{\wtbbeta}_{\bfZ_i}, \bfZ_i)  \right] \\
	& \;\;\;\; \times  \left[  {E}(\balpha(\bfX_i,\bfZ_i) | \bfX_i^\transp{\wtbbeta}_{\bfZ_i}, \bfZ_i) - \widehat{E}(\balpha(\bfX_i,\bfZ_i) | \bfX_i^\transp{\wtbbeta}_{\bfZ_i}, \bfZ_i)  \right]. \\
	\end{align*}
	\endgroup
	
	By Conditions (C1)-(C5), Lemmas \ref{thm:lemma1} and \ref{thm:lemma2} and similar arguments as in \citet{Ma2012}, the last three terms in the above are $o_p(1)$. Thus $n^{-1/2}\hat{\Psi}_n(\wtbbeta) = n^{-1/2}{\Psi}_n(\wtbbeta) + o_p(1)$. Define
	\begin{align*}
	{\boldsymbol G}(\coef) = {} & E\left\{ \frac{\partial}{\partial \coef} \left[  Y - {E}(Y | \bfX^\transp{\coef}_{\bfZ}, \bfZ)  \right]\left[  \balpha(\bfX,\bfZ) - {E}(\balpha(\bfX,\bfZ) | \bfX^\transp{\coef}_{\bfZ}, \bfX)  \right] \right\} \\
	\widehat{{\boldsymbol G} }(\coef) = {} & n^{-1}\frac{\partial}{\partial \coef} \hat{\Psi}_n(\coef).
	\end{align*}
	
	Then by Conditions (C1)-(C5), the first order conditions implied by the definition of $\wtbbeta$, and Slutsky's theorem, we have 
	\begin{align*}
	0 = {} & n^{-1/2}\widehat{{\boldsymbol G}}(\wtbbeta)^\transp\boldsymbol W_n \hat{\Psi}_n(\wtbbeta) \\
	= {} & n^{-1/2}\{{\boldsymbol G}(\bbeta^*)^\transp + o_p(1) \} \boldsymbol W \hat{\Psi}_n(\wtbbeta) \\
	= {} & n^{-1/2}\{{\boldsymbol G}(\bbeta^*)^\transp + o_p(1) \} \boldsymbol W {\Psi}_n(\bbeta^*) \\
	& + \{{\boldsymbol G}(\bbeta^*)^\transp + o_p(1) \} \boldsymbol W \{{\boldsymbol G}(\bbeta^*) + o_p(1) \} n^{1/2}(\wtbbeta - \bbeta^*) + o_p(1) \\
	= {} & n^{-1/2}{\boldsymbol G}(\bbeta^*)^\transp \boldsymbol W {\Psi}_n(\bbeta^*) \\
	& + {\boldsymbol G}(\bbeta^*)^\transp \boldsymbol W{\boldsymbol G}(\bbeta^*)n^{1/2}(\wtbbeta - \bbeta^*)  + o_p(1).
	\end{align*}
	Thus $$ n^{1/2}(\wtbbeta - \bbeta^*) = - n^{-1/2}\{{\boldsymbol G}(\bbeta^*)^\transp \boldsymbol W {\boldsymbol G}(\bbeta^*)\}^{-1} {\boldsymbol G}(\bbeta^*)^\transp \boldsymbol W{\Psi}_n(\bbeta^*) + o_p(1),$$  completing the proof.

\end{proof}

\begin{proof}[Proof of Proposition \ref{thm:achieving_hier_central}]
	We prove the result for two binary factors. The proof for more than 2 binary factors can be done in the same fashion, though it is more tedious. Imagine there exist parameters  $\bsbeta^\ddagger_{\bfz}$ for $\bfz \in \{00, 10, 01, 11\}$ of smallest dimension such that \eqref{eqn:mean_models} and \eqref{eqn:hierarchical_assumption_none} and \eqref{eqn:none_assumption2} hold but that \eqref{eqn:equality_condition} does not hold. Consider the case that $d_{00} < d_{10}, d_{01} < d_{11}$, since the re-parameterizations in \eqref{eqn:equality_condition} are trivial for equal dimension cases (e.g. $d_{00} = d_{10}$). Since \eqref{eqn:equality_condition} does not hold, there are no other matrices  $\bsbeta^\dagger_{00}$ and $\bsbeta^\dagger_{10}$ that can be represented as $\bsbeta^\dagger_{10} = (\bsbeta^\dagger_{00}, \bnu^\dagger_{10\dagger})$ such that $\text{span}(\bsbeta^\ddagger_{10}) = \text{span}(\bsbeta^\dagger_{10})$  and $\text{span}(\bsbeta^\ddagger_{00}) = \text{span}(\bsbeta^\dagger_{00})$. But then this implies $\text{span}(\bsbeta^\ddagger_{00}) \not\subseteq \text{span}(\bsbeta^\ddagger_{10})$, implying a contradiction. Similar arguments can be made for $\bsbeta^\ddagger_{11}$ and $\bsbeta^\ddagger_{01}$.
\end{proof}

\begin{proof}[Proof of Theorem \ref{thm1}]
	
	Recall that our proposed estimator is given by the solution $\widehat{\bbeta} = ({\widehat{\bbeta}_\bfz}: \bfz \in \mathcal{Z}^C) $ to
	\begin{align}
	\argmin_{\bbeta} {} & \frac{1}{2n}\hat{\Psi}_{n}(\bbeta)^\transp\boldsymbol W_n\hat{\Psi}_{n}(\bbeta) \nonumber \\
	& \mbox{such that } \boldsymbol \bsC^\transp \vecl(\bsbeta) = \boldsymbol 0,  \label{eqn:constr_min}
	\end{align}
	where $\boldsymbol W_n$ is a weight matrix possibly chosen for efficiency improvements.

	Similar to the proof of Theorem  \ref{thm:thm1a}, we drop the use of $\vecl$ and $\vec$ and henceforth always refer to the vectorized versions of $\wtbbeta$, $\whbbeta$, and $\bbeta^0$. The solution $\widehat{\bbeta}$ of (\ref{eqn:constr_min}) has the following first order conditions:
	\begin{align*}
	\frac{1}{n}\left(\frac{\partial}{\partial\bbeta}\hat{\Psi}_n(\widehat{\bbeta})\right)^\transp\boldsymbol W\hat{\Psi}_n(\widehat{\bbeta}) - \bsC\widehat{\blambda} & {} = \bzero \\
	\mbox{and  }\bsC^\transp\widehat{\bbeta} & {} = \bzero
	\end{align*}
	
	Denote ${\boldsymbol G}(\bbeta)$  as $\boldsymbol G_2 $ but evaluated at a given $\bbeta$  and denote $\widehat{\boldsymbol G}(\bbeta)$ as in the proof of Theorem \ref{thm:thm1a} but accounting for the fact that upper identity block constraints are different in the estimator under question in Theorem \ref{thm1}.
	
	Similar to the proof of Theorem \ref{thm:thm1a}, we take a Taylor expansion of the first order conditions around $\bbeta^0$, multiply both the first condition and the second first condition by $n^{1/2}$. By Conditions (C1)-(C5), Lemmas \ref{thm:lemma1} and \ref{thm:lemma2}, we have
	\begin{align*}
	0 = {} & n^{-1/2}\widehat{\boldsymbol G}(\whbbeta)^\transp\boldsymbol W_n \hat{\Psi}_n(\whbbeta) - n^{-1/2}\bsC\widehat{\blambda} \\
	= {} & n^{-1/2}\{{\boldsymbol G}(\bbeta^0)^\transp + o_p(1) \} \boldsymbol W \hat{\Psi}_n(\whbbeta) - n^{-1/2}\bsC\widehat{\blambda} \\
	= {} & n^{-1/2}\{{\boldsymbol G}(\bbeta^0)^\transp + o_p(1) \} \boldsymbol W {\Psi}_n(\bbeta^0) \\
	& + \{{\boldsymbol G}(\bbeta^0)^\transp + o_p(1) \} \boldsymbol W \{{\boldsymbol G}(\bbeta^0) + o_p(1) \} n^{1/2}(\whbbeta - \bbeta^0) - n^{-1/2}\bsC\widehat{\blambda} + o_p(1) \\
	= {} & n^{-1/2}{\boldsymbol G}(\bbeta^0)^\transp \boldsymbol W {\Psi}_n(\bbeta^0) \\
	& + {\boldsymbol G}(\bbeta^0)^\transp \boldsymbol W{\boldsymbol G}(\bbeta^0)n^{1/2}(\whbbeta - \bbeta^0) - n^{-1/2}\bsC\widehat{\blambda} + o_p(1) \\
	\mbox{and } & \bsC^\transp n^{1/2}(\widehat{\bbeta} - \bbeta^0) = o_p(1).
	\end{align*}
	
	Thus 
	\begin{align}
	n^{1/2}(\whbbeta - \bbeta^0) = {} &  \{{\boldsymbol G}(\bbeta^0)^\transp \boldsymbol W{\boldsymbol G}(\bbeta^0)\}^{-1}n^{-1/2}\bsC\widehat{\blambda} \nonumber \\
	& - \{{\boldsymbol G}(\bbeta^0)^\transp \boldsymbol W{\boldsymbol G}(\bbeta^0)\}^{-1}n^{-1/2}{\boldsymbol G}(\bbeta^0)^\transp \boldsymbol W {\Psi}_n(\bbeta^0)  + o_p(1). \label{eqn:represent_with_lambda}
	\end{align}
	
	From $\bsC^\transp n^{1/2}(\widehat{\bbeta} - \bbeta^0)$, multiplying \eqref{eqn:represent_with_lambda} by $\bsC^\transp$, we have
		\begin{align*}
	\bsC^\transp n^{1/2}(\whbbeta - \bbeta^0) = o_p(1) = {} &  \bsC^\transp\{{\boldsymbol G}(\bbeta^0)^\transp \boldsymbol W{\boldsymbol G}(\bbeta^0)\}^{-1}n^{-1/2}\bsC\widehat{\blambda}  \\
	& + \bsC^\transp\{{\boldsymbol G}(\bbeta^0)^\transp \boldsymbol W{\boldsymbol G}(\bbeta^0)\}^{-1}n^{-1/2}{\boldsymbol G}(\bbeta^0)^\transp \boldsymbol W {\Psi}_n(\bbeta^0)  + o_p(1)
	\end{align*}
	
and thus 
\begin{align*}
n^{-1/2}\widehat{\blambda}  = -\left( \bsC^\transp \{{\boldsymbol G}(\bbeta^0)^\transp \boldsymbol W{\boldsymbol G}(\bbeta^0)\}^{-1}\bsC \right)^{-1}\bsC^\transp n^{-1/2}{\boldsymbol G}(\bbeta^0)^\transp \boldsymbol W {\Psi}_n(\bbeta^0) + o_p(1).
\end{align*}

	Combining the above with \eqref{eqn:represent_with_lambda} completes the proof.

\end{proof}

\begin{proof}[Proof of Corollary \ref{thm:corollary1}]
	Similar to the proof of Theorem \ref{thm1}, it can be shown that 
	\begin{align*}
	\bsC^\transp \whbbeta =  \boldsymbol 0 = {} &  \bsC^\transp\{\bbeta^0 - n^{-1}\{{\boldsymbol G}(\bbeta^0)^\transp \boldsymbol W{\boldsymbol G}(\bbeta^0)\}^{-1}{\boldsymbol G}(\bbeta^0)^\transp \boldsymbol W {\Psi}_n(\bbeta^0)\} \\
	&  + \bsC^\transp \{{\boldsymbol G}(\bbeta^0)^\transp \boldsymbol W{\boldsymbol G}(\bbeta^0)\}^{-1}\bsC n^{-1}\widehat{\boldsymbol \lambda} + o_p(1)
	\end{align*}
	where $$ n^{-1}\widehat{\boldsymbol \lambda}  = \bsC^\transp\wtbbeta +\{  \bsC^\transp \{{\boldsymbol G}(\bbeta^0)^\transp \boldsymbol W{\boldsymbol G}(\bbeta^0)\}^{-1}\bsC\}^{-1} \bsC^\transp\wtbbeta + o_p(1).$$
	Since $\whbbeta = \bbeta^0 - n^{-1}[ \{{\boldsymbol G}(\bbeta^0)^\transp \boldsymbol W{\boldsymbol G}(\bbeta^0)\}^{-1}{\boldsymbol G}(\bbeta^0)^\transp \boldsymbol W {\Psi}_n(\bbeta^0) + \{{\boldsymbol G}(\bbeta^0)^\transp \boldsymbol W{\boldsymbol G}(\bbeta^0)\}^{-1}\bsC\widehat{\boldsymbol \lambda}]$,
	the bias result then holds by rearranging terms and an application of Slutsky's theorem. The proof is completed by plugging in the above value for $n^{-1/2}\widehat{\boldsymbol \lambda}$ into \eqref{eqn:represent_with_lambda} and adding $n^{1/2}\boldsymbol P \coef^0$ to both sides.
\end{proof}

\begin{proof}[Proof of Theorem \ref{thm:vic_consistency}]
	Similar to the proof of Theorem \ref{thm:thm1a}, it can be shown that 
	\begin{equation}\label{eqn:gamma_tilde_expand}
	n^{-1/2}\hat{\Psi}_n(\whbbeta_{(\boldsymbol d)}(v)) = n^{-1/2}{\Psi}_n(\whbbeta_{(\boldsymbol d)}(v)) + o_p(1).
	\end{equation}
	Following the arguments of the proof of Theorem \ref{thm1} 
	we have
	\begin{align}
	n^{1/2}(\whbbeta_{(\boldsymbol d)} - \bbeta^0_{(\boldsymbol d)}) = {} & - n^{1/2}\{{\boldsymbol G}(\bbeta^0_{(\boldsymbol d)})^\transp \boldsymbol W {\boldsymbol G}(\bbeta^0_{(\boldsymbol d)})\}^{-1} {\boldsymbol G}(\bbeta^0_{(\boldsymbol d)})^\transp \boldsymbol W{\Psi}_n(\bbeta^0_{(\boldsymbol d)}) \nonumber \\
	& + o_p(1)n^{-1/2}\Psi_n( \bbeta^0_{(\boldsymbol d)})  + o_p(1), \label{eqn:beta_hat_expand_careful}
	\end{align}
	where 
	 ${\boldsymbol G}(\bbeta^0_{(\boldsymbol d)})$ is defined as $ \boldsymbol G_2$ in Theorem \ref{thm1} but allowing for differing dimensions of $\bbeta^0_{(\boldsymbol d)}$ from the true structural dimensions. 
	 Unlike in the main text, for clarity we denote the constituent components of $\whbbeta_{(\boldsymbol d)}(v)$ and $\bbeta^0_{(\boldsymbol d)}(v)$ as $\whbbeta_{\bfz(\boldsymbol d)}(v)$ and $\bbeta^0_{\bfz(\boldsymbol d)}(v)$, respectively, such that the notation explicitly shows their dimensions.  
	Similar to the proof of \citet{Ma2015}, due to the definitions of $\whbbeta_{(\boldsymbol d)}(v)$ and $\bbeta^0_{(\boldsymbol d)}(v)$, there exist 
	$$\left\{(p -1 - k_{\bfz})(k_{\bfz} + 1 ) + \sum_{\bfz':\bfz' \in \mathcal{Z}^C, \bfz' \neq \bfz}(p - k_{\bfz'})k_{\bfz'} \right\}\times \left\{\sum_{\bfz:\bfz \in \mathcal{Z}^C}(p - k_{\bfz})k_{\bfz}\right\}$$ 
	matrices $\boldsymbol L_{(\boldsymbol d)\bfz}(v)$ such that $\vecl(\whbbeta_{\bfz(\boldsymbol d)}(v) - \bbeta^0_{\bfz(\boldsymbol d)}(v))  = \boldsymbol L_{(\boldsymbol d)\bfz}(v)\vecl(\whbbeta_{\bfz(\boldsymbol d)} - \bbeta^0_{\bfz(\boldsymbol d)})$. Hence there exists a matrix $\boldsymbol L_{(\boldsymbol d)}(v)$ such that $\vecl(\whbbeta_{(\boldsymbol d)}(v) - \bbeta^0_{(\boldsymbol d)}(v)) = \boldsymbol L_{(\boldsymbol d)}(v)\vecl(\whbbeta_{(\boldsymbol d)} - \bbeta^0_{(\boldsymbol d)})$. Now by taking a Taylor expansion, using Conditions (C1)-(C5) with $\bstheta = 
	\bbeta^0_{(\boldsymbol d)}(v) $, and \eqref{eqn:beta_hat_expand_careful} above, we can express
	\begin{align}
	n^{-1/2}\Psi_n(\whbbeta_{(\boldsymbol d)}(v)) = {} & n^{-1/2}\Psi_n(\bbeta^0_{(\boldsymbol d)}(v)) + n^{1/2} {\boldsymbol G}(\bbeta^0_{(\boldsymbol d)}(v))\vecl(\whbbeta_{(\boldsymbol d)}(v) - \bbeta^0_{(\boldsymbol d)}(v)) + o_p(1) \nonumber \\
	= {} & n^{-1/2}\Psi_n(\bbeta^0_{(\boldsymbol d)}(v)) + n^{1/2} {\boldsymbol G}(\bbeta^0_{(\boldsymbol d)}(v)) \boldsymbol L_{(\boldsymbol d)}(v)\vecl(\whbbeta_{(\boldsymbol d)} - \bbeta^0_{(\boldsymbol d)}) + o_p(1) \nonumber \\
	= {} & n^{-1/2}\Psi_n(\bbeta^0_{(\boldsymbol d)}(v))  - n^{-1/2} \boldsymbol M_{(\boldsymbol d)}(v) \Psi_n(\bbeta^0_{(\boldsymbol d)}) + o_p(1)n^{-1/2}\Psi_n( \bbeta^0_{(\boldsymbol d)})  + o_p(1), \label{eqn:gamma_tilde_expand_careful}
	\end{align}
	where ${\boldsymbol G}(\bbeta^0_{(\boldsymbol d)}(v))$ is defined in the same way as ${\boldsymbol G}(\bbeta^0_{(\boldsymbol d)})$ and $$\boldsymbol M_{(\boldsymbol d)}(v) = {\boldsymbol G}(\bbeta^0_{(\boldsymbol d)}(v)) {\boldsymbol L}_{(\boldsymbol d)}(v) \{{\boldsymbol G}(\bbeta^0_{(\boldsymbol d)})^\transp \boldsymbol W {\boldsymbol G}(\bbeta^0_{(\boldsymbol d)})\}^{-1} {\boldsymbol G}(\bbeta^0_{(\boldsymbol d)})^\transp \boldsymbol W.$$
	
	Then gathering \eqref{eqn:gamma_tilde_expand} and \eqref{eqn:gamma_tilde_expand_careful}, we can express
	\begin{align}
	n^{-1/2}\hat{\Psi}_n(\whbbeta_{(\boldsymbol d)}(v)) = {} & n^{-1/2}\Psi_n(\bbeta^0_{(\boldsymbol d)}(v))  - n^{-1/2} \boldsymbol M_{(\boldsymbol d)}(v) \Psi_n(\bbeta^0_{(\boldsymbol d)}) \nonumber\\ 
	& + o_p(1)n^{-1/2}\Psi_n( \bbeta^0_{(\boldsymbol d)})  + o_p(1) \\
	= {} & n^{-1/2}\left\{  \Psi_n(\bbeta^0_{(\boldsymbol d)}(v)) + n \Psi_0(\bbeta^0_{(\boldsymbol d)}(v)) -  \boldsymbol M_{(\boldsymbol d)}(v) \Psi_n(\bbeta^0_{(\boldsymbol d)})   \right\}  \nonumber \\
	& + o_p(1)n^{-1/2}\left\{ \Psi_n(\bbeta^0_{(\boldsymbol d)}) - n \Psi_0(\bbeta^0_{(\boldsymbol d)})\right\} \nonumber \\
	& + n^{1/2}\Psi_0(\bbeta^0_{(\boldsymbol d)}(v)) + o_p(1)n^{1/2}\Psi_0(\bbeta^0_{(\boldsymbol d)}) + o_p(1). \label{eqn:gamma_tilde_expand_careful2}
	\end{align}
	The first term in \eqref{eqn:gamma_tilde_expand_careful2} converges to a mean zero normal random variable and the second term is $o_p(1)$. However the behavior of the third and fourth terms depends on the dimensions $\boldsymbol d$. When $\boldsymbol d = \boldsymbol d^0$, $\Psi_0(\bbeta^0_{(\boldsymbol d)})$ is clearly zero and hence the fourth term is zero. Since the terms $\bbeta^0_{\bfz(\boldsymbol d)}(v)$ are linear combinations of $\bbeta^0_{\bfz(\boldsymbol d)}$, $\Psi_0(\bbeta^0_{(\boldsymbol d)}(v))$ can also be shown to be zero due to the properties of conditional expectation. Thus $\text{VIC}(\boldsymbol d^0) = \log(n)p\sum_{\bfz \in \mathcal{Z}^C}d_\bfz + O_p(1)$. For vectors $\boldsymbol v$ and $\boldsymbol v'$, we say $\boldsymbol v \prec \boldsymbol v'$ if $v_j < v'_j$ for all $j$ and similarly $\boldsymbol v \preceq \boldsymbol v'$ if $v_j \leq v'_j$ for all $j$.
	Now, when $\boldsymbol d \prec \boldsymbol d^0$, $\Psi_0(\bbeta^0_{(\boldsymbol d)}(v))$ is not in general $\boldsymbol 0$ unless by chance $\bbeta^0_{(\boldsymbol d)}(v) = \bbeta^0_{(\boldsymbol {d+1})}$. When $\bbeta^0_{(\boldsymbol d)}(v) \neq \bbeta^0_{(\boldsymbol {d+1})}$, we have $\Psi_0(\bbeta^0_{(\boldsymbol d)}(v)) = \boldsymbol c(v) \neq 0$ , hence the third term in \eqref{eqn:gamma_tilde_expand_careful2}  is of order $n^{1/2}$ and the fourth term is $o_p(n^{1/2})$. Thus, when $\boldsymbol d \prec \boldsymbol d^0$, $\text{VIC}(\boldsymbol d) = r^{-1}n\{ \sum_{j = 1}^r{\boldsymbol c(v_j)}^\transp \boldsymbol c(v_j)  \} + o_{p}(n) + \log(n)p\sum_{\bfz \in \mathcal{Z}^C}d_\bfz$, which clearly dominates $\text{VIC}(\boldsymbol d^0)$ as $n \to \infty$. Simply due to the magnitude of $p\sum_{\bfz \in \mathcal{Z}^C}d_\bfz$, it is straightforward to see that for $\boldsymbol d$ such that $\boldsymbol d^0 \prec \boldsymbol d$,  $\text{VIC}(\boldsymbol d) > \log(n)p\sum_{\bfz \in \mathcal{Z}^C}d^0_\bfz$ and thus $\text{Pr}(\text{VIC}(\boldsymbol d) > \text{VIC}(\boldsymbol d^0)) \to 1$ as $n \to\infty$ . The result is now shown. 
\end{proof}

\section{Additional simulation studies and results}

\subsection{Outcome models for simulation in main text}

The outcome models used in the simulation studies in the main text are as follows.

\resizebox{0.8\linewidth}{!}{
	\begin{minipage}{\linewidth}
		\begin{eqnarray*}
			\begin{matrix}
				& 	\textbf{Model 1}  & 	\textbf{Model 2} & 	\textbf{Model 3}  \\
				\ell_{00}= & 2(\bfX^\transp\bbeta^0_{00})^2                                                 & 2(\bfX^\transp\bbeta^0_{00})^2                                                             &  2(\bfX^\transp\bbeta^0_{00})^2  \\  
				\ell_{10}= & 0.5(  \bfX^\transp\bbeta^0_{00} )^3                                            & \dfrac{1}{0.1 + 0.5(\bfX^\transp\bbeta^0_{00})^2 }  -0.5(\bfX^\transp\bbeta^0_{00})^2       &  0.5(  \bfX^\transp\bbeta^0_{00} )  ( \bfX^\transp\bnu^0_{10} )^2   \\   
				\ell_{01}= & 2\bfX^\transp\bbeta^0_{00}                                                     & 2(\bfX^\transp\bbeta^0_{00})^2                                                             &  \exp\{  \bfX^\transp\bbeta^0_{00}  \}  \\  
				\ell_{11}= & \dfrac{\bfX^\transp\bbeta^0_{00}}{0.5 + (\bfX^\transp\bbeta^0_{00} + 1.5) ^ 2}  & 2(\bfX^\transp\bbeta^0_{00})^2 + (\bfX^\transp\bbeta^0_{00}) (\bfX^\transp\bnu^0_{11}) &  5\exp\{-(\bfX^\transp\bbeta^0_{00})^2\} (\bfX^\transp\bnu^0_{10}) \\  \\
			\end{matrix}
		\end{eqnarray*}
	\end{minipage}
}

The conditions \eqref{eqn:hierarchical_assumption_none} and \eqref{eqn:none_assumption2} are satisfied for all the three data generating models. Throughout the simulations, all of $\bbeta^0_{00}, \bnu^0_{10}$, and $\bnu^0_{11}$  are $p\times 1$ vectors. Half of the elements of each of $\bbeta^0_{00}, \bnu^0_{10}$, and $\bnu^0_{11}$ are from a uniform distribution on $[-0.25, 0.25]$ and the rest from a uniform distribution on $[-0.5, 0.5]$. The true dimensions are $\boldsymbol d^0 = (d^0_{00}, d^0_{10}, d^0_{01}, d^0_{11}) = (1, 1, 1, 1)$  for Model 1, $\boldsymbol d^0 = (1, 1, 1, 2)$  for Model 2, and  $\boldsymbol d^0 = (1, 2, 1, 2)$  for Model 3. For $n_\bfz$ and $p$, we consider scenarios with all combinations of $n_\bfz\in \{200, 600, 1200\}$ and $p\in\{10,20\}$.

We consider the following data-generating scenario for the covariates. Like the setup in \citet{Ma2012}, for $\bfz \in \{ 00, 10, 01, 11\}$, the first $p/2$ columns of $\bfX$ are generated from a normal distribution with mean ${\bm 0}$ and variance-covariance matrix $(\sigma_{ij})_{(p/2)\times(p/2)}$ where $\sigma_{ij} = 0.5^{|i-j|}$. The next $\lfloor p/4\rfloor$ columns $X_{p/2 + i}$, $i = 1, \dots, \lfloor p/4\rfloor$ are generated from Bernoulli distributions with success probabilities $1 / \{1 + \exp(-X_i)\}$ where $\lfloor \cdot \rfloor$ is the floor function. Finally, the last $\lfloor p/4\rfloor$ columns $X_{\lfloor 3p/2 \rfloor + i}$, $i = 1, \dots, \lfloor p/4\rfloor$ are generated as $X_{\lfloor 3p/2\rfloor + i} = X_{\lfloor p/4\rfloor + i} ^ 2 + N(0, 1)$. These covariates do not meet the linearity condition or the constant variance condition required by many SDR estimation methods. 

\subsection{Structural dimension determination results}

To evaluate the performance of the VIC criterion for structural dimension determination for the ``hier sPHD VIC" approach, we considered all combinations of dimensions such that the largest total dimension is less than 5. Note that when there are multiple subpopulations, there are many possible choices of structural dimensions that at least encompass the true underlying structural dimensions. Hence performance in estimating the subspaces may not be harmed much by the inclusion of an extra dimension for some subpopulations.

\begin{figure}[!ht]
	\centering
	\includegraphics[width=0.65\textwidth]{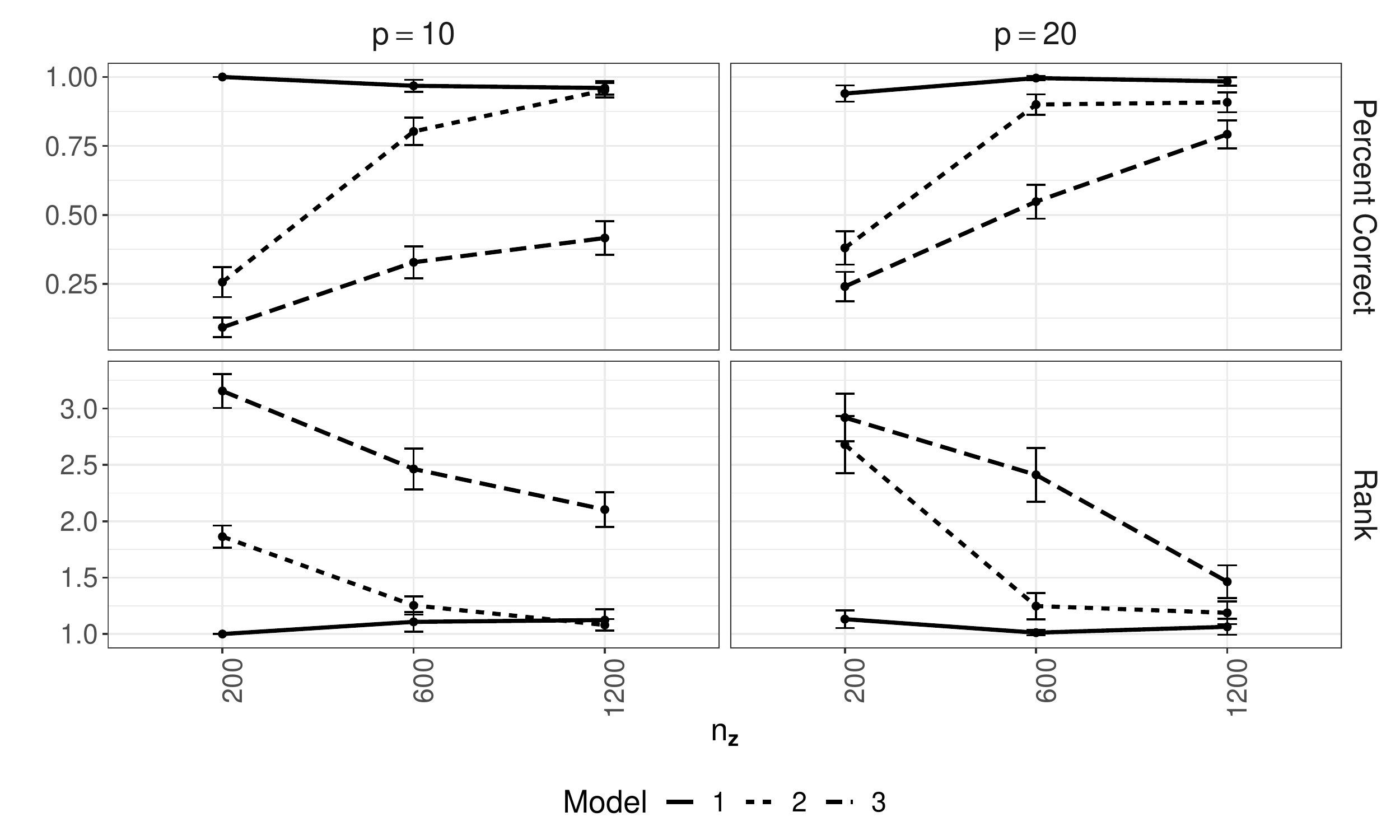}
	\caption{Displayed on the top are the proportion of times over the simulation runs that the exact set of dimensions was correctly selected. Displayed on the bottom are average ranks of the VIC values of the true set of dimensions among the candidate dimension possibilities. }
	\label{fig:sim_dimsel_prob_none}
\end{figure}

In using \eqref{eqn:vic_criterion}, we used $v_1=-1, v_2=-1/2, v_3=0,v_4=1/2$, and $v_5=1$. This value $r = 5$ and the choices for $v_j$ were somewhat arbitrarily, as their particular values are not critically-important and $r=5$ works well in practice.  Results corresponding to the probability of correct selection of the structural dimensions and the rank of the VIC value of the true structural dimensions are displayed in Figure \ref{fig:sim_dimsel_prob_none}. 

In essentially all scenarios, the probability of correct determination of all of the  structural dimensions increases with sample size and the rank of the VIC corresponding to the set of true structural dimensions approaches 1. 

\subsection{Simulation results under hierarchy misspecification}

In the next set of simulations, data were generated from a modification of Model 1 of the simulation setup in the main text. The modification is such that the hierarchical assumptions \eqref{eqn:hierarchical_assumption_none} and \eqref{eqn:none_assumption2} do not hold. We consider three scenarios of varying degrees of hierarchy misspecification.  In all scenarios we allow the hierarchical assumption to hold approximately by setting $\bbeta^0_{\bfz} = \bbeta^0_{00}  +N(0, \tau^2)$ for $\bfz \in \{10, 01, 11\}$.  In Scenario 1 we set $\tau = 0.05$, in Scenario 2 we set $\tau = 0.1$, and in Scenario 3 we set $\tau = 0.25$, thus the scenarios represent increasingly severe misspecification of hierarchy.

\begin{figure}[H]
	\centering
	\includegraphics[width=0.9\textwidth]{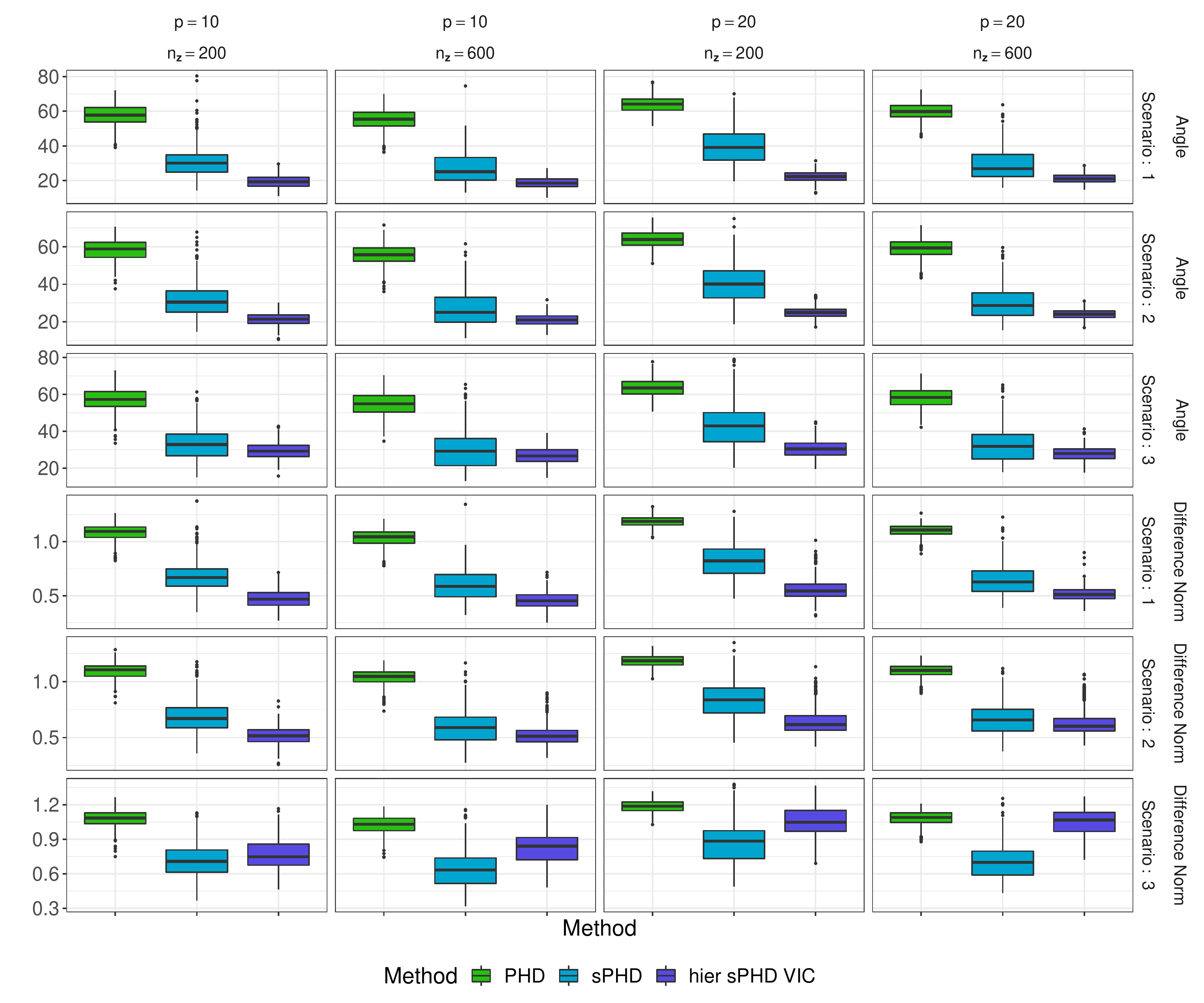}
	\caption{Displayed are the difference norms and angles between true and estimated subspaces for each method over a variety of simulation settings over 250 datasets when the hierarchical assumption is misspecified. All results are for the mixed discrete and continuous covariates setting.}
	\label{fig:sim_model_123_none_norm_missp2}
\end{figure}

We use the proposed VIC dimension selection approach for selecting the structural dimensions and evaluate the resulting estimation performance. The results for the mixed discrete and continuous covariate setting in terms of the difference norm metric and the angle metric are displayed in Figure \ref{fig:sim_model_123_none_norm_missp2}. Since the difference norm $\frac{1}{2^C}\sum_{\bfz \in \mathcal{Z}^C}|| \widehat{\bbeta}_\bfz (\widehat{\bbeta}_\bfz^\transp\widehat{\bbeta}_\bfz)^{-1}\widehat{\bbeta}_\bfz^\transp - \bbeta^0_\bfz({\bbeta^0}_\bfz^\transp\bbeta^0_\bfz)^{-1}{\bbeta^0}_\bfz^\transp  ||_2.$ and the angle between subspaces do not require the dimensions of $\bbeta^0_\bfz$ and $\widehat{\bbeta}_{\bfz}$ to be the same, they can be used to evaluate estimation performance of models with estimated structural dimensions.

In terms of the angle metric, ``hier sPHD VIC" performs the best in all scenarios even under the strongest misspecification setting, demonstrating its capability to reconstruct the true underlying central mean subspaces even if it requires more dimensions than the underlying truth. Despite requiring a larger dimension, ``hier sPHD VIC" is able to recover the true subspace more accurately and with lower variance. Thus, even if the hierarchical assumption is violated to  a certain degree, ``hier sPHD VIC" still performs well. However, since the difference norm penalizes solutions with incorrect dimensions, ``hier sPHD VIC" performs worse in terms of the difference norm metric than the angle metric.

\begin{figure}[H]
	\centering
	\includegraphics[width=0.9\textwidth]{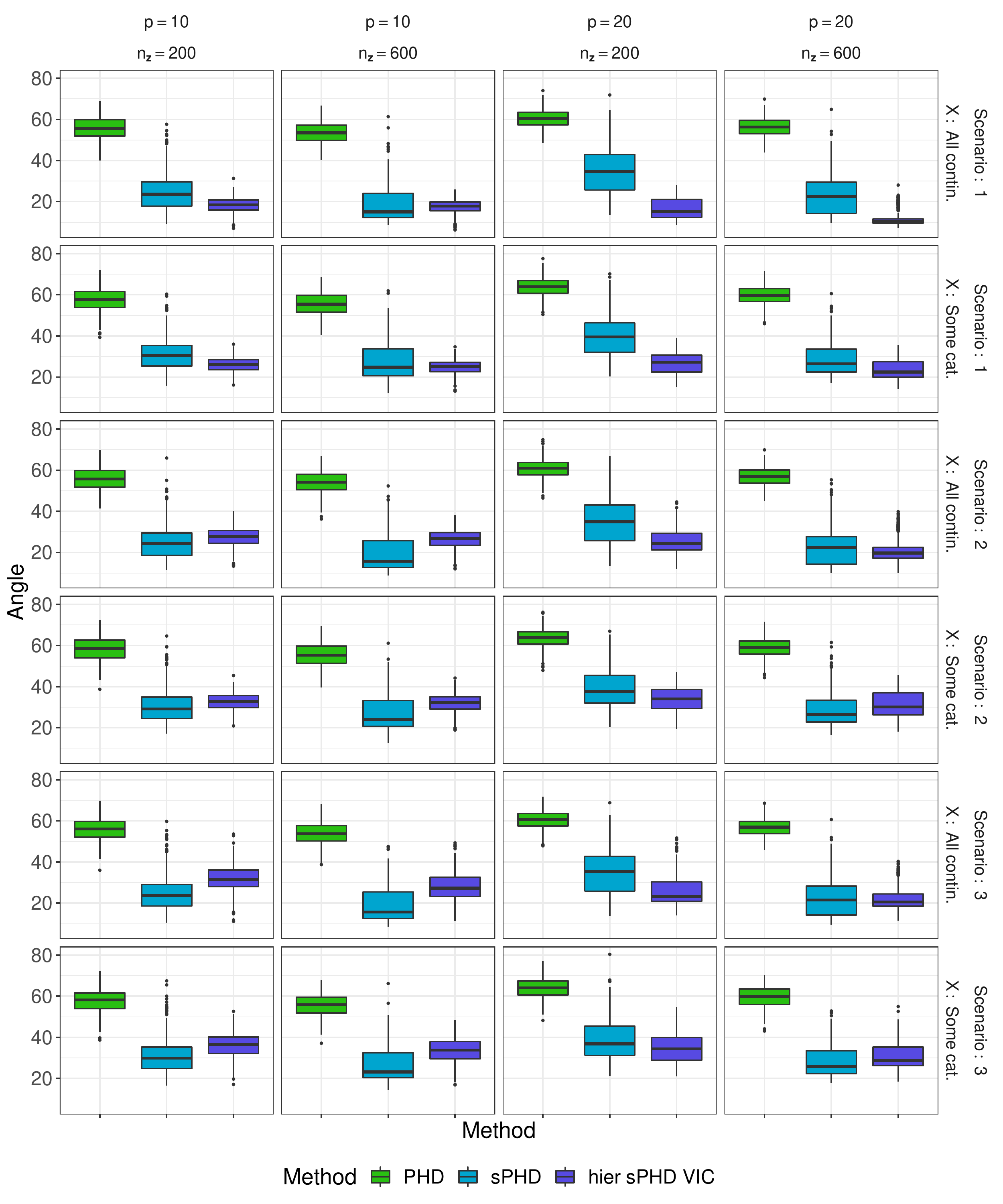}
	\caption{Displayed are the angles between truth and estimate for each method over a variety of simulation settings over 500 datasets when \eqref{eqn:hierarchical_assumption_none} and \eqref{eqn:none_assumption2} are misspecified. }
	\label{fig:sim_model_123_none_angle_missp}
\end{figure}

\begin{figure}[H]
	\centering
	\includegraphics[width=0.9\textwidth]{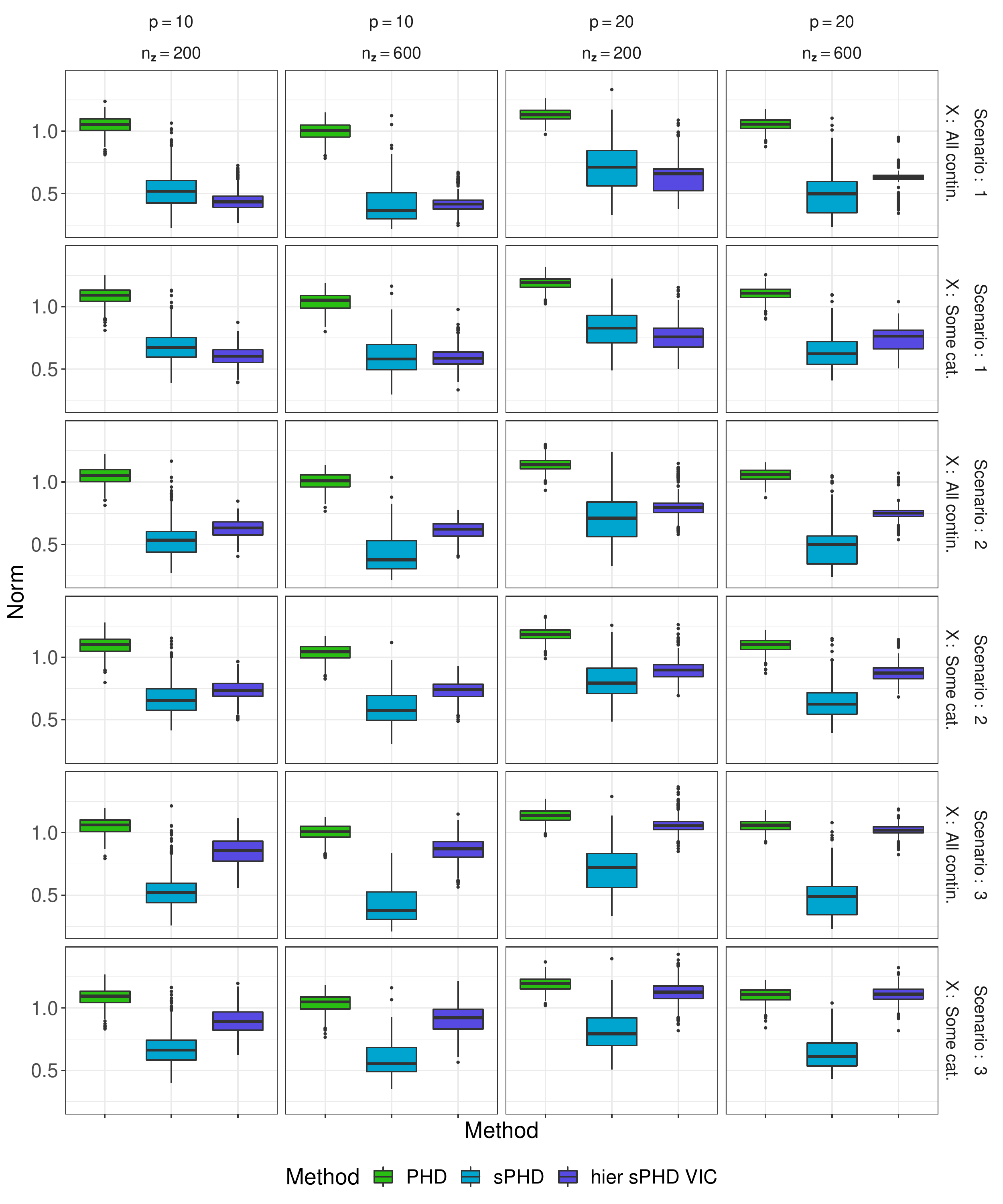}
	\caption{Displayed are the difference norms for each method over a variety of simulation settings over 500 datasets when \eqref{eqn:hierarchical_assumption_none} and \eqref{eqn:none_assumption2} are misspecified. }
	\label{fig:sim_model_123_none_norm_missp}
\end{figure}

Now we consider scenarios with a different form of hierarchy misspecification. This misspecification is more severe than the previously considered hierarchy misspecifications. 
All scenarios are generated under Model 1 of the main text. In Scenario 1, we set $\bbeta^0_{10}$ and $\bbeta^0_{11}$ to be equal to $\bbeta^0_{00}$ and left $\bbeta^0_{01}$ to be completely different. In this scenario the hierarchical assumption is not satisfied. In Scenario 2, we set $\bbeta^0_{10}$  to be equal to $\bbeta^0_{00}$ and left $\bbeta^0_{01}$ and $\bbeta^0_{11}$, to be completely different, resulting in a more severe hierarchy misspecification than Scenario 1. While $\bbeta^0_{01}$ and $\bbeta^0_{11}$ are generated independently of each other, there may exist some overlap of their column spaces as they are from the same distribution and not forced to be orthogonal to each other.  In Scenario 3 all parameters are completely independent of each other,  thus hierarchy does not even hold  approximately. The results in terms of the angle between estimated and true subspaces and the difference norms are displayed in Figures \ref{fig:sim_model_123_none_angle_missp} and \ref{fig:sim_model_123_none_norm_missp}, respectively. For Scenario 3, hier sPHD VIC performs better than sPHD when $p=20$ under the angle metric. Even though the estimated dimensions are larger than the true structural dimensions, the subspaces are still recovered well.

\subsection{Subpopulation specific simulation results}

\begin{figure}[!ht]
	\centering
	\includegraphics[width=1\textwidth]{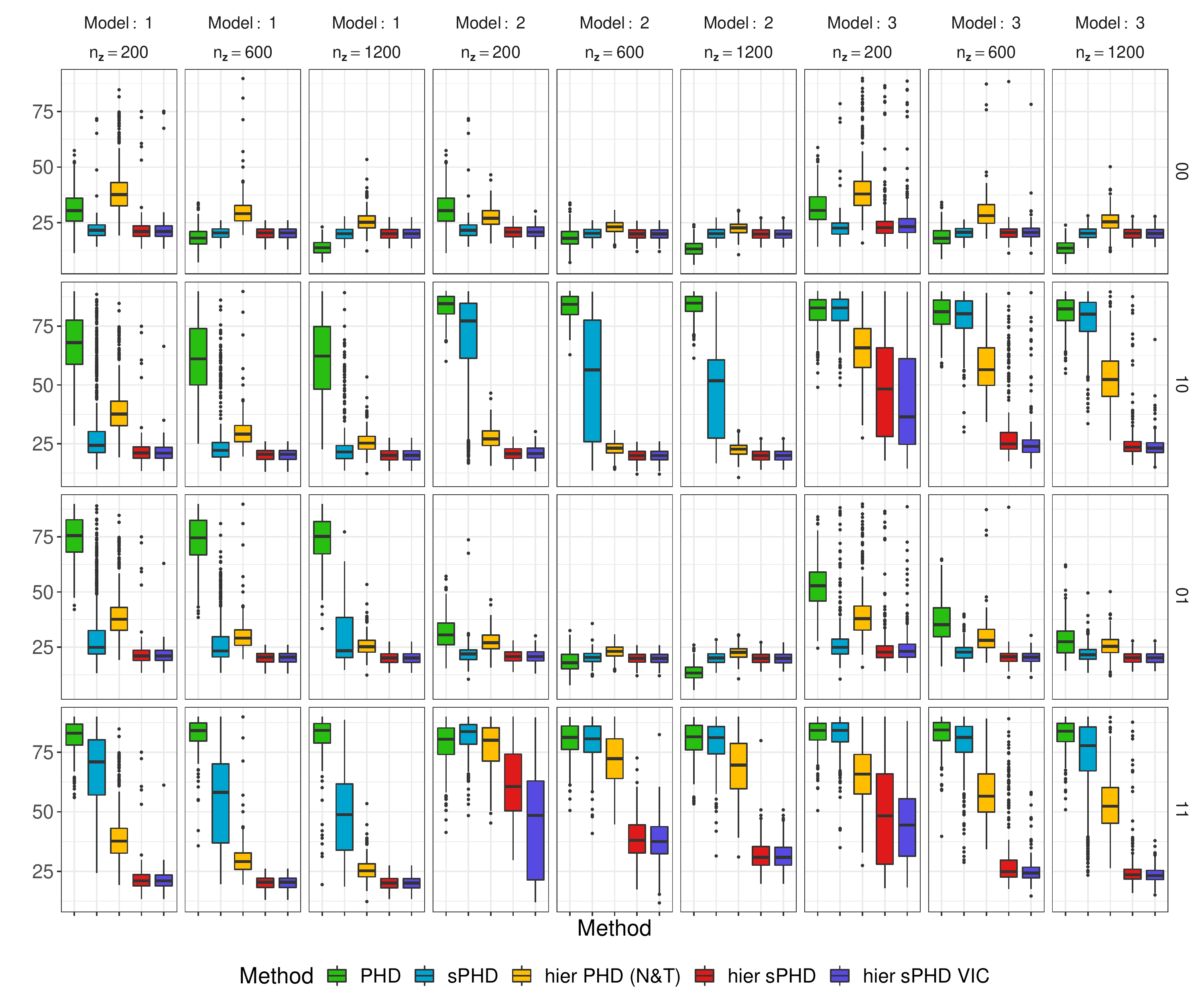}
	\caption{Displayed are the angles between estimated and true subpopulation-specific subspaces for each method over the simulation settings from the main text with $p=20$ over 250 datasets under Models 1, 2, and 3 when the covariates are generated under the all continuous setting. }
	\label{fig:sim_model_123_angle_none_categ_subpop}
\end{figure}

\begin{figure}[!ht]
	\centering
	\includegraphics[width=1\textwidth]{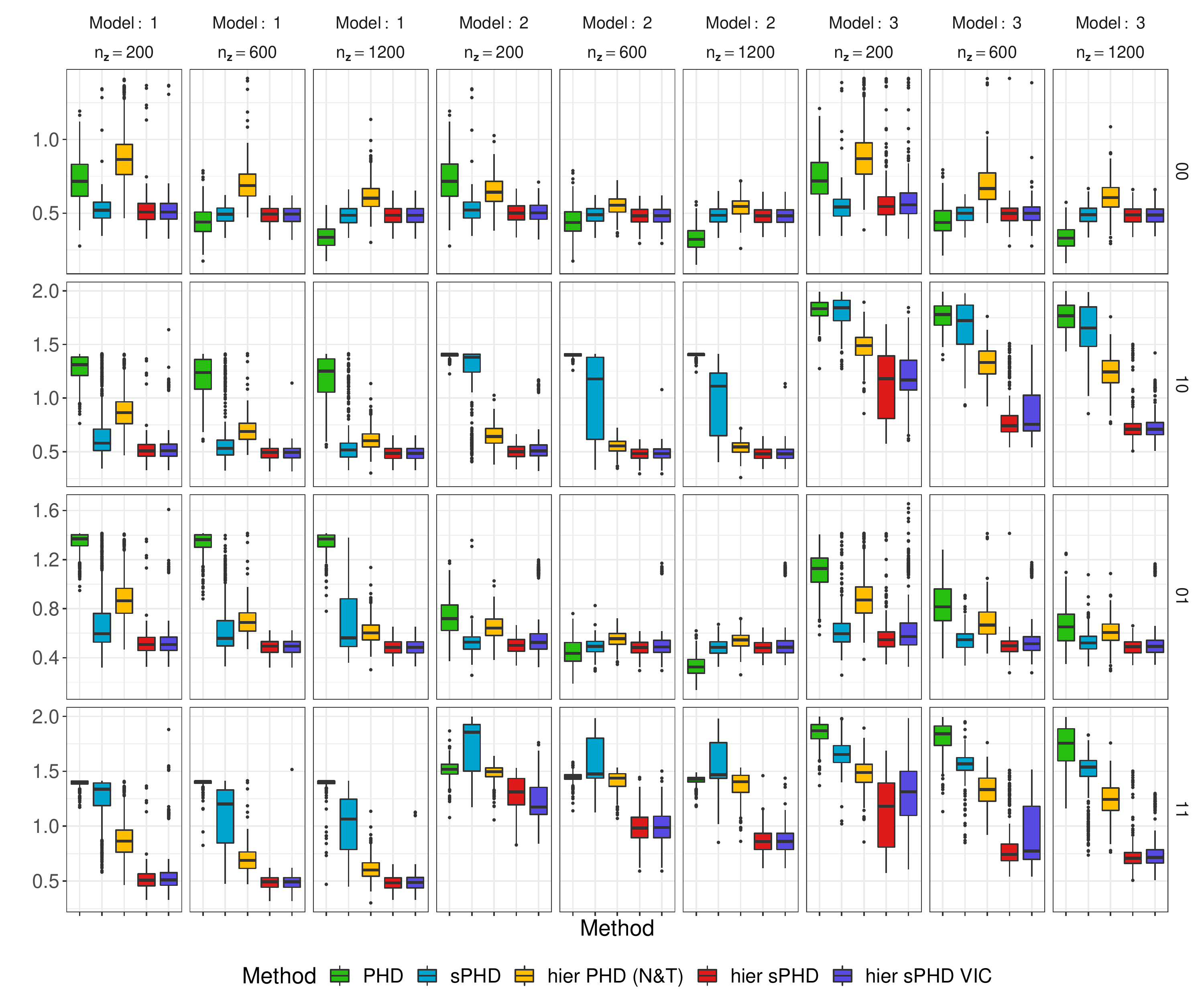}
	\caption{Displayed are the difference norms comparing the estimated and true subpopulation-specific subspaces for each method over the simulation settings from the main text with $p=20$ over 250 datasets under Models 1, 2, and 3 when the covariates are generated under the all continuous setting. }
	\label{fig:sim_model_123_nnorm_none_categ_subpop}
\end{figure}

In this section we provide more detailed information about the simulation from the main text. In particular, instead of computing the norm distance and the angle averaged across the subpopulations, we consider subpopulation-specific difference norms and angles for the simulation settings considered in the main text with dimension $p=20$. We display the results in terms of the angles between the estimated and true subpopulation-specific central mean subspaces in Figure \ref{fig:sim_model_123_angle_none_categ_subpop}. We display the same results in terms of the difference norm metric in Figure \ref{fig:sim_model_123_nnorm_none_categ_subpop}. Except for the ``00'' subpopulation, where the PHD approach works best in several settings, the hier sPHD approaches with known and estimated dimensions generally have far better estimation performance for most subpopulations, especially for subpopulation ``11'' for all three models. The sPHD approach performs reasonably well for the ``01'' subpopulation across all models, but not well for subpopulation ``10'', whereas the N\&T approach works reasonably well for ``10'' but not nearly as well for any of the other subpopulations.

\subsection{Additional simulation results with all continuous covariates}

\begin{figure}[!ht]
	\centering
	\includegraphics[width=0.9\textwidth]{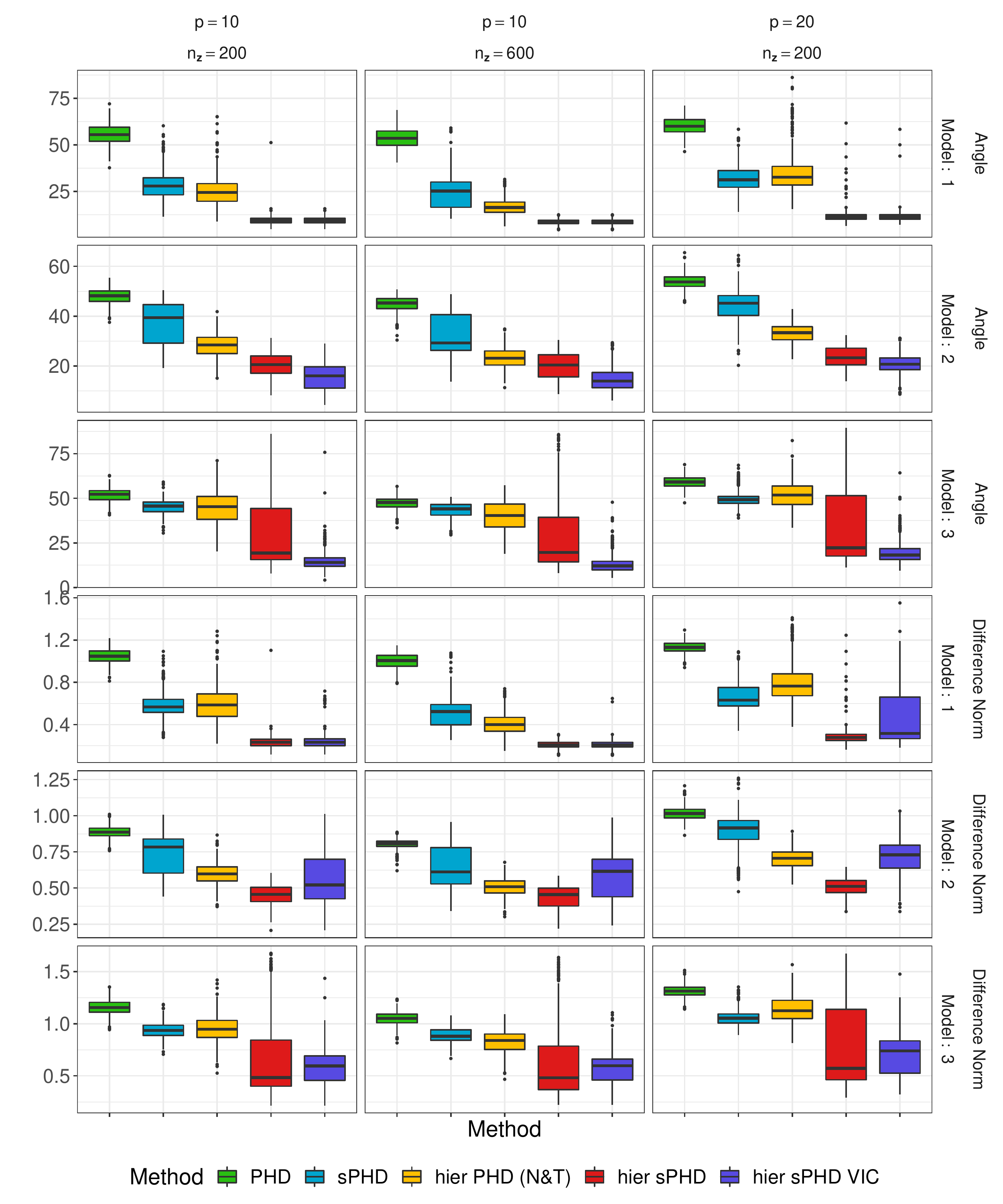}
	\caption{Displayed are the difference norms and angles between estimated and true subspaces for each method over a variety of simulation settings over 250 datasets under Models 1, 2, and 3 when the covariates are generated under the all continuous setting.  All metrics are averages over all subpopulations.}
	\label{fig:sim_model_123_norm_none_contin}
\end{figure}

The results in this section mirror the simulations from the main text except that covariates are  all continuous. Figure \ref{fig:sim_model_123_norm_none_contin} displays the corresponding estimation results. The results are consistent with these when some covariates are discrete and thus we omit further discussion.

\subsection{Simulation with more complex heterogeneity and outcome models}

Our simulation setup in this experiment is identical to the setup of the simulation experiments in the main text, however in this section we explore an additional outcome model that exhibits additional complexity in the structural dimensions and outcome models. In this setting, there is more heterogeneity across subpopulations. The model is displayed below.

\begin{figure}[!ht]
	\centering
	\includegraphics[width=0.95\textwidth]{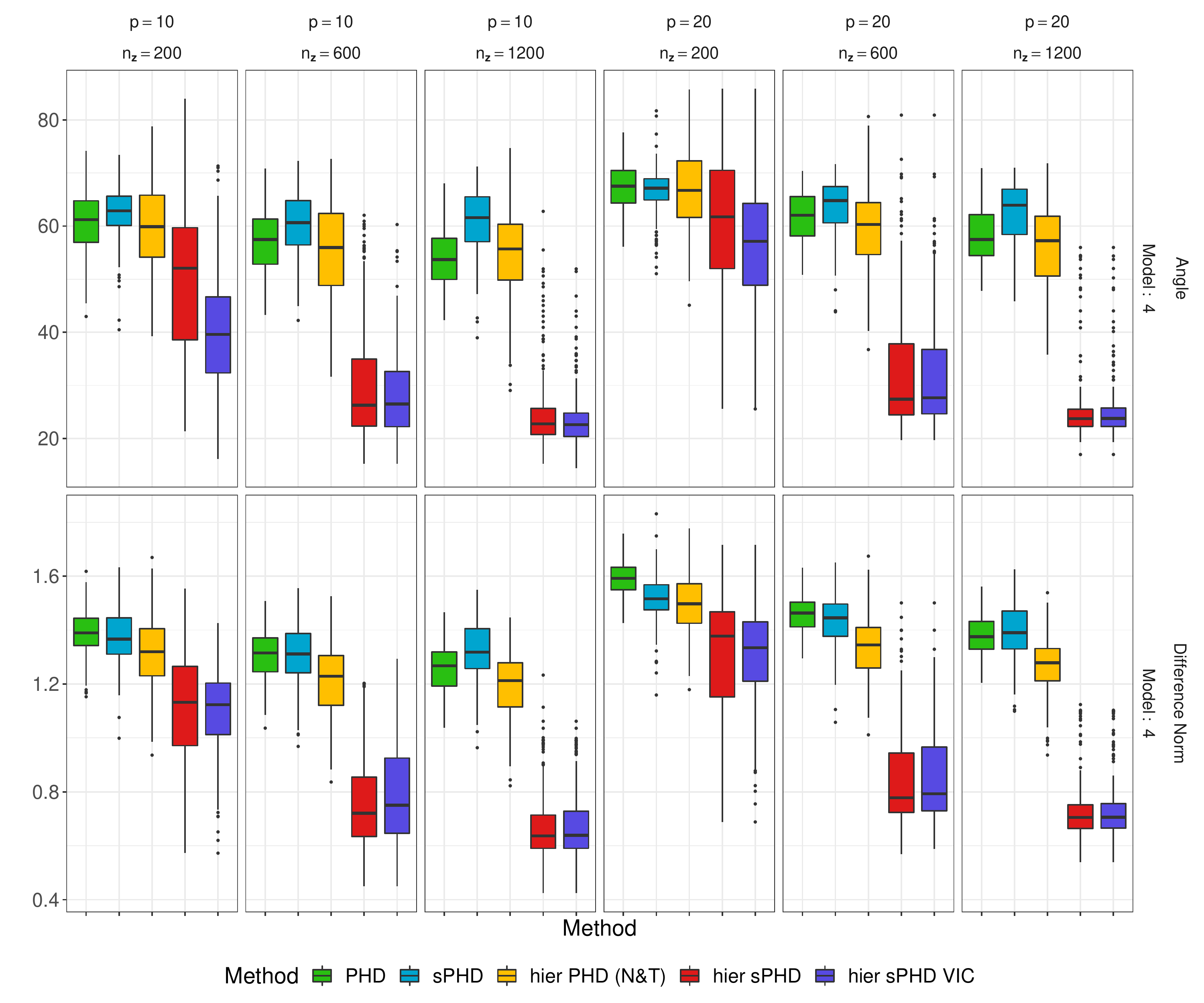}
	\caption{Displayed are the difference norms and angles between estimated and true subspaces for each method over a variety of simulation settings over 250 datasets under Model 4 when the covariates are generated under the mixed discrete and continuous setting. All metrics are averages over all subpopulations.}
	\label{fig:sim_model_4_norm_none}
\end{figure}

\begin{figure}[!ht]
	\centering
	\includegraphics[width=1\textwidth]{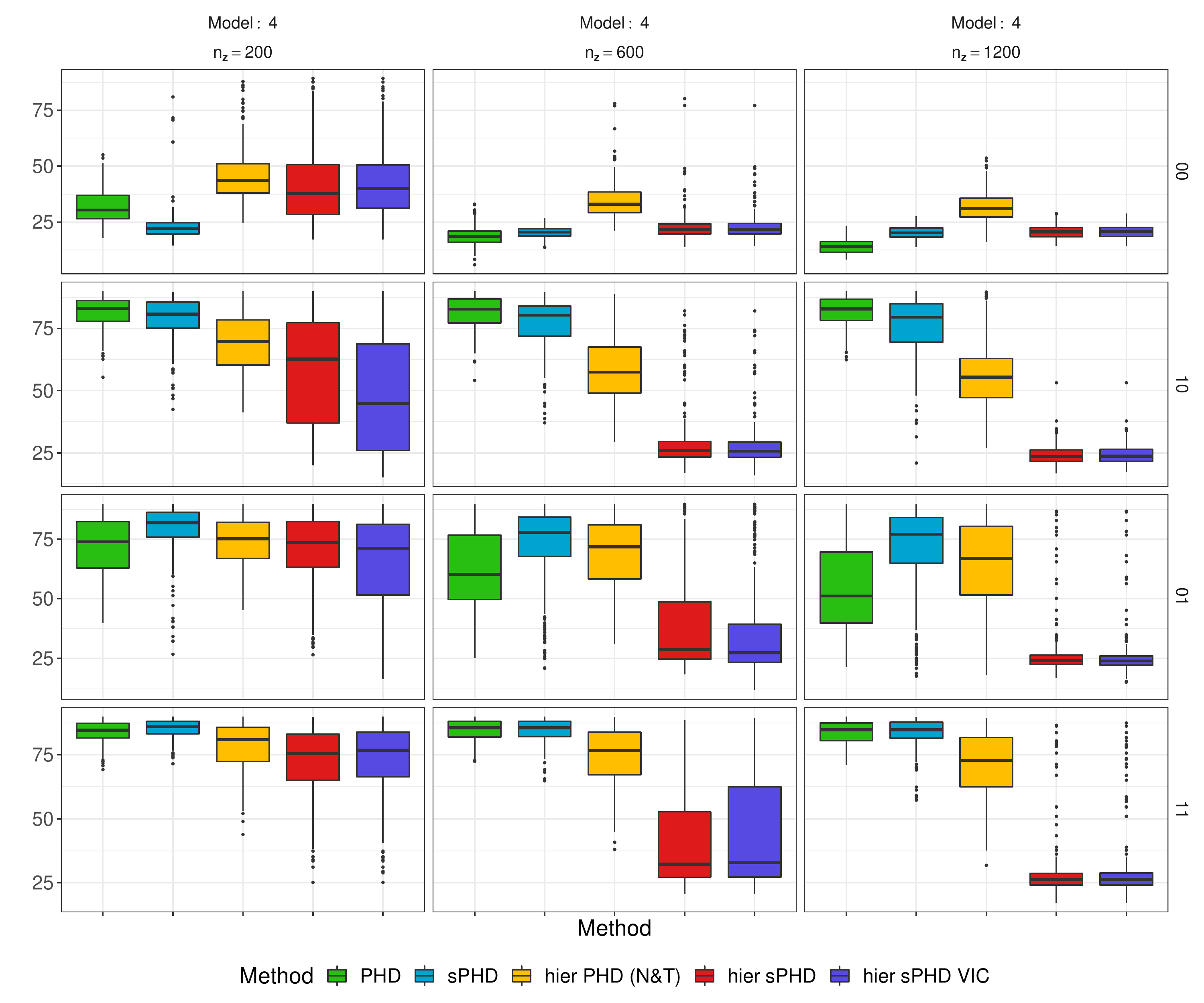}
	\caption{Displayed are the angles between estimated and true subpopulation-specific subspaces for each method over the simulation settings from the main text with $p=20$ over 250 datasets under Model 4 when the covariates are generated under the all continuous setting. }
	\label{fig:sim_model_4_angle_none_categ_subpop}
\end{figure}

\begin{figure}[!ht]
	\centering
	\includegraphics[width=1\textwidth]{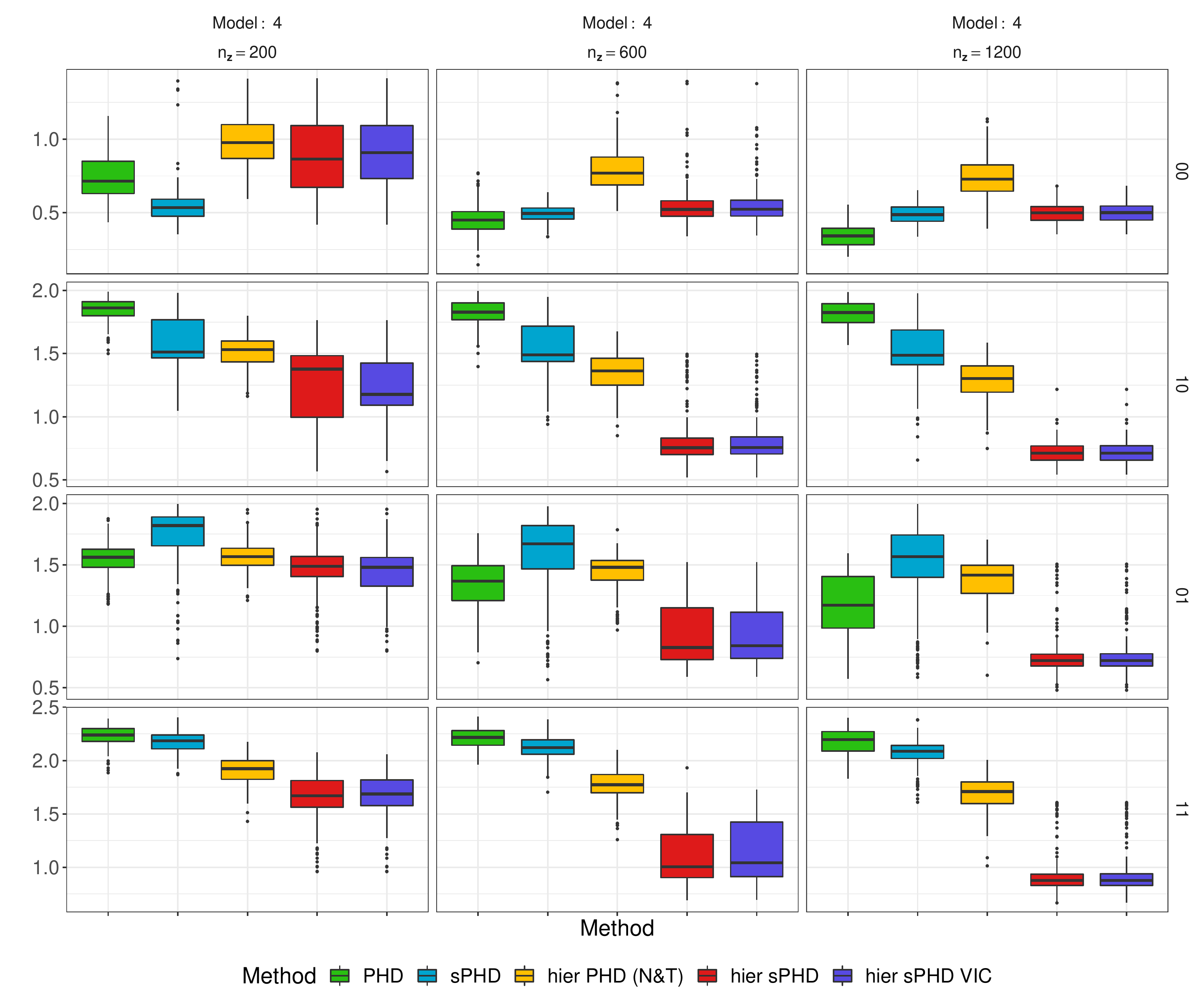}
	\caption{Displayed are the difference norms comparing the estimated and true subpopulation-specific subspaces for each method over the simulation settings from the main text with $p=20$ over 250 datasets under Model 4 when the covariates are generated under the all continuous setting. }
	\label{fig:sim_model_4_nnorm_none_categ_subpop}
\end{figure}

\begin{minipage}{\linewidth}
	\begin{eqnarray*}
		\begin{matrix}
			& 	\textbf{Model 4}  \\
			\ell_{00}= &  2(\bfX^\transp\bbeta^0_{00})^2  \\  
			\ell_{10}= &   0.5(  \bfX^\transp\bbeta^0_{00} )  ( \bfX^\transp\bnu^0_{10} )^2 + 5 \bfX^\transp\bnu^0_{10}   \\   
			\ell_{01}= &   \exp\{  \bfX^\transp\bbeta^0_{00}  \}+( \bfX^\transp\bnu^0_{01} )^2  \\  
			\ell_{11}= &   5\exp\{-(\bfX^\transp\bbeta^0_{00})^2\} (\bfX^\transp\bnu^0_{10}) + 5\sin(2\pi \bfX^\transp \bnu^0_{01}) \\  \\
		\end{matrix}
	\end{eqnarray*}
\end{minipage}
In this model, the true structural dimensions are $\boldsymbol d^0 = (1, 2, 2, 3)$. The average signal-to-noise ratio under this model is around 6.5. The results over 250 simulation replications are displayed below. From Figure \ref{fig:sim_dimsel_prob_none_model4}, we can see that the structural dimensions are recovered frequently and that the correct recovery probability tends to 1 with increasing sample size. From Figure \ref{fig:sim_model_4_norm_none}, which displays the results in terms of the angle and difference norm  averaged over all the subpopulations, we can see that the proposed approach works quite well with both true and estimated structural dimensions and outperforms other approaches. The subpopulation-specific results are displayed in Figures \ref{fig:sim_model_4_angle_none_categ_subpop} and \ref{fig:sim_model_4_nnorm_none_categ_subpop}. From these results, we can see that while the hier sPHD method does not perform best for the ``none'' subpopulation, it is substantially better than other approaches for all sample sizes for the other three subpopulations.

\begin{figure}[!ht]
	\centering
	\includegraphics[width=0.85\textwidth]{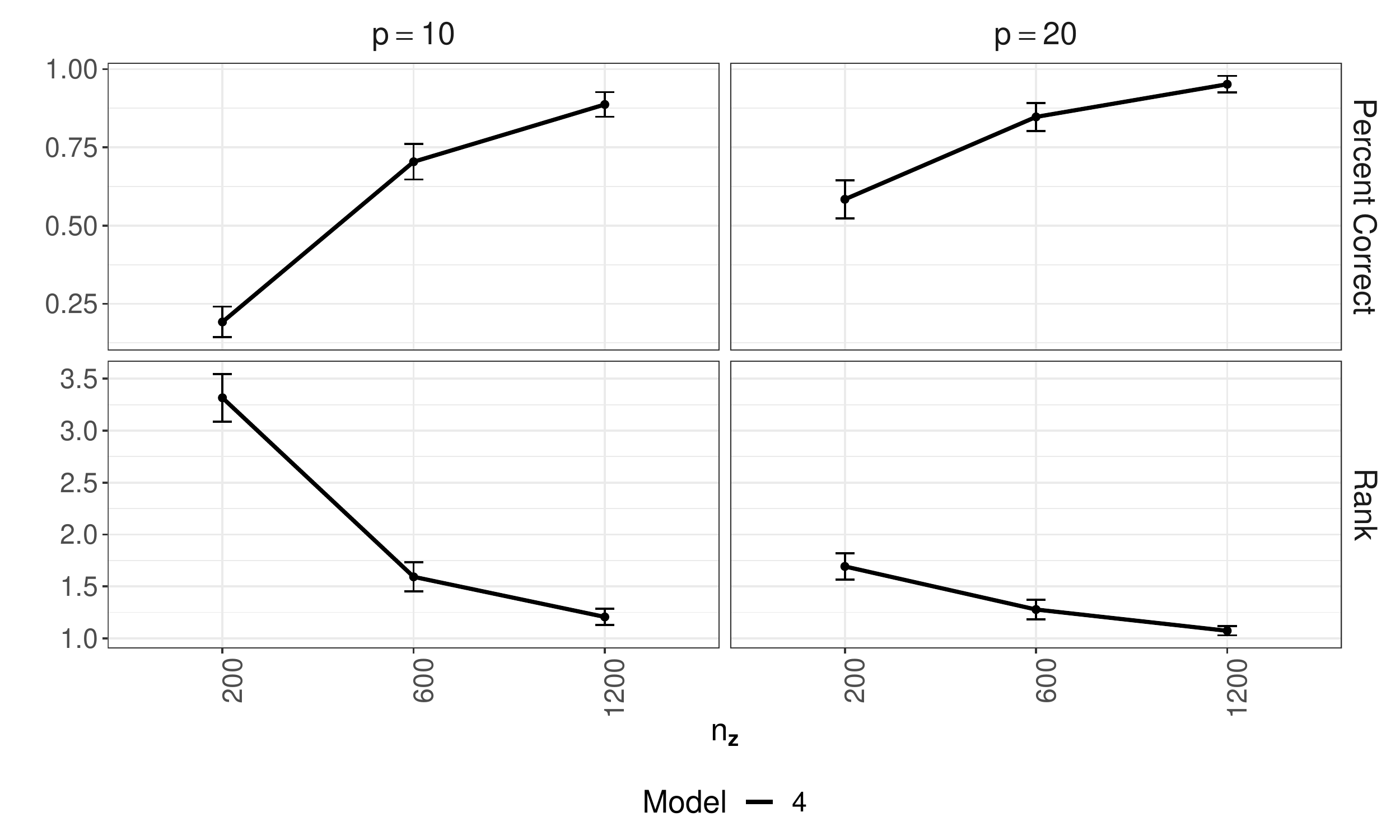}
	\caption{Displayed on the top are the proportion of times over the simulation runs that the exact set of dimensions was correctly selected. Displayed on the bottom are average ranks of the VIC values of the true set of dimensions among the candidate dimension possibilities. }
	\label{fig:sim_dimsel_prob_none_model4}
\end{figure}

\clearpage

{
 \bibliographystyle{apalike}
 \bibliography{sdr_hierarchical_main_biom}
 }

\makeatletter\@input{xx.tex}\makeatother